\documentclass[preprints,article,accept,moreauthors,pdftex]{Definitions/mdpi} 

\firstpage{1} 
\pubvolume{1}
\issuenum{1}
\articlenumber{0}
\pubyear{2021}
\copyrightyear{2021}
\datereceived{} 
\dateaccepted{} 
\datepublished{} 
\hreflink{https://doi.org/} 

\usepackage{bbm}
\usepackage{braket}
\usepackage{hyperref}
\usepackage{longtable}
\usepackage{mathtools}
\usepackage{physics}
\usepackage{xcolor} 
\usepackage{ytableau}

\usepackage{longtable}

\Title{Quantum Foundations of Classical Reversible Computing}

\TitleCitation{Quantum Foundations of Classical Reversible Computing}

\Author{Michael P. Frank $^{1,\dagger,}$*\orcidA{} and Karpur Shukla $^{2,\dagger,}$*\orcidB{}}

\AuthorNames{Michael P. Frank and Karpur Shukla}

\AuthorCitation{Frank, M.P.; Shukla, K.}

\address{%
$^{1}$ \quad Center for Computing Research, Sandia National Laboratories; mpfrank@sandia.gov\\
$^{2}$ \quad Department of Electrical and Computer Engineering, Brown University; karpur\_shukla@brown.edu}

\corres{Correspondence: mpfrank@sandia.gov, karpur\_shukla@brown.edu; Tel.: +1-505-284-4103 (M.F.), +1-646-580-5277 (K.S.)}

\firstnote{These authors contributed equally to this work.} 

\abstract{The reversible computation paradigm aims to provide a new foundation for general classical digital computing that is capable of circumventing the thermodynamic limits to the energy efficiency of the conventional, non-reversible digital paradigm. However, to date, the essential rationale for and analysis of classical reversible computing (RC) has not yet been expressed in terms that leverage the modern formal methods of non-equilibrium quantum thermodynamics (NEQT). In this paper, we begin developing an NEQT-based foundation for the physics of reversible computing. We use the framework of Gorini-Kossakowski-Sudarshan-Lindblad dynamics  (a.k.a.\ \emph{Lindbladians}) with multiple asymptotic states, incorporating recent results from resource theory, full counting statistics, and stochastic thermodynamics. Important conclusions include that, as expected: (1) Landauer's Principle indeed sets a strict lower bound on entropy generation in traditional non-reversible architectures for deterministic computing machines when we account for the loss of correlations; and (2) implementations of the alternative {\em reversible} computation paradigm can potentially avoid such losses, and thereby circumvent the Landauer limit, potentially allowing the efficiency of future digital computing technologies to continue improving indefinitely. We also outline a research plan for identifying the fundamental minimum energy dissipation of reversible computing machines as a function of speed.}
\keyword{non-equilibrium quantum thermodynamics; thermodynamics of computing; Landauer's principle; Landauer limit; reversible computing; resource theory of quantum thermodynamics; Gorini-Kossakowski-Sudarshan-Lindblad dynamics; Lindbladians; von Neumann entropy; Rényi entropy; open quantum systems}

\newcommand{\mypara}[1]{\paragraph{\arabic{section}.\arabic{subsection}.\arabic{subsubsection}.\arabic{paragraph}. #1}}

\begin{document}




\section{Introduction}
\label{sec:intro}
The concept of \emph{reversible computation}, or computation without information loss (even locally), played a centrally important role in the historical development of the thermodynamics of computation \cite{Lan61,Ben73,Ben82,BL85,Lan87,Ben88,Ben03}. It remains critically important today in the field of quantum computing, where it is necessary for maintaining coherence in quantum algorithms \cite{NC00}. However, the original motivation for reversible computation, which was to circumvent the $k_{\mathrm{B}}T\ln 2$ \emph{Landauer limit}\footnote{In this expression, $k=k_\mathrm{B}\simeq 1.38\times 10^{-23}\;\mathrm{J}/\mathrm{K}$ is Boltzmann's constant, which is the natural logarithmic unit of entropy \cite{Fra05}, and $T$ is temperature.} on energy dissipation in \emph{classical} digital computing, is less often remembered today. Some authors have critiqued the original arguments for Landauer's limit and reversible computing as relying on equilibrium assumptions (\emph{e.g.}, \cite{Wol19a}), but in fact, no such assumption beyond the existence of an \emph{external} heat sink at some temperature $T$ is required. When properly stated and interpreted, Landauer's limit holds regardless of whether the computing system is at (or even \emph{close} to) equilibrium \emph{internally}. This statement follows directly from elementary statistical physics and information theory \cite{Fra18}. 

Indeed, Landauer's limit has also been derived directly for systems \emph{out} of equilibrium \cite{Goold15, GCGPVP17}. This nonequilibrium limit is expressed purely in terms of the non-unitality of the quantum channel evolving the system and heat bath. In other words, Landauer's limit has been derived \emph{solely} as a consequence of thermal operations (as defined in NEQT) acting on the joint quantum mechanical evolution of a system and a bath. This directly reinforces the motivation for reversible computing, which is to avoid the Landauer cost of ejecting correlated information into the environment. The free energy\footnote{The free energy referred to here is the $\alpha = 1$ free energy in particular; \emph{i.e.}, the (nonequilibrium) Helmholtz free energy. In the words of \cite{Mueller18} directly, that work shows that the most general possible type of catalytic thermal operation (and thus the most general type of transition possible in quantum thermodynamics) ``restores the distinguished role of the Helmholtz free energy.''} cost of operations that do \emph{not} eject correlated information can be made arbitrarily small, a fact rigorously proven using resource theoretic techniques in NEQT \cite{Mueller18}. We discuss these connections in some detail in later sections. Further, the enterprise of recasting the classic understanding of the thermodynamics of computing in more modern terms offers other benefits. In particular, it allows the theoretical apparatus of the modern NEQT formalism to be brought to bear on the problem of analyzing the potential capabilities of, and fundamental limits on, classical reversible computational processes.

This problem is of far more than just academic interest. Today, an increasingly serious concern is that the conventional non-reversible paradigm for general digital computation is approaching firm limits to its energy efficiency and cost efficiency. These limits ultimately trace back to the $k_{\mathrm{B}}T$ thermal energy scale. Since reversible computing is, broadly speaking, the only non-conventional computing paradigm that can potentially offer a sustainable path forward capable of circumventing the efficiency limits associated with that energy scale in \textit{general digital} computing, it is therefore critically important to the prospects for medium- and long-term improvement in the efficiency and economic utility of general digital computing to determine exactly what the potentialities and limitations of reversible computational mechanisms may be, according to fundamental theory.

In this paper, we aim to carry out the essential groundwork for this enterprise, laying down low-level theoretical foundations upon which a comprehensive NEQT-based treatment of physical mechanisms for reversible computation may be based. It is essential for any such effort to identify the most appropriate definitions for key concepts, and we do this throughout, taking special care with the definitions of the appropriate physical concepts corresponding to classical digital computational states and operations. On this ground, we advocate for our position that the most appropriate understanding of Landauer's Principle is to view it as comprising, most essentially, a statement about the strict entropy {\em increase} that is required due to the loss of mutual information that necessarily occurs whenever (nontrivially) deterministically computed (ergo correlated) bits are thermalized in isolation. There are other, oft-cited forms of the Principle that deal only with a {\em transfer} of entropy between computational and non-computational forms; but we instead refer to these as {\em The Fundamental Theorem of the Thermodynamics of Computation} to avoid confusion, since it has long been known that simple transfers of entropy between different forms can occur in a thermodynamically reversible way \cite{Ben03}. The inappropriate conflation of Landauer's Principle proper, as we identify it, with the Fundamental Theorem is what we believe has been the root cause of a great deal of confusion in the thermodynamics of computing field. As we suggest, simply appropriately distinguishing these concepts permits the straightforward resolution of many long-standing controversies.

Another central aim of this work is to go beyond discussions of Landauer's limit, to develop a first-principles model of classical RC operations using information-theoretic techniques in nonequilibrium quantum thermodynamics. These techniques allow us to understand the fundamental quantum mechanical expressions of, and restrictions on, classical RC operations in several ways. From resource theory and fluctuation theorems \cite{Mueller18,FUS18}, these techniques provide us with a way of understanding the overall limitations of state transitions, including those that correspond to classical RC operations. In addition to constraints, the framework of Gorini-Kossakowski-Sudarshan-Lindblad operators (GKSL operators, also known as Lindbladians) with multiple asymptotic states \cite{Albert14,ABFJ16,Albert18} offer a framework by which \emph{explicit} nonequilibrium quantum thermodynamic expressions of classical RC operations can be realized. As such, these techniques offer a natural language for expressing the dynamics of RC operations, and provide us with an understanding of the fundamental quantum mechanical restrictions on the way these operations can manifest in physical systems. Fundamental bounds on quantities of interest, such as the dissipation of an operation as a function of its speed, will necessarily have to arise from NEQT.

Here, we provide a description of RC operations via the theory of open quantum systems. In this formulation, we can examine the joint evolution of the computing system with a thermal environment (a.k.a.\ heat bath), using the machinery of completely positive trace preserving maps (CPTP maps, a.k.a.\ quantum channels). In particular, we rely on the framework of GKSL operators with multiple asymptotic states \cite{Albert14,ABFJ16,Albert18} to develop representations of classical reversible information processing operations. Quite powerfully, this framework can directly give us bounds on dissipation quantities of interest not only for RC operations, but for quantum computation (QC) operations as well---since we express RC operations in terms of quantum channels, the results we derive for RC operations can be directly extended to QC operations in future work.

Nonequilibrium quantum thermodynamics is, we believe, a natural and proper language for understanding the fundamental principles of reversible computing, and for expressing RC operations. As such, a broader aim of this work is to bridge outstanding gaps between the NEQT and RC communities. By providing RC practitioners a feel for some of the modern tools used in NEQT and expressing familiar RC concepts in this language, and by providing NEQT practitioners a sense of how RC principles arise naturally from familiar NEQT frameworks, we hope to achieve this synthesis. As such, this presentation is intended to be a brief and self-contained exposure to some of the basic concepts of NEQT. For further reading, \cite{DC19} provides a highly pedagogical introduction, while \cite{BCGAA18} gives a clear and comprehensive overview of current major topics.

Our framework of expressing RC in NEQT relies on the quantum information perspective of thermodynamics (comprehensively reviewed in \cite{GHRDS16}), the resource theory of quantum thermodynamics (comprehensively reviewed in \cite{NW18,Lostalgio19,CG19}), and the theory of open quantum systems (reviewed in \cite{Alicki07}, comprehensively discussed in \cite{Breuer07,Banerjee18}, extended to multiple asymptotic states in \cite{Albert14,ABFJ16,Albert18}). The concepts of quantum speed limits and shortcuts to adiabaticity are not discussed here, but will appear in future work; these are comprehensively reviewed in \cite{DC17} and \cite{GORKTMGM19}, respectively. Readers interested in greater detail on this framework are highly encouraged to read these references. Readers unfamiliar with quantum information theory are also encouraged to refer to \cite{NC00,NRS07,Wolf12,Attal14,Wilde17,Preskill19}.

Also, we would like to emphasize that in this paper, we are manifestly taking a stance towards the thermodynamics of information that treats systems as evolving \textit{autonomously}, that is, without invoking the concept of an ``observer'' outside the system that is performing measurements on and/or controlling the system. This is necessary in order to construct a complete, coherent treatment of self-contained physical computing systems. Other examples of work that takes a self-contained/autonomous perspective to the physics of information include \cite{DJ13,BS14,SSBE17}.

The structure of the remainder of the paper is as follows. Section~\ref{sec:mat-meth} describes materials and methods, including outlining a broad theoretical framework in \S\ref{ssec:found}, relating that broad framework to the more detailed tools and methods of NEQT in \S\ref{ssec:neqt}, and reviewing a variety of existing and proposed physical implementation technologies for reversible computing in \S\ref{ssec:tech}. Section~\ref{sec:res} presents our early results, specifically reviewing how a few classic theorems can be easily proven in our framework. These include (\S\ref{ssec:fun-thm}) The Fundamental Theorem of the Thermodynamics of Computing, which we distinguish from (\S\ref{ssec:lan-prin}) Landauer's Principle (properly stated); (\S\ref{ssec:rc-thms}) fundamental theorems of traditional and generalized reversible computing; and  (\S\S\ref{ssec:ctos-rc}--\ref{ssec:rc-op-rep}) the representation of classical reversible computational operations via the frameworks of catalytic thermal operations and GKSL dynamics. Section~\ref{sec:disc} gives some general discussion of results and outlines our research plan looking forwards, and Section~\ref{sec:conc} concludes.

\section{Materials and Methods}
\label{sec:mat-meth}

As this article presents theoretical, rather than experimental work, there is no laboratory apparatus to speak of; however, we provide a brief review of some of the existing and proposed physical implementation technologies for reversible computing in \S\ref{ssec:tech}. But first, we present the key foundational definitions of our broad theoretical picture in \S\ref{ssec:found}. Please note that this presentation roughly follows, but expands upon, that given in \cite{Fra18,Fra17a}. Then in \S\ref{ssec:neqt}, we tie this broad picture to the much more detailed theoretical apparatus of NEQT.

\subsection{Broad Theoretical Foundations}
\label{ssec:found}

In this subsection, we present and review a number of important low-level definitions that form the broad foundation upon which our overall approach to the physics of reversible computing is built. This includes (\S\ref{sssec:open-q}) our overall picture based on a framework of open quantum systems; (\mbox{\S\ref{sssec:comp-states}}) our definition of classical digital computational states and their physical representation, which invokes what we call a {\em proto-computational basis}, which may in general be time-dependent; (\S\ref{sssec:comp-oper}) our definitions for classical computational operations, and different types of operations, which are expressible in terms of (\S\ref{sssec:comp-trans}) primitive computational state transitions; and (\mbox{\S\ref{sssec:corresp}}) the appropriate definition of what it means for a given unitary quantum time evolution to \textit{implement} a classical computational operation.

\subsubsection{Open Quantum Systems Framework}
\label{sssec:open-q}

In this subsection, we briefly review the broad outlines of the open quantum systems based picture that we are using in this paper. Further details are developed in \S\ref{ssec:neqt}.

\mypara{System and environment.}\label{par:sys-env}
We begin with a fairly conventional picture of a physical computer system based on an open quantum systems perspective. At the highest level, we assume that the model universe $\mathfrak{U}$ under study can be described as a composition of two subsystems $\mathfrak{S},\mathfrak{E}$, where $\mathfrak{S}$ is the physical computer system in question, and $\mathfrak{E}$ is its external environment (\emph{i.e.}, the rest of $\mathfrak{U}$, outside of $\mathfrak{S}$).
As an example, one could define the ``computer system'' $\mathfrak{S}$ as consisting of everything (\emph{i.e.}, all quantum fields) encompassed within some region of (3+1)D spacetime circumscribed by some closed (2+1)-dimensional bounding surface. For simplicity, one could think of a spatial boundary that is unchanging in time over some interval.
Typically, our analyses will treat the environment $\mathfrak{E}$ as an (effectively infinitely large) uniform thermal reservoir (heat bath) that is internally at thermal equilibrium, at some (effectively constant over time) temperature $T$\@. The temperature may be treated as effectively constant when the environment is large enough that its temperature is negligibly affected by heat transferred from $\mathfrak{S}$.\footnote{Of course, this model is already somewhat of an idealization, since a typical real environment would attain a nonuniform temperature profile under a steady-state thermal flow with constant power output from $\mathfrak{S}$, but it can nevertheless be considered an adequate model for an initial study.}

\begin{figure}[t] 
\centerline{\includegraphics[width=5 cm]{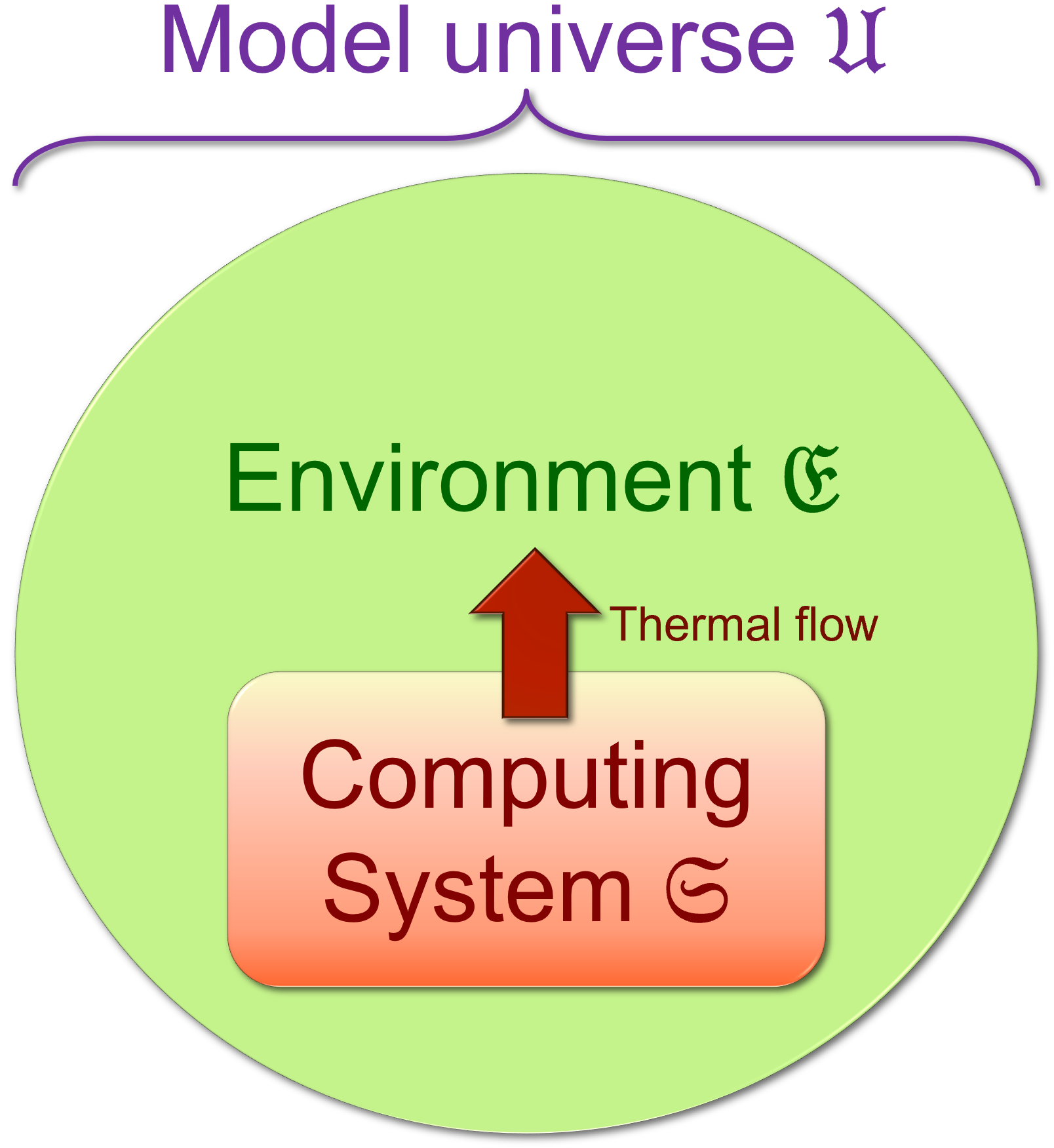}}
\caption{Simplified picture of our model universe $\mathfrak{U}$ in an open quantum systems framework. Power supplies and waste heat removal mechanisms are assumed to be included within the physical computer system $\mathfrak{S}$. In general, we may assume there is a flow of waste heat from the system out to an (assumed very large) external heat bath $\mathfrak{E}$.\label{fig:model-u}}
\end{figure}

Meanwhile, we will treat the computer system $\mathfrak{S}$ as an (in general) \emph{non-equilibrium} system which includes its own internal supply of free energy (\emph{e.g.}, this could be a battery or a fuel reservoir). This is just a simplification of the overall picture, for our present purposes, to avoid the need to explicitly represent a flow of work or free energy in from a separate ``power supply'' system; \emph{i.e.}, the power supply is treated as internal to the computer.
However, we will allow that the system $\mathfrak{S}$ is able to exchange thermal energy (and entropy) with its environment $\mathfrak{E}$ through (all or some portion of) its boundary. Typically, the computer would be assumed to expel waste heat to its external environment $\mathfrak{E}$ during operation in order to maintain its ($\mathfrak{S}$'s) own internal operating temperature (which will generally be non-uniform) within some reasonable bounds. The mechanisms for managing the needed thermal flows are generally assumed to be contained within $\mathfrak{S}$. See Figure~\ref{fig:model-u}.

\mypara{Decoherence model.}\label{par:decoher}
A further important simplifying assumption is that we, as modelers, cannot effectively track any (classical or quantum) correlations between the detailed states of $\mathfrak{S}$ and $\mathfrak{E}$, or internal correlations between different parts of $\mathfrak{E}$, on any practical timescales. Note, this is not to say that such correlations do not exist physically, as they do under unitary time evolution, but rather just that they are not reflected in our state of knowledge as modelers. A typical assumption, which we adopt, is that it is also safe, or reasonable, to ignore any such correlations that may exist.\footnote{Whether this is, in fact, a valid assumption is a broad question about the applicability of this open-systems perspective which we do not address in this paper.}

Stated slightly more formally, we first assume, for simplicity, that the Hilbert space $\mathcal{H}_\mathfrak{U}$ of the model universe $\mathfrak{U}$ factorizes neatly into separate Hilbert spaces for the system $\mathfrak{S}$ and environment $\mathfrak{E}$, that is,
\begin{equation}
    \label{eq-1}
    \mathcal{H}_\mathfrak{U} = \mathcal{H}_\mathfrak{E} \otimes \mathcal{H}_\mathfrak{S}.
\end{equation}
Given this, we can imagine that, within a negligible thermalization timescale subsequent to the emission of any small increment $\Delta Q$ of waste heat out of $\mathfrak{S}$, the mixed state $\rho_\mathfrak{U}$ of the model universe would quickly degrade, for all practical purposes, into a (correlation-free) product state $\rho_\mathfrak{U} = \rho_\mathfrak{E} \otimes \rho_\mathfrak{S}$, where $\rho_\mathfrak{E}$ is the maximum-entropy (equilibrium) mixed state of energy $Q_\mathfrak{E}$ which includes the heat increment $\Delta Q$ after it has diffused into the environment, and $\rho_\mathfrak{S}$ is a reduced density matrix for the mixed state of the computer system after one has traced out any lingering correlations it may have had with the environment initially upon emission of the waste heat. Note that in the absence of totally separable dynamics (\emph{i.e.}, a Hamiltonian over $\mathcal{H}_{\mathfrak{U}}$ given at all times by $\widehat{H}_{\mathfrak{U}} = \widehat{H}_{\mathfrak{E}}\otimes \widehat{H}_{\mathfrak{S}}$), a strict entropy increase is implied by taking the trace over $\mathfrak{E}$ (compared to the entropy of an immediately-prior joint state $\rho^\prime_\mathfrak{U}$ briefly entangling the environment with the system) in the instant just after the emission of the heat $\Delta Q$. Thus, simply performing this state reduction results in global entropy increases in the model universe even when all other dynamics (including the internal dynamics within $\mathfrak{S}$) is taken to be unitary. This state reduction process models the effective decoherence of the system $\mathfrak{S}$ as a result of its interaction with the (modeled as thermal) environment $\mathfrak{E}$ \cite{Zur03}.

A slightly more general model (with weaker assumptions) can be provided by stipulating that we take the trace over $\mathfrak{E}$ only once, at the very \emph{end} of an evolution of interest, rather than continuously after each incremental emission $\Delta Q$ of heat into this environment. Postponing the state reduction allows for the possibility that correlations/entanglements between the system $\mathfrak{S}$ and environment $\mathfrak{E}$, and within $\mathfrak{E}$ may persist for some period of time, and affect the evolution to some extent. However, it is not expected that this change will make very much difference in practice.\footnote{As a thought experiment, consider a computer system $\mathfrak{S}$ in deep space, such that any thermal photons emitted from the system into the environment would be expected to mostly just propagate to infinity, with only an astronomically tiny probability of reflecting off of interplanetary gas or dust in such a way as to convey correlated quantum information back into the system. The analysis of such cases, at least, would clearly be only insignificantly affected by treating state reduction as a continuous process.}


Modeling the allowed thermal transformations of open quantum systems in detail is the topic of the \textit{resource theory of quantum thermodynamics} (RTQT), which we review briefly in \S\ref{sssec:rtqt} below. But first, we continue outlining our broad framework for studying the physics of RC.

\subsubsection{Computational States and the Proto-Computational Basis}
\label{sssec:comp-states}

We now discuss how to formally model, in both abstract and physical terms, the digital computational states of a computer system $\mathfrak{S}$.
Note that our emphasis, in this paper, is on {\em classical}, not quantum, reversible computation. Furthermore, we wish the scope of our model to include the usual case in real engineered digital computing systems, which is that digital computational states may be encoded by extended physical objects, whose detailed microstate is, in general, not fully determined by the computational state being represented. As an example of this, consider a logic node (connected conductor) within a digital CMOS circuit, where typically the digital symbols '\texttt{0}' and '\texttt{1}' may nominally be represented by node voltages within certain pre-specified, non-overlapping low-to-high ranges $[V_{\mathtt{0}\mathrm{L}},V_{\mathtt{0}\mathrm{H}}]$ and $[V_{\mathtt{1}\mathrm{L}},V_{\mathtt{1}\mathrm{H}}]$, respectively.

We will make the above, informally stated notion (regarding the correspondence between the abstract digital state, and the more detailed physical microstates that are interpreted as encoding it) more formal and precise in \S\ref{par:basis-sets} below, and then in \S\ref{par:c-vs-nc} we will show how this formal structure allows us to systematically subdivide any given computing system into what we call \textit{computational} versus \textit{non-computational} subsystems. These formalize what Bennett \cite{Ben03} refers to as the \textit{information-bearing} versus \textit{non-information-bearing} degrees of freedom within the system. (However, we will avoid using the latter terminology in this paper, since we consider the non-computational subsystem to still contain \textit{physical} information.)

\mypara{Designated times.}\label{par:desig-times}
Given that we wish to model active computing machines in which the abstract computational state of the machine \emph{changes} over the course of some time interval, we can expect to encounter the difficulty that the classical digital state of the machine, which, as a discrete entity, takes on values that range over some merely countable set, may not be well-defined (in traditional terms, at least) at all moments during the (physically continuous) transition from one state to the next. 
To avoid this difficulty, while maintaining  simplicity in our model, we will declare, for purposes of the present paper, that there exists some countable set $\set{\tau_\ell}$ of time points $\tau_\ell\in\mathbb{R}$, labeled with integers $\ell\in\mathbb{Z}$, which we will call the \textit{designated times} at which the classical digital computational state of the machine is well-defined. 

Note that this model is somewhat oversimplified, since a real engineered computing system is typically not monolithic, but is broken down into subsystems, and it may be the case that the larger system is globally \emph{asynchronous}, in the sense that some subsystems may be in the middle of state transitions while others are in well-defined states. Indeed, depending on the system architecture, there may be \emph{no} moments at which the \emph{entire} machine is simultaneously in a nominally well-defined digital state. However, we will postpone elaboration upon methods to handle this more general case to a later time, as it does not affect anything essential in the present paper.\footnote{However, to just briefly preview one way in which a resolution of this problem can work, we can augment the concept of a \emph{well-defined state} of a classical computation with that of a \emph{well-defined state transition}, as we do in \S\ref{sssec:comp-trans}; this can be meaningful even for non-reversible and/or stochastic operations. Then, at any moment across an extended, asynchronous machine, we can say that each local subsystem is either in a well-defined computational state, or is partway through a well-defined state transition.}

\mypara{Computational states correspond to sets of orthogonal microstates.}\label{par:basis-sets}
Regardless of what precise physical encoding of digital computational states is used, we take it as fundamental to the concept of a classical digital computational state that at any designated time $t=\tau_\ell \in \mathbb{R}$, there exists some set $\boldsymbol{C}(t)=\set{c_i(t)}$ of abstract entities comprising all of the possible alternative {\em well-defined computational states} $c_i(t)$ of the computer system $\mathfrak{S}$ that the machine could occupy at the time $t$, and further, that, for any given such state $c = c_i(t)$, there exists some corresponding set $\boldsymbol{B}_c \subset \mathcal{H}_\mathfrak{S}$ of mutually orthogonal, normalized basis vectors $\vec{b}\in \boldsymbol{B}_c$, each of which represents a pure quantum state of $\mathfrak{S}$ that is unambiguously interpretable as representing the state $c$. In other words, there is some orthonormal basis $\boldsymbol{\mathcal{B}} \supset \boldsymbol{B}_c$ for $\mathcal{H}_\mathfrak{S}$ such that, if one were to hypothetically perform a complete projective measurement\footnote{Please note that this definition of computational states does not require us to actually be able to do these complete projective measurements in practice; it is sufficient, for purposes of the definition, that they could be done {\em in principle}, by (we can imagine) applying a suitable abstract operator that measures some complete set of commuting observables of the system.} of the state of the entire computer system $\mathfrak{S}$ down onto the basis $\boldsymbol{\mathcal{B}}$ at time $t$, and the measured state $\ket{\psi}$ in that basis were found to be one of the basis states $\vec{b}\in \boldsymbol{B}_c$, then the computational state is unambiguously interpreted to be $c$. It follows from this that any superposition of the $\vec{b}\in \boldsymbol{B}_c$ must also be unambiguously interpreted as $c$, since such a superposition is not distinguishable from the members of $\boldsymbol{B}_c$; if any such superposition state were measured in the basis $\boldsymbol{\mathcal{B}}$, the projected state would necessarily be contained in the set $\boldsymbol{B}_c$.

We note that, when the physical state of a machine is dynamically evolving over (continuous) time, the basis states making up the set $\boldsymbol{B}_c$ could also generally be evolving; for example, consider an information-bearing signal pulse propagating down a transmission line, which may be convenient to represent in terms of a basis that propagates down the line along with the pulse. When discussing such cases, we can write $\boldsymbol{B}_c(t)$ to explicitly denote the possible time-dependence of the physical basis-set representation of a given computational state.

However, as mentioned above, we will often assume, for simplicity, that there exists some discrete set of designated time points $\tau_\ell \in \mathbb{R}$ (where $\ell\in\mathbb{Z}$) at which the computational states are well-defined, and focus our attention on those. This will then allow us to characterize non-reversible and stochastic computational evolutions in between those designated time points, in which there is merging or splitting of computational states, at a more abstract level in our model, without having to specify all details of the transition process, such as when, exactly, the computational states split or merge (and indeed, physically, these transitions will in general not be sharp). 

Further, it follows from the assumption that, at designated time points $t=\tau_\ell$, each $\boldsymbol{B}_c(\tau_\ell)$ identifies $c(t)$ unambiguously, that, at least at these times, all of the $\boldsymbol{B}_c$ are mutually orthogonal to each other, and thus can be taken to be disjoint subsets of a single ``master'' orthonormal basis $\boldsymbol{\mathcal{B}}(t)$; \emph{i.e.,} $\forall c \in \boldsymbol{C}(t):\boldsymbol{B}_c(t) \subset \boldsymbol{\mathcal{B}}(t)$. We call such a master basis a {\em proto-computational basis} for the system $\mathfrak{S}$; ``proto'' because the basis states unambiguously determine the computational state, but are not in general uniquely determined by the computational state. They are lower-level, physical entities, defined prior to the computational state itself.

\begin{figure}[t] 
\centerline{(a) \includegraphics[width=5 cm]{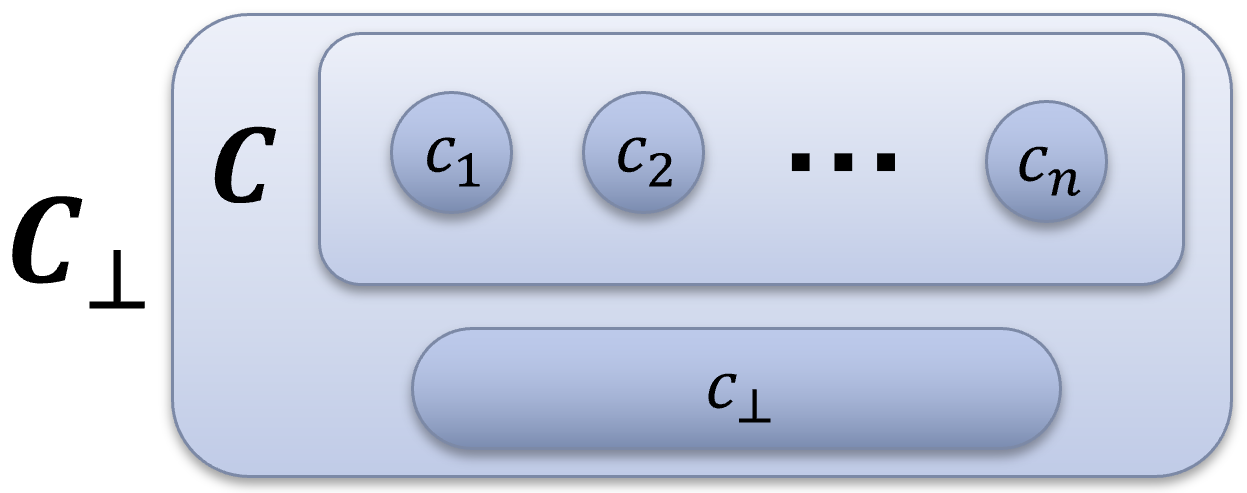}
\hspace{1 cm}(b) \includegraphics[width=5 cm]{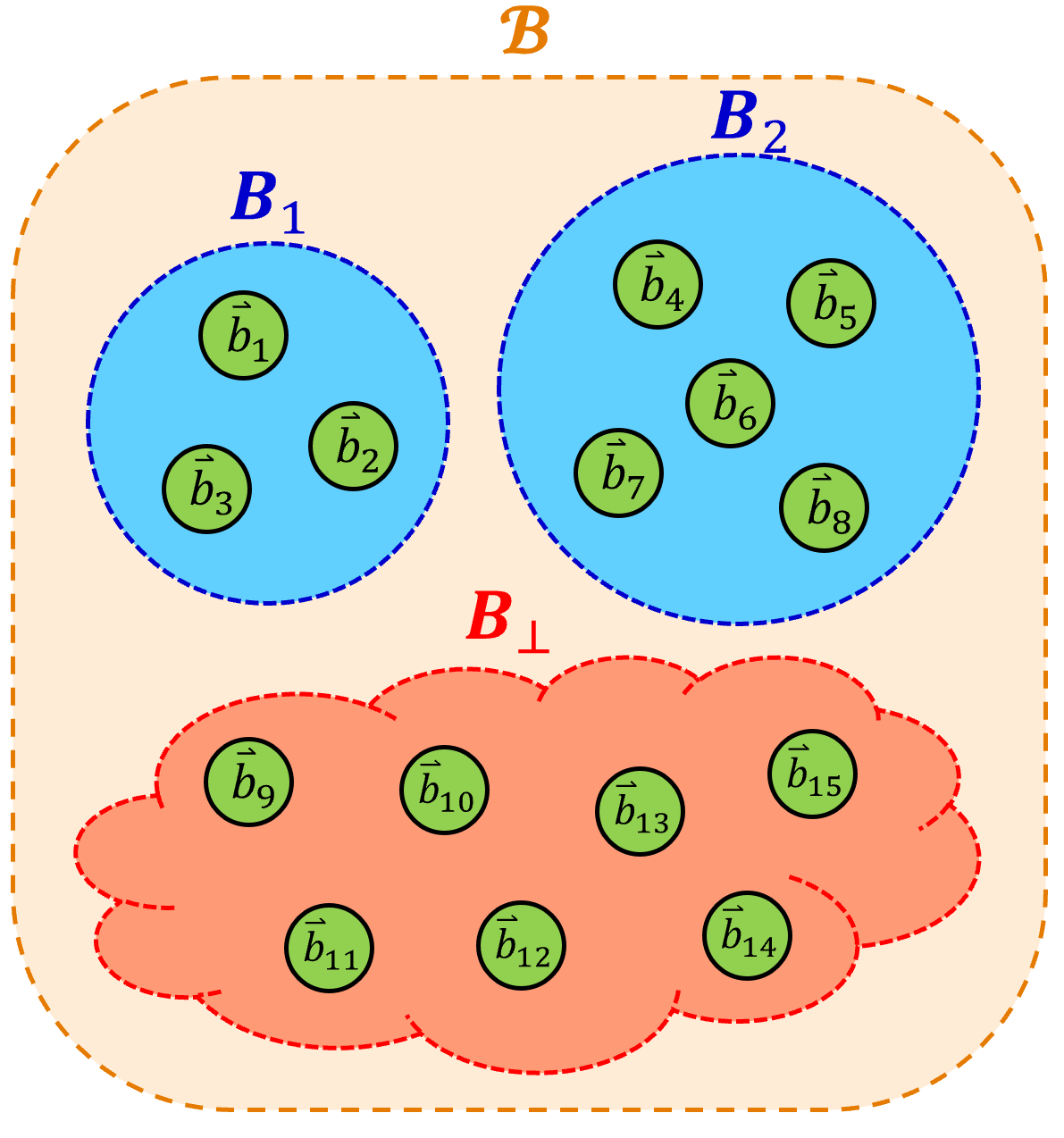}}
\caption{Model of classical digital computational states (at some particular time $t = \tau_\ell \in \mathbb{R}$). (\textbf{a}) Abstract computational states of a physical computer system $\mathfrak{S}$ with $n$ distinct states. The ``catch-all'' state $c_\bot$ represents the condition that the physical state of $\mathfrak{S}$ is such that the computational state is not otherwise well-defined. (\textbf{b}) Basis sets $\boldsymbol{B}_i$ corresponding to the computational states $c_i$, where $i\in\set{\bot,1,\ldots,n}$. Here, $n=2$. These basis sets partition the complete proto-computational basis $\boldsymbol{\mathcal{B}}$.\label{fig:comp-states}}
\end{figure}

Note that any particular proto-computational basis $\boldsymbol{\mathcal{B}}(t)$ at a designated time point $t=\tau_\ell$, since it is defined to be a {\em complete} basis for the Hilbert space $\mathcal{H}_\mathfrak{S}$ of the physical system comprising our computer, may in general include some basis states $\vec{b}$ that do not fall into {\em any} of the sets $\boldsymbol{B}_c(t)$. These are microstates of the system that do not correspond to well-defined computational states. Such microstates could arise in practice for any number of reasons; \emph{e.g.}, such as if the machine has not yet been powered on and initialized, or it has broken down, or simply gone out of spec. Regardless of the cause, we will group these ``invalid'' basis states together into a special set $\boldsymbol{B}_\bot = \boldsymbol{\mathcal{B}} - \bigcup_{c\in \boldsymbol{C}} \boldsymbol{B}_c$ meaning that the computational state is undefined; for convenience, we can also define an extra ``dummy'' computational state $c_\bot$ representing this undefined condition, and an augmented computational state set $\boldsymbol{C}_\bot = \boldsymbol{C} \cup \set{c_\bot}$, so that then we can say that the system $\mathfrak{S}$ always has \emph{some} computational state $c \in \boldsymbol{C}_\bot$, although it may be the undefined state $c_\bot$. With this change, note that the set $\set{\boldsymbol{B}_c}$ of basis sets corresponding to all of the computational states $c\in \boldsymbol{C}_\bot$ (in the augmented set) now corresponds to a proper set-theoretic \emph{partition} of the full proto-computational basis $\boldsymbol{\mathcal{B}}$. See Figure~\ref{fig:comp-states}.

Note that the foregoing treatment of computational states is really no different, fundamentally, from the case of identifying any other (potentially macroscale) classical discrete state variable. In other words, a classical computational state, in our formulation, can be viewed as simply corresponding to a discrete physical macrostate that we happen to consider as carrying some informational significance within a computational system.

\mypara{Computational and non-computational subsystems.}\label{par:c-vs-nc}
As an additional, but inessential assumption that will be useful in some derivations, we can suppose that the Hilbert space $\mathcal{H}_\mathfrak{S}$ of the system can be factored as a product of subspaces corresponding to what we call {\em computational} and {\em non-computational} subsystems $\mathfrak{C},\mathfrak{N}$ of the computer system $\mathfrak{S}$. That is, we write $\mathcal{H}_\mathfrak{S} = \mathcal{H}_\mathfrak{C} \otimes \mathcal{H}_\mathfrak{N}$, with the idea being that the computational states $c$ correspond to basis vectors of $\mathcal{H}_\mathfrak{C}$, which are tensored with the basis vectors of $\mathcal{H}_\mathfrak{N}$ to obtain the protocomputational basis $\boldsymbol{\mathcal{B}}$ for the entire system $\mathfrak{S}$. 

\begin{figure}[t] 
\centerline{(a) \includegraphics[width=5 cm]{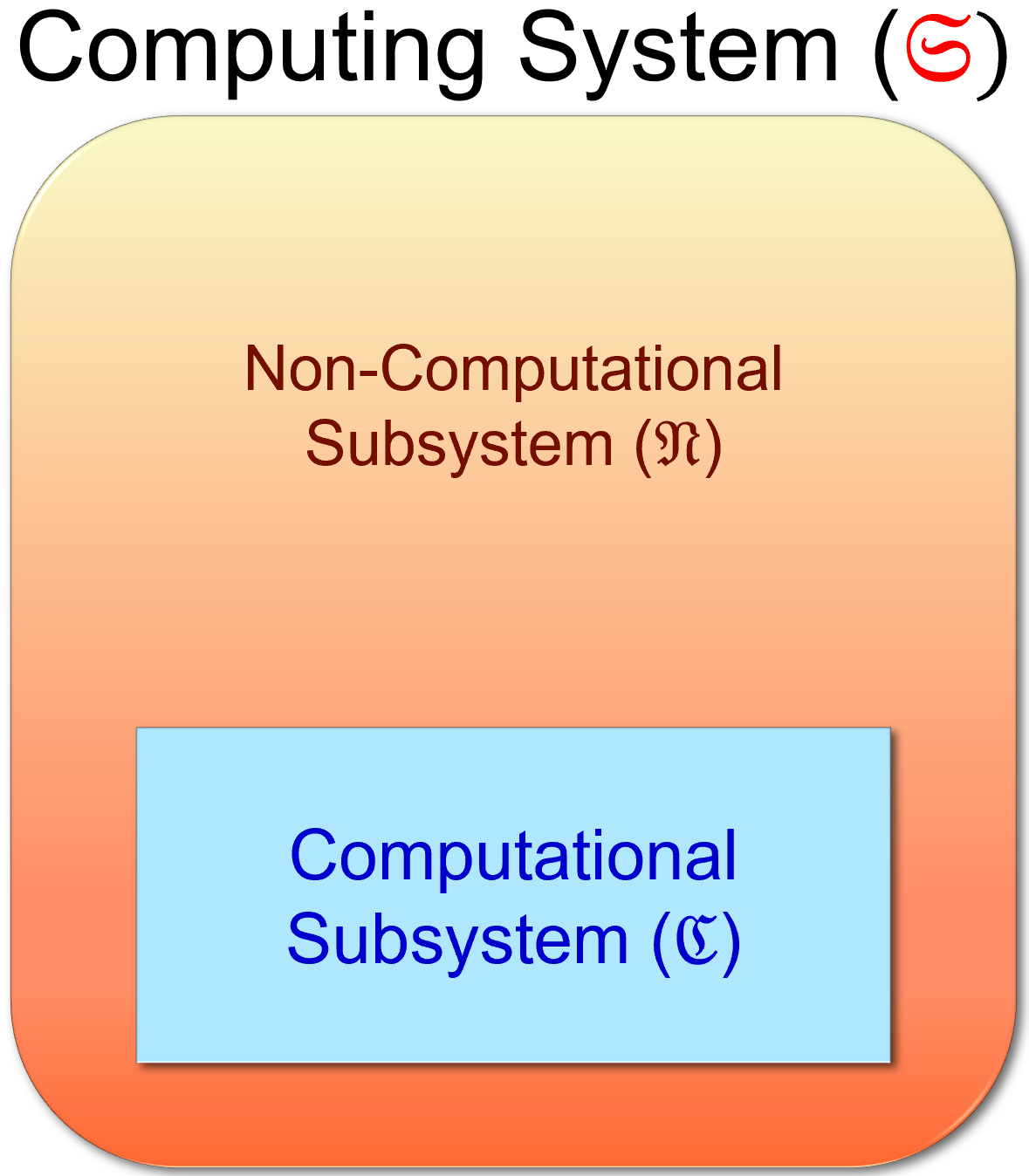}\hspace{1cm}(b) \includegraphics[width=4 cm]{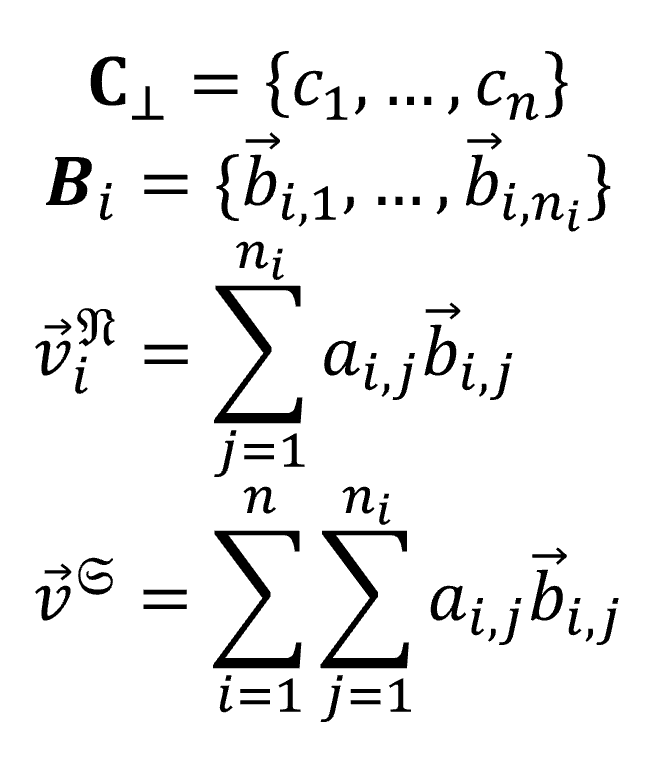}}
\caption{\textbf{(a)} Breakdown of the computating system $\mathfrak{S}$ into computational ($\mathfrak{C}$) and non-computational ($\mathfrak{N}$) subsystems. The Hilbert space of $\mathfrak{S}$ can be expressed as either a product $\mathcal{H}^\mathfrak{S} = \mathcal{H}^\mathfrak{C} \otimes \mathcal{H}^\mathfrak{N}$ of subsystem subspaces, or more generally as a subspace sum $\mathcal{H}^\mathfrak{S} = \bigoplus_{c\in\boldsymbol{C}_\bot} \mathcal{H}_c^\mathfrak{N}$. \textbf{(b)} A general vector $\vec{v}^{\,\mathfrak{S}}$ in a system Hilbert space $\mathcal{H}^\mathfrak{S}$ that is defined as a sum of the non-computational Hilbert spaces $\mathcal{H}_i^\mathfrak{N}$ corresponding to the $n$ individual computational states $c_i\in\mathbf{C}_\bot$ can be obtained by simply summing general vectors $\vec{v}_i^{\,\mathfrak{N}}$ within the individual subspaces $\mathcal{H}_c^\mathfrak{N}$, treating the basis vectors $\vec{b}_{i,j}$ across all $n$ subspaces as mutually orthogonal. Note that the dimensionalities $n_i$ of the component subspaces do not all have to be the same in general.\label{fig:c-vs-n}}
\end{figure}

However, this factorizability assumption is really just a special case, which only holds when the basis sets $\boldsymbol{B}_c$ of the whole system are identically sized. More generally, we can express $\mathcal{H}_\mathfrak{S}$ as a {\em subspace sum}:
\begin{equation}
    \label{eq:subspace-sum}
    \mathcal{H}_\mathfrak{S} = \bigoplus_{c\in\boldsymbol{C}_\bot} \mathcal{H}^c_\mathfrak{N},
\end{equation}
where $\mathcal{H}^c_\mathfrak{N}$ denotes the Hilbert space of the non-computational subsystem $\mathfrak{N}$, when restricted to the case that the computational state is $c$; that is, it is the subspace spanned by the basis vectors in $\boldsymbol{B}_c$. See Figure~\ref{fig:c-vs-n}.

\mypara{Rapid collapse of superpositions.}\label{par:rapid-collapse}
In the course of the real physical evolution of the system $\mathfrak{S}$, it is of course possible, in general, that quantum states could arise that are superpositions of basis states $\vec{b}$ from \emph{different} basis sets $\boldsymbol{B}_c$ and exist in the system briefly, yielding an indeterminate computational state at such moments. However, since our primary focus, in the present paper, is on the analysis of machines that are not even designed to carry out quantum computing, it is reasonable to suppose that such superposition states will spontaneously decohere on very short timescales, as they would naturally tend to do anyway in most large-scale systems. In other words, we expect that, most of the time, our computational subsystem $\mathfrak{C}$ will be living in a {\em decoherence-free subspace} (DFS), such that the computational states are naturally stable, as part of the system's ``pointer states'' \cite{Zur03} towards which the system is continually being decohered by its interactions with its environment. (In this context, the environment of the computational subsystem $\mathfrak{C}$ can include portions of the non-computational subsystem $\mathfrak{N}$ of the computer system, as well as the machine's external environment $\mathfrak{E}$.)

Thus, at least in the case of a logically deterministic (non-stochastic) computational process starting from a well-defined initial computational state $c(\tau_0)$, we assume that each of the system's pointer states will, at any designated time $\tau_\ell$ (for $\ell>0$), have all (or nearly all) of its probability mass concentrated within a single well-defined computational state $c(\tau_\ell)$. And, even in a stochastic computation, we will obtain a classical statistical mixture of computational states, not a quantum superposition over them.
The challenge, in reversible computing, is then to arrange for the system's already naturally stable pointer states to (notwithstanding their stability) still remain subject to undergoing a physically natural (if engineered) dynamics in which they will evolve, relatively quickly over time, translating themselves (eventually) one-to-one into new computational states $c(\tau_{\ell+1})$, which may bear new semantic interpretations. Such a system could thereby carry out useful computations at useful speeds. 

We will discuss computational operations, and their physical correlates, in more detail in \S\S\ref{sssec:comp-oper}--\ref{sssec:corresp}. These can be very directly embedded in open quantum systems exhibiting GKSL dynamics, which we discuss in detail in \S\ref{ssec:rc-op-rep}. 
However, even just the above definitions already suffice to prove what we call \emph{The Fundamental Theorem of the Thermodynamics of Computing}; this is discussed in \S\ref{ssec:fun-thm}.

\mypara{Timing variables.}\label{par:timing-vars}
One nearly ubiquitous feature of engineered physical computing systems is the concept of a \emph{timing variable}, that is, some non-computational, non-equilibrium degree of freedom that influences \emph{when} transitions between computational states will occur, and possibly how long they will take. As an example, an ordinary synchronous digital computer normally includes at least one \emph{clock oscillator}, which outputs a periodic \emph{clock} waveform at a prespecified or controllable frequency which is used to control the timing of digital operations. In such a situation, we can take the phase $\theta$ of the oscillator as a timing variable. In adiabatic circuits (see \S\ref{sssec:ra-cmos}), not only the frequency but also the \textit{speed} (quickness) of digital state transitions is controlled by the clock speed $\omega = \mathrm{d}\theta/\mathrm{d}t$.
Furthermore, even non-synchronous computing systems typically still have physical degrees of freedom that influence the timing of transitions. For example, in the novel BARC (Ballistic Asynchronous RC) computing paradigm being developed at Sandia, discussed further in \S\ref{sssec:barcs} below, individual bits propagate ballistically as flux solitons (\textit{fluxons}) traveling along interconnect lines between devices; the position $x$ of a given fluxon (of a given velocity) along the length of its interconnect can be considered a timing variable.

It is important to note that, while the values of timing variables are not digitally discretized, they are also generally \textit{not} entirely random, or uncorrelated to other parts of the machine, unlike thermal state variables. Thus, timing variables will be the one common exception, in digital computating systems, to our general rule that non-computational degrees of freedom will be assumed to rapidly thermalize.

Next, we define some key concepts of classical computational operations.

\subsubsection{Computational Operations}
\label{sssec:comp-oper}

In order to discuss in detail the thermodynamic implications and limits of performing classical digital computational operations (including reversible operations), we first present some basic terminology and definitions relating to such operations in this subsection.

As mentioned, in general the set $\boldsymbol{C}(\tau_\ell)$ of well-defined computational states could be different at different designated time points $\tau_\ell$, but, to simplify our presentation, we will temporarily focus on the case where it is unchanging, \emph{i.e.}, $\forall \ell\in\mathbb{Z}: \boldsymbol{C}(\tau_\ell) = \mathbf{C}$. 

To permit treatment of stochastic (randomizing) computational operations, we define some related notation. Let $\mathcal{P}(\mathbf{C})$ denote the set of all (normalized) probability distributions over $\mathbf{C}$. For simplicity, to avoid having to deal with normalizability issues, we can assume that $\mathbf{C}$ is finite.\footnote{Indeed, due to the holographic bound \cite{Bek00}, if the minimal bounding surface of $\mathfrak{S}$ has finite area, then the Hilbert space $\mathcal{H}_\mathfrak{S}$, and therefore also $\mathcal{H}_\mathfrak{C}$, must be finite-dimensional in any event.} 

Then, a (possibly stochastic) \emph{computational operation $O$ on $\mathbf{C}$} simply refers to some arbitrary function $O:\mathbf{C} \rightarrow \mathcal{P}(\mathbf{C})$ mapping each \emph{initial state} $c_\mathrm{I}\in \mathbf{C}$ to a probability distribution over the possible \emph{final states} $c_\mathrm{F}\in\mathbf{C}$. For a given initial computational state $c_i$, we can write $O(c_i)=P_i\in \mathcal{P}(\mathbf{C})$ where $P_i:\mathbf{C}\rightarrow [0,1]$ denotes the resulting probability distribution over final states. We can also allow $O$ to be a partial function, \emph{e.g.},\ when discussing operations that are not defined over all states $c\in\mathbf{C}$, which can be useful if the operation will only ever be applied to states $c\in\mathrm{dom}[O]\subset\mathbf{C}$.

Note that it is sufficient, for our present purposes, to use probabilities in the above definition instead of complex amplitudes, since, for classical reversible computing systems, we are going to assume that the system is highly decoherent in any case; any superposition over different computational states would soon decohere to a classical statistical mixture.\footnote{However, in the future, we anticipate that the present line of work may usefully be extended to explore dissipation limits for quantum computation as well; at that point, it would be appropriate to replace the probability distribution $P_i(c)$ with a more general density operator $\rho_i$.}

\mypara{Deterministic operations.}\label{par:det-ops} A particular computational operation $O$ is called \emph{(fully) deterministic} (meaning, non-stochastic) if and only if all of its final-state distributions $P_i$ have zero entropy, that is, $\forall c\in\mathbf{C}: H(O(c))=0$, where here we reference the standard (Shannon) entropy functional $H(\cdot):\mathcal{P}(\mathbf{C})\rightarrow\mathbb{R}^{0+}$, \textit{i.e.},
\begin{equation}
    H(p) = -\sum_{c\in \mathbf{C}} p(c)\log p(c),
\end{equation}
in generic logarithmic units \cite{Fra05}. (Note this is 0 only in the limit of a point distribution.)

If an operation is not fully deterministic, we say it is \emph{stochastic}. We could also have that $O$ is \emph{deterministic over a subset} $\boldsymbol{A}\subset\mathbf{C}$ of initial states, whilst not being deterministic over the entire set $\mathbf{C}$. Such an $O$ can also be called \emph{conditionally deterministic under the precondition that} the initial state $c\in\boldsymbol{A}$.

\mypara{Reversible operations.}\label{par:rev-ops} We say that an operation $O$ is \emph{(unconditionally, logically, fully) reversible} if and only if there is no state $c_k\in \mathbf{C}$ such that for two different $i,j$ ($i\neq j$), both $P_i(c_k) > 0$ and $P_j(c_k) > 0$. Otherwise, we say that $O$ is \emph{logically irreversible}. We say that $O$ is \emph{conditionally (logically) reversible under the precondition that} $c \in \boldsymbol{A}$, for some $\boldsymbol{A}\subset\mathbf{C}$, if and only if there is no state $c_k\in\mathbf{C}$ such that, for two different $i,j$ ($i\neq j$) with $c_i,c_j\in\boldsymbol{A}$, it is the case that $P_i(c_k) > 0$ and $P_j(c_k) > 0$. In such a case, we could also say that $O$ is \emph{reversible over} $\boldsymbol{A}$.

\mypara{Time-dependent case.}\label{par:time-dep} Note that it is easy to generalize the above definitions to situations in which the set $\boldsymbol{C}$ of computational states may be different at different designated times. Let $s=\tau_\ell$, $t=\tau_m$ be two different designated times, with $s<t$; then we can write $O_s^t$ to denote a computational operation being performed over the time interval from start time $s$ to end time $t$. Then we have that $O_s^t:\boldsymbol{C}(s)\rightarrow\mathcal{P}(\boldsymbol{C}(t))$, and the remaining definitions (for determinism, reversibility, {\em etc.}) also change accordingly, in the natural way.
For an operation taking place between times $s$ and $t$ ($s<t$), we can define $d=t-s$ as the \textit{delay} or latency of the operation, and $q=d^{-1}$ as its \textit{quickness} or speed.

\vspace{3mm}
The above definitions are illustrated in Figure~\ref{fig:op-types} below. 

\subsubsection{Computational State Transitions}
\label{sssec:comp-trans}

\begin{figure}[h]
\widefigure
\centerline{\includegraphics[width=10 cm]{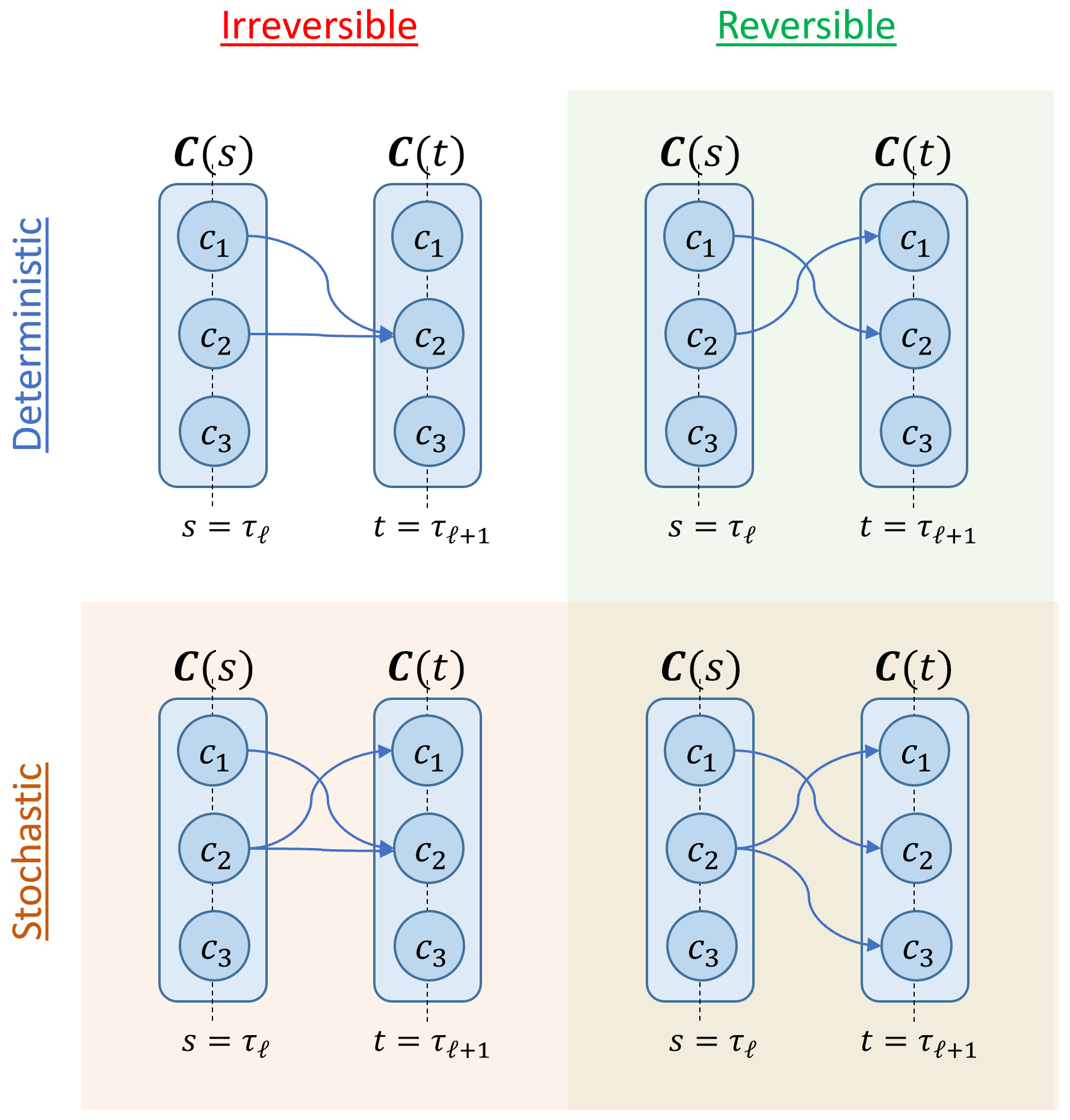}}
\caption{Illustration of different types of computational operations $O_s^t$ on a set $\boldsymbol{C}$ of 3 computational states. Examples shown here are partial functions---$O(c_3)$ is not defined. At upper-left is a conventional (deterministic but non-reversible) computational operation which merges two initially distinct computational states. At upper-right is a deterministic, reversible operation which is injective (one-to-one) over the subset $\boldsymbol{A} = \set{c_1,c_2}$ of initial states for which it is defined. At lower-right is a stochastic but reversible operation which does not merge any states, but splits the state $c_2$ (with some nonzero probability to transition to either $c_1$ or $c_3$). Finally, at lower-left is a stochastic, irreversible operation which includes both splits and merges.\label{fig:op-types}}
\end{figure}

We can describe the computational operations from the previous subsection as a combination of various primitive {\em computational state transitions}, such as those illustrated in Figure~\ref{fig:transitions} below. Other types of transitions may be described as combinations of these. For example, the lower-left operation in Figure~\ref{fig:op-types} includes a transition of $\set{c_1,c_2}$ from time $s$ to $t$ that exhibits both splits and merges. But, as suggested by the arrows in the diagram, it could be decomposed into a sequence of a split of $c_2$ into two (unlabeled) states, followed by a merge of $c_1$ into one of those states. 

\end{paracol}

\begin{figure}[h]	
\widefigure
(a) \includegraphics[width=4.6 cm]{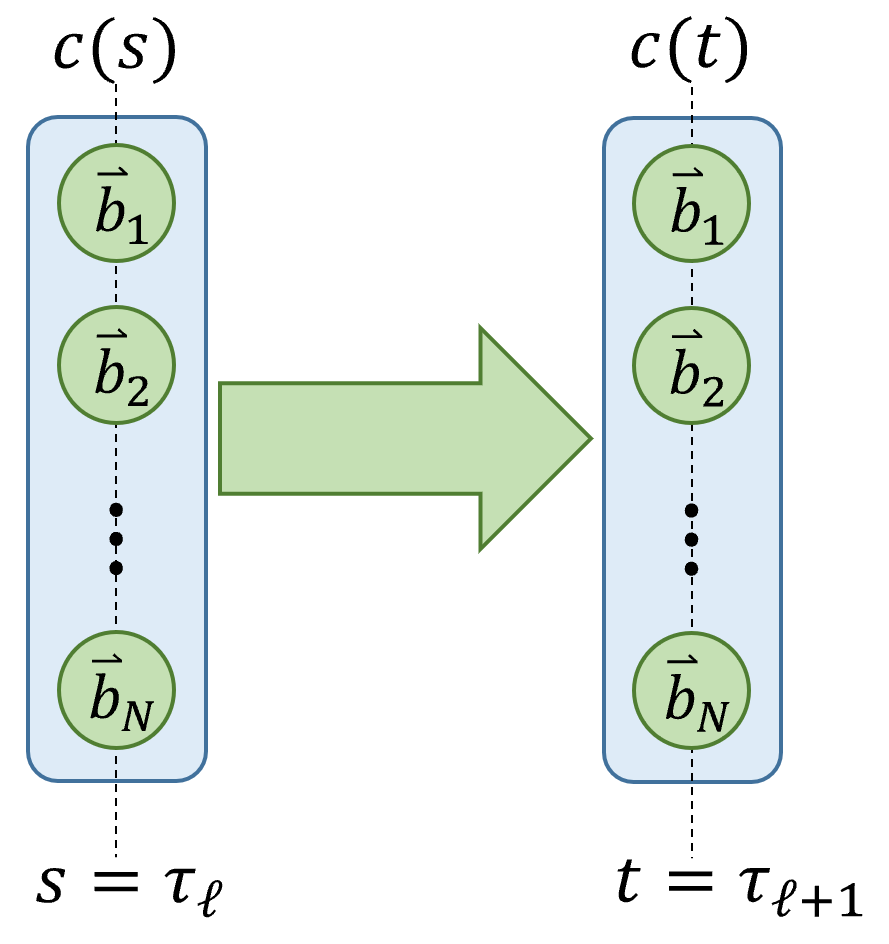}
\hspace{0.6 cm}(b) \includegraphics[width=4.6 cm]{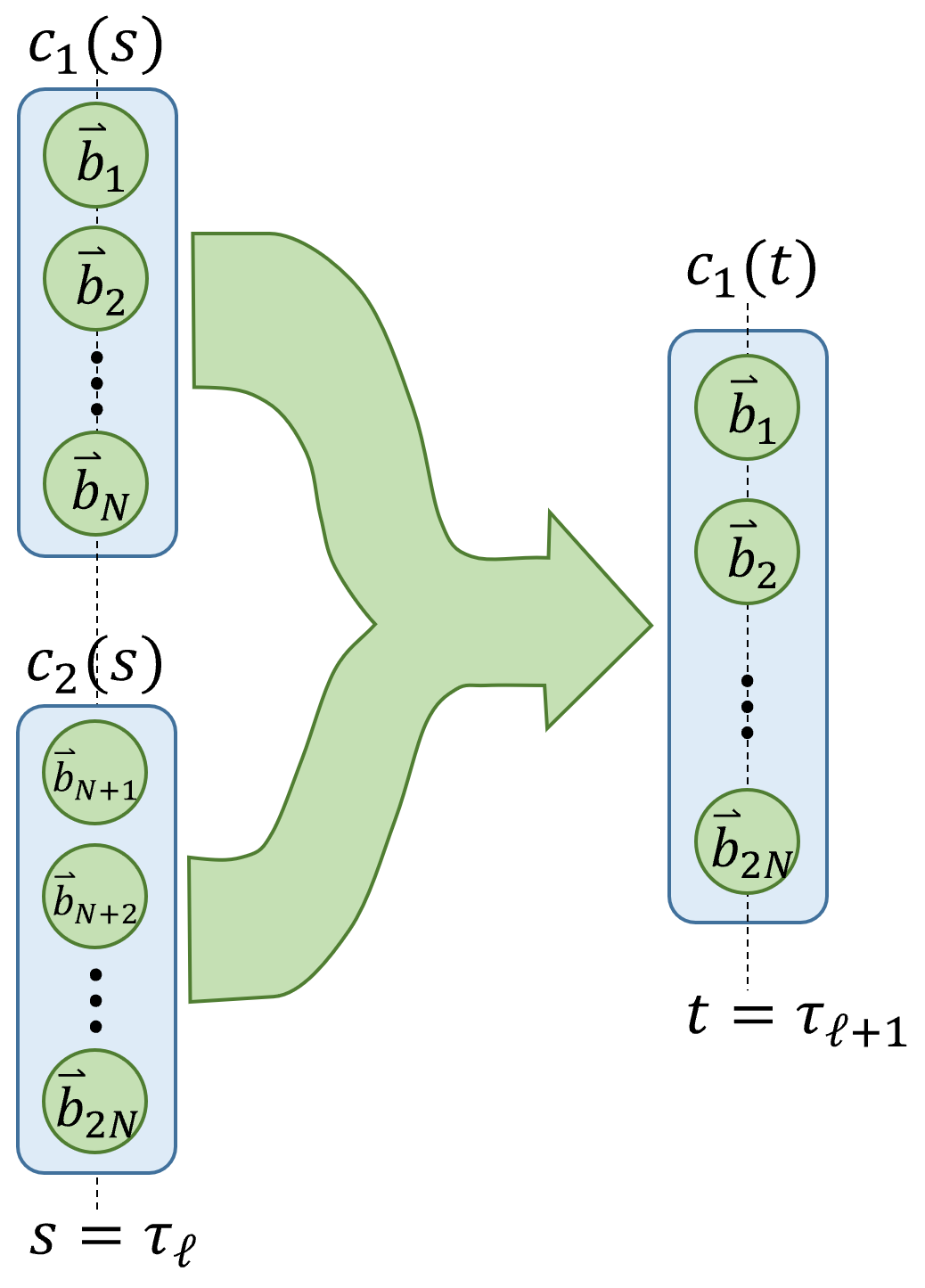}
\hspace{0.6 cm}(c) \includegraphics[width=4.6 cm]{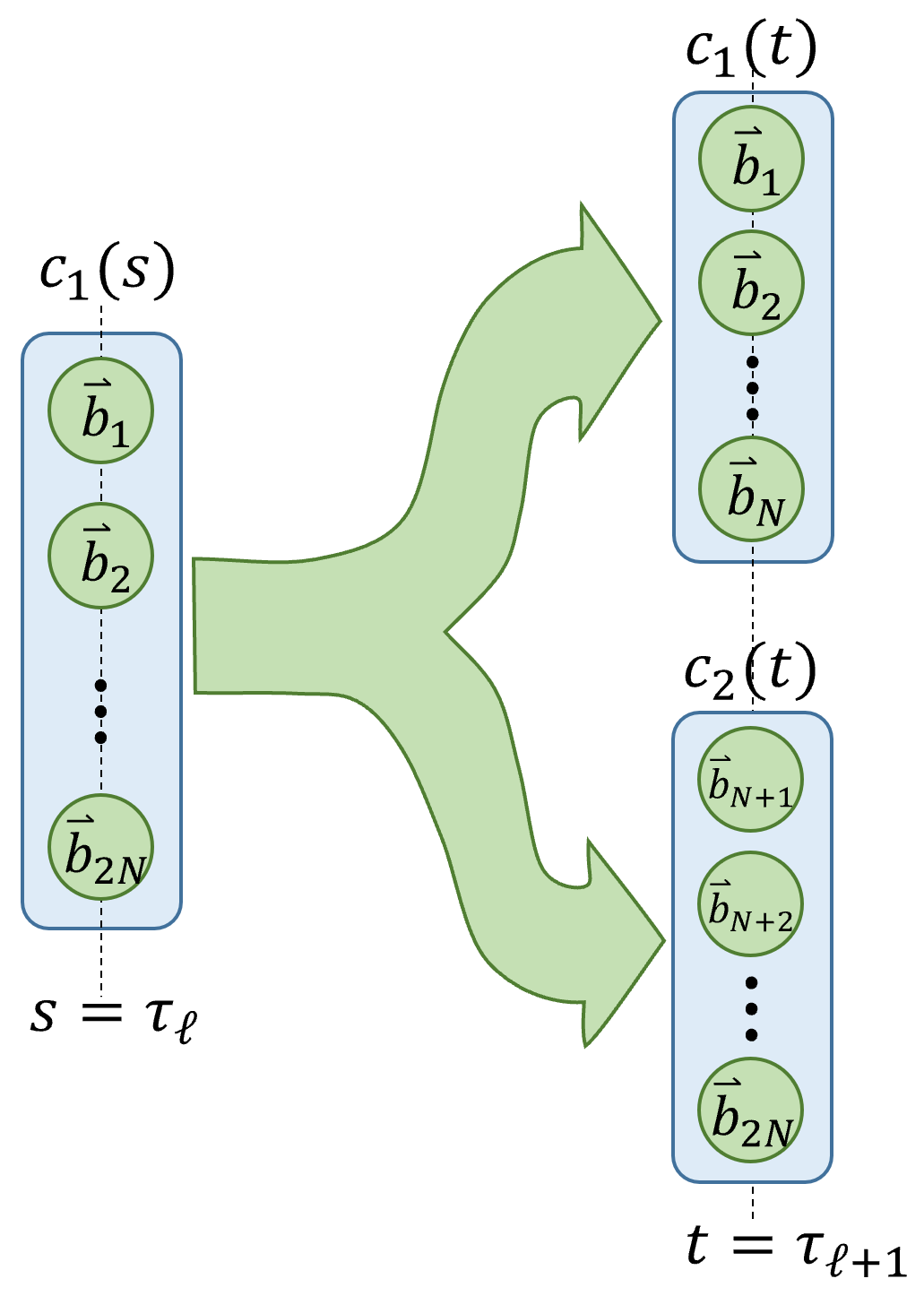}
\caption{Illustration of different types of computational state transitions between two designated time points $s$ and $t$, showing the protocomputational basis state sets $\boldsymbol{B}_c$ corresponding to computational states $c$. All of the individual basis states $\vec{b} = \vec{b}(\tau)$ are also implicitly time-dependent in general. Reversible computational operations include only one-to-one transitions such as \textbf{(a)}. In such transitions, the initial and final basis state sets may be the same size $N$, even in a closed system. Irreversible computational operations include at least some examples of many-to-one transitions (merges), such as \textbf{(b)}. In a closed system, the basis state set resulting from a merge must be (at least) as large as the sum of the merged sets, due to unitarity. Stochastic operations include one-to-many transitions (splits), such as \textbf{(c)}. Each of the basis state sets resulting from a split may be smaller than the original set, even in a closed system, although their aggregate size must be at least as large.\label{fig:transitions}}
\end{figure}  
\begin{paracol}{2}
\switchcolumn

\subsubsection{Correspondence Between Classical Operations and Quantum Evolution}
\label{sssec:corresp}

In this subsection, we give a general theoretical picture regarding how a real (ergo, quantum-mechanical) physical process may effectively implement classical computational state transitions, and computational operations, such as described above.

\mypara{Unitary dynamics.}\label{par:unitary}

As before, we focus our attention on a computational process taking place between two designated time points $s=\tau_\ell$ and $t=\tau_{\ell+1}$ (where $t>s$). Consider, now, the joint Hilbert space $\mathcal{H}_\mathfrak{U}$ of the model universe (the environment together with the computer). Whatever is happening physically in the universe over the time interval $[s,t]$ (including the performance of the computational operation) will be encompassed, in a theoretical perspective assuming perfect knowledge of the universe's dynamics, by the overall time evolution operator, which we will denote $\widehat{U}=\widehat{U}_s^t(\mathfrak{U})$, that applies between those times in $\mathcal{H}_\mathfrak{U}$. Formally, if we describe the initial quantum state of the model universe as a mixed state using an initial density matrix $\rho_s$, then the final density matrix $\rho_t$ is given by
\begin{equation}
    \label{eq-3}
    \rho_t = \widehat{U}\rho_s \widehat{U}^\dagger.
\end{equation}
This overall time evolution process includes activities such as the dynamical details of the computation process itself, together with the incremental delivery of some needed free energy from the power supply (\emph{e.g.}, battery) into that process, and the transport of some incremental amount of dissipated energy (waste heat) away from that process, or more precisely, the incremental progression of a continuous \emph{flow} of waste heat that is propagating away from the computational mechanism, and out towards the environment $\mathfrak{E}$---since in general, the waste heat that resulted from prior operations will still be traveling outwards when subsequent operations occur. We call this picture the \emph{open system} case.

Now, let us restrict attention temporarily to the subspace of $\mathcal{H}_\mathfrak{U}$ that is the Hilbert space $\mathcal{H}_\mathfrak{S}$ of the closed spacelike hypersurface (slice of spacetime volume) enclosing the computer system. Ignoring, for the moment, the flow of waste heat through the system's bounding surface, let us pretend for a moment that the dynamics within the surface itself can also be described by a unitary time-evolution operator $\widehat{U}_s^t(\mathfrak{S})$ over $\mathcal{H}_\mathfrak{S}$, the quantum subsystem contained within the boundary. We call this the \emph{closed system} case.

Of course, the closed-system picture is a simplification, since in reality, no thermal isolation is perfect, and so there will also be interactions across the surface, to transport heat out. However, we expect that theoretical developments for the closed-system case can generally be preserved when re-expanding the model to include the outward thermal flow, since the net effect of that flow will just be to maintain a reasonable temperature inside the boundary, by exporting excess thermal entropy to the environment.

An easy way to see that this translation from the closed-system to the open-system case ought to work, in general, is simply to note that if the bounding surface of $\mathfrak{S}$ is taken to be extremely distant to begin with, then there will be negligible \emph{practical} difference between the open-system and closed-system cases. \emph{I.e.,} a real computer with an internal power supply would operate just fine, at least for a while, even if enclosed in a very large, but finite, perfectly thermally insulated box.

Thus, henceforth in this subsection, we will take the time-evolution $\widehat{U}_s^t$ to be the one for the computer system $\mathfrak{S}$, in the closed-system picture, while remembering that we can revert to the open-system view when necessary.

Earlier, we noted that the protocomputational basis $\boldsymbol{\mathcal{B}}$ may, in general, be time-dependent, so that the two bases $\boldsymbol{\mathcal{B}}(s)$ and $\boldsymbol{\mathcal{B}}(t)$ may not correspond to exactly the same set of physical quantum states. However, the effect of any change in the protocomputational basis $\boldsymbol{\mathcal{B}}$ between times $s$ and $t$ can also just be represented as a unitary operator, which we denote $\widehat{U}_{\boldsymbol{\mathcal{B}}(s)}^{\boldsymbol{\mathcal{B}}(t)}$. Then, we can define a suitably ``basis-corrected'' version of $\widehat{U}_s^t(\mathfrak{S})$ as:
\begin{equation}
    \label{eq-4}
    \widehat{U}_s^t(\mathfrak{S},\boldsymbol{\mathcal{B}}) = \widehat{U}_{\boldsymbol{\mathcal{B}}(s)}^{\boldsymbol{\mathcal{B}}(t)} \cdot \widehat{U}_s^t(\mathfrak{S}).
\end{equation}

\mypara{Quantum statistical operating contexts.}\label{par:qsoc}

Next, we need to define a \textit{computational process} in a statistically-contexualized form. Earlier, we abstractly defined computational state transitions and computational operations, but this definition said nothing whatsoever about the statistics of the initial state (either computational or physical) before the operation was performed. We require a formalism for describing such information in order to speak meaningfully about the informational or thermodynamic effect of performing a computational operation within a particular, statistically-defined scenario. 
Note that the following presentation just generalizes the discussion of (classical)  statistical operating contexts that can be found in, \emph{e.g.}, \cite{Fra18}, to a quantum context.

Since we want to produce a quantum-mechanical model of classical computation (including reversible computation), we require a quantum statistical picture. Thus, let us define $\rho_s$ to be a mixed quantum state (\emph{i.e.}, a statistical mixture of orthogonal pure states, in some diagonal basis) that encompasses all of our uncertainty, as modelers, regarding what the initial quantum state of the physical computational system $\mathfrak{S}$ is at time $s$, prior to performing the desired computational operation $O_s^t$.

We further require that $\rho_s$ must have a \emph{block-diagonal} structure in the initial protocomputational basis $\boldsymbol{\mathcal{B}}(s)$, such that the blocks correspond to the partition $\set{\boldsymbol{B}_c}$ of basis vectors corresponding to the (augmented) initial computational state set $\boldsymbol{C}_\bot(s)$. Stated more formally, the density matrix representation of $\rho_s$ in the $\boldsymbol{\mathcal{B}}(s)$ basis must not include any nonzero, off-diagonal terms between basis states $\vec{b}_p,\vec{b}_q \in \boldsymbol{\mathcal{B}}(s)$ such that $\vec{b}_p \in \boldsymbol{B}_i$ and $\vec{b}_q \in \boldsymbol{B}_j$ where $\boldsymbol{B}_i,\boldsymbol{B}_j$ are the subsets of $\boldsymbol{\mathcal{B}}(s)$ corresponding to two distinct computational states $c_i,c_j \in \boldsymbol{C}_\bot(s)$, \emph{i.e.}, with $i\neq j$. See Figure~\ref{fig:block-diag}.
This block-diagonal structure models our assumption, mentioned earlier, that a classical computer is highly \emph{decoherent}; thus, there are no quantum coherences between the blocks corresponding to different computational states. (In the terminology coined by Zurek \cite{Zur81}, the digital computational states would be considered natural \textit{pointer states} of the computing apparatus.)
However, note that it is permissible for coherences to exist \textit{within} blocks. This is just another way of saying that the choice of protocomputational basis vectors is completely arbitrary within the subspace corresponding to each block; the sub-basis for any block can be freely rotated within its subspace, and we still will have a valid protocomputational basis for time $s$.

\end{paracol}

\begin{figure}[t] 
\widefigure
\centerline{\includegraphics[width=12 cm]{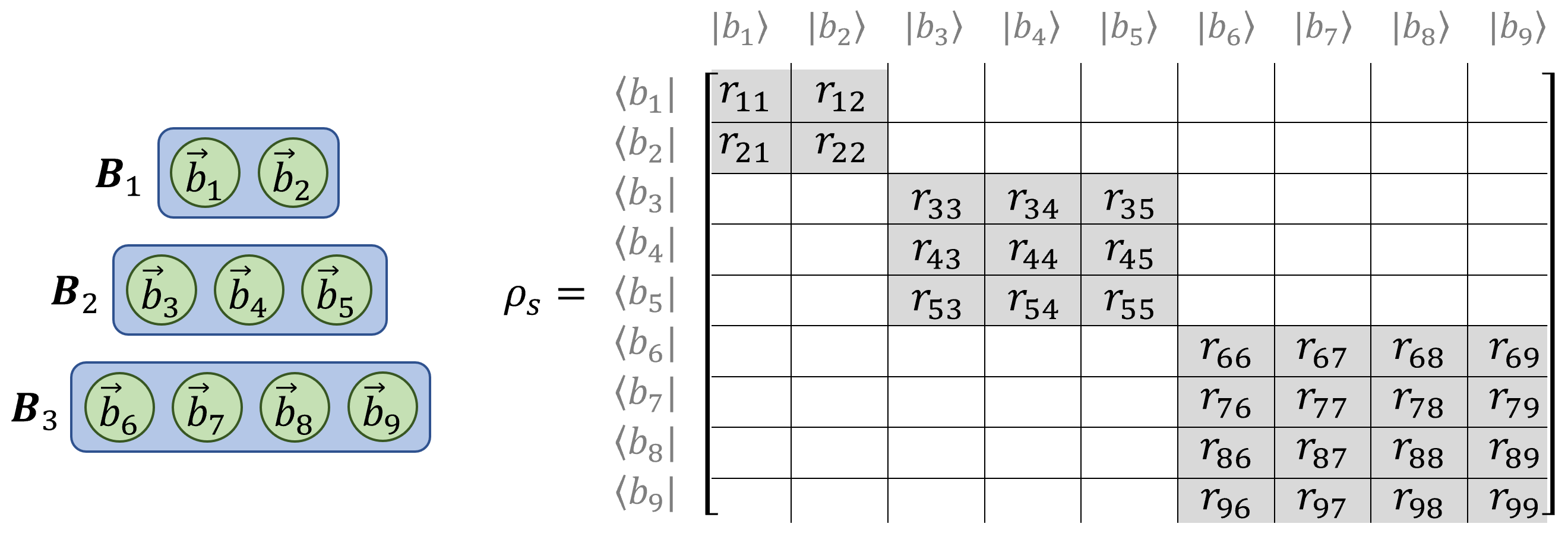}\hspace{1 cm}\includegraphics[width=4 cm]{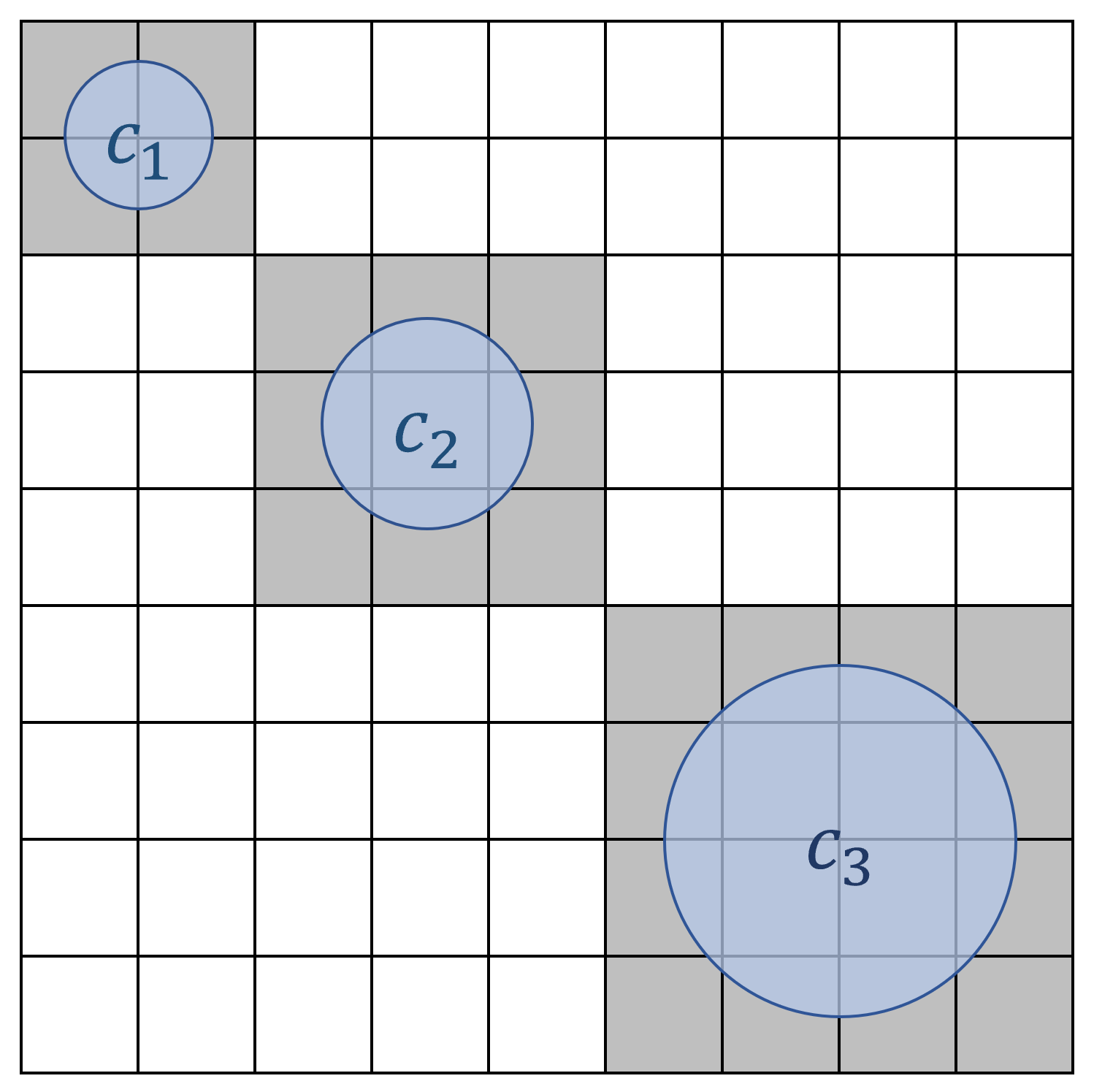}}
\caption{\textbf{Block-diagonal density matrix for an initial quantum statistical operating context $\rho_s$.} In this example, we imagine there are 3 computational states $c_1,c_2,c_3$ (and let $c_\bot=c_3$, say) with corresponding basis state sets $\boldsymbol{B}_1,\boldsymbol{B}_2,\boldsymbol{B}_3$ (left). At the center, we illustrate a corresponding block-diagonal \textit{quantum statistical operating context} or initial density matrix $\rho_s$. Rows and columns are labeled in gray with the corresponding basis vectors; note that $\Braket{b_i|\rho_s|b_j}=r_{ij}$. This is an Hermitian matrix, so $r_{ij}=r_{ji}^*$. Also $\mathrm{Tr}\left[\rho_s\right]=1$. Matrix entries left blank are 0. At right is a simplified depiction of $\rho_s$.\label{fig:block-diag}}
\end{figure}   
\begin{paracol}{2}
\switchcolumn

\mypara{Quantum contextualized computations.}\label{par:qc-comp}

Now that we have defined the quantum version of a statistical operating context, we can define what a ``quantum-contextualized computation'' means. This generalizes the discussion of (statistically contextualized) computations that can be found in \cite{Fra18}.

A \textit{quantum-contextualized computational process} or just \textit{quantum-contextualized computation}, denoted $\mathcal{C}_s^t(O_s^t,\rho_s)$, refers to the act of carrying out a specified computational operation $O_s^t$ from time $s$ to $t>s$ within the computer system ($\mathfrak{S}$) in a quantum statistical operating context wherein the initial mixed state of $\mathfrak{S}$ at time $s$ is given by $\rho_s$; where $\rho_s$ meets the conditions (\emph{i.e.}, block-diagonal structure) described above, given the protocomputational basis $\boldsymbol{\mathcal{B}}(s)$ and computational state set $\boldsymbol{C}(s)$. (Note that $\boldsymbol{\mathcal{B}}(s)$ and $\boldsymbol{C}(s)$ are left implicit in the $\mathcal{C}_s^t(\cdot)$ notation for brevity.)

\mypara{Implementation of classical computation by unitary dynamics.}\label{par:unitary-comp}

Given the above definitions, we can now formally define what it means for a system's unitary dynamics to implement a given (classical) computation.

We say that the basis-adjusted time-evolution operator $\widehat{U}_s^t(\mathfrak{S},\boldsymbol{\mathcal{B}})$ \textit{implements} the quantum contextualized computational process $\mathcal{C}_s^t(O_s^t,\rho_s)$, written
\begin{equation}
    \label{eq-5}
    \widehat{U}_s^t(\mathfrak{S},\boldsymbol{\mathcal{B}}) \Vdash \mathcal{C}_s^t(O_s^t,\rho_s),
\end{equation}
if and only if the final density matrix
\begin{equation}
    \label{eq-6}
    \rho_t = \widehat{U}_s^t(\mathfrak{S},\boldsymbol{\mathcal{B}})\rho_s \widehat{U}_s^t(\mathfrak{S},\boldsymbol{\mathcal{B}})^\dagger
\end{equation}
generated by applying $\widehat{U}_s^t(\mathfrak{S},\boldsymbol{\mathcal{B}})$ to $\rho_s$ has the property that, for any initial computational state $c_i(s)\in\boldsymbol{C}(s)$ that has nonzero probability under $\rho_s$, if we were to zero out all elements of $\rho_s$ outside of the rows/columns corresponding to $c_i(s)$'s basis set $\boldsymbol{B}_i(s)$ and renormalize, and then apply $\widehat{U}_s^t(\mathfrak{S},\boldsymbol{\mathcal{B}})$ to this restricted $\rho_s^\prime$, the resulting final mixed state $\rho_t^\prime$ would imply the same probability distribution $P_i(t)$ over final computational states in $\boldsymbol{C}(t)$ as is specified by applying the stochastic map $O_s^t$ to the initial computational state, that is, $P_i = O_s^t(c_i(s))$.

One can see by inspection that this is a very straightforward and natural definition. Since, by assumption, the initial quantum statistical operating context $\rho_s$ has no coherences between different initial computational states, it is impossible for the transition amplitudes from initial to final basis states to interfere with each other in ways that would disrupt the overall probability distribution over final computational states from what one would obtain by simply combining the results from treating the initial computational states separately.\footnote{As discussed in \ref{par:qsoc}, each computational state corresponds to an orthogonal subspace; indeed, for \emph{classical} computing, each computational state \emph{must} be an orthogonal subspace, to 
ensure distinguishability of 
different computational states. Then, this statement is a direct consequence of the fact that the full space can be decomposed into the sum of projectors over all of its orthogonal subspaces.}

Note that the above definition \emph{doesn't} by itself immediately require that the unitary evolution $\widehat{U}_s^t(\mathfrak{S},\boldsymbol{\mathcal{B}})$ can't introduce any \emph{immediate} coherences between different computational states $c_i(t),c_j(t)$, where $i\neq j$, but, \emph{this is unproblematic}, since one of our background assumptions throughout this treatment is that the system will naturally decohere very quickly to a definite computational state, so, any off-diagonal matrix elements between different computational states that may arise will naturally decay by themselves very quickly. This can happen via the usual Zurek process~\cite{Zur03}, wherein the decoherent state variables entangle with nearby non-computational degrees of freedom, which then---at least, in the open-system version of this treatment---carry the associated quantum information out to the external thermal environment $\mathfrak{E}$. Once it's in that environment, taking the trace over the environment state to reflect our ignorance about the environment's detailed evolution then effectively erases the entanglement between the system $\mathfrak{S}$ and the environment, and decays the coherences between the different naturally-stable ``pointer states'' of the computer. In Zurek's terms, the natural interaction between the computer system and its environment effectively ``observes'' the state of the system, and this effective measurement of the system by the environment collapses the system down to (what is then effectively just a classical statistical mixture of) the observably distinct classical computational states.

At this point, having described what it means, in quantum physical terms, to perform classical digital computational operations, our problem in building quantum physical models of reversible computing has now been reduced to:

\begin{enumerate}
    \item Finding specific closed-system time-evolution unitaries             $\widehat{U}_s^t(\mathfrak{S},\boldsymbol{\mathcal{B}})$ that meet the above definition of the ``implements'' operator $\Vdash$ for the case of desired \emph{reversible} (and/or conditionally reversible) operations $O_s^t$ in specific physical setups---and, it's easy to see that there's no essential loss of generality in starting with the closed-system case, since, for large enough systems, closed-system evolution should work just fine for a while, until the system runs out of effective free energy\footnote{Technically speaking, free energy in closed systems is conserved, but if part of $\mathfrak{S}$ treated thermally, there can be an increase in effective entropy.} or overheats.
    
    \item Showing that the closed-system definition of $\widehat{U}_s^t(\mathfrak{S})$ can then be extended appropriately to the open-system case where there may be a heat flow out from the system's bounding surface, for consistency with the existence of a global unitary evolution $\widehat{U}_s^t(\mathfrak{U})$ for the model universe that includes the process of heat outflow to the environment---but this part is expected to be a relatively easy formal technicality. And finally,
    
    \item Showing that some such unitaries can indeed be implemented via realistic, buildable physical computing mechanisms. Of these three steps, this one is expected to be the most difficult one to accomplish in practice.
\end{enumerate}

However, the supposition that the above physical picture of classical reversible computing \emph{can}, in fact, be realistically implemented is supported by the illustration of a number of existing and proposed examples of concrete physical implementation technologies which appear to accomplish this, which are briefly reviewed in \S\ref{ssec:tech} below.

But first, we now review some relevant tools and methods from NEQT which can be used to flesh out the general theoretical framework presented above in more detail.

\subsection{Tools and Methods from Non-equilibrium Quantum Thermodynamics}
\label{ssec:neqt}

In this section, we review some key theoretical tools and methods from non-equilibrium quantum thermodynamics (NEQT) that we believe will prove to be invaluable in the effort to arrive at a more complete understanding of the physics of reversible computing, and relate them to the more general picture presented above in \S\ref{ssec:found}.

\subsubsection{Resource Theory of Quantum Thermodynamics}
\label{sssec:rtqt}

First, we review several theoretical tools relating to what is known as \textit{the resource theory of quantum thermodynamics} (RTQT), in order to relate them to the broad framework presented above.

\mypara{Stinespring dilation theorem and thermomajorization.}
\label{par:Stinespring}

We briefly summarized the overall open quantum systems perspective in \S\ref{sssec:open-q} earlier. The rules of quantum thermodynamics let us turn the broad intuitions summarized there into specific statements about the types of transformations allowable on the system $\mathfrak{S}$. The evolution of a general density matrix $\rho$ is given by a \emph{completely positive trace-preserving \emph{(CPTP)} map} $\rho \mapsto \Lambda_{t}\left[\rho\right]$, also known as a \emph{quantum channel} or \emph{(quantum) dynamical map}. The map $\Lambda_{t}$ maps the initial density matrix to a final density matrix. (Here, $t$ represents the time interval from the initial time $t_{0}$ to the final time $t_\mathrm{f}$.\footnote{Most generally, a $t$ subscript for a time interval can be taken to specify an ordered pair $\left(t_{0},t_\mathrm{f}\right)$ of start and end times: \textit{i.e.}, the notation doesn't need to assume time translation invariance \emph{a priori}; but, in the usual case when it is present, for $t$ to specify just the time difference $d= t_{\mathrm{f}} - t_{0}$ is sufficient.})

$\Lambda_{t}\left[\rho\right]$ represents the most generic type of transformation that we can apply to $\rho$. In general, the density matrices $\rho$ and $\Lambda_{t}\left[\rho\right]$ aren't required to be taken over the same Hilbert space. In our setup, however, we stipulate that the initial and final Hilbert spaces are the same (namely, $\mathcal{H}_\mathfrak{S}$). As the name ``CPTP map'' suggests, in order for the map $\rho \mapsto \Lambda_{t}\left[\rho\right]$ to satisfy the laws of quantum mechanics, we need $\Lambda_{t}$ to preserve the trace of $\rho$, to preserve the positivity of $\rho$, and to preserve the positivity of $\rho$ even when $\rho$ is part of a larger system (which $\Lambda_{t}$ acts on as a whole). Furthermore, we need $\Lambda_{t}$ to be Hermitian, to be linear, and finite under the trace norm.

The \emph{Stinespring dilation theorem} \cite{Stinespring55} provides a very natural representation of this channel. From this theorem, the action of $\Lambda_{t}\left[\rho\right]$ can always be represented by embedding $\rho$ in a larger Hilbert space, where the dynamics corresponding to $\Lambda_{t}\left[\rho\right]$ is now unitary, and then tracing out the auxiliary part of the larger space. Then, we can express the evolution of an open quantum system $\mathfrak{S}$ in terms of the unitary joint evolution of the system and the environment together, which together comprise the entire universe (\emph{i.e.}, $\mathfrak{U} = \mathfrak{SE}$). If $\mathfrak{S}$ starts in the state $\rho_{\mathrm{in},\mathfrak{S}}$ and $\mathfrak{E}$ starts in the state $\rho_{\mathfrak{E}}$, the evolution appears as:
\begin{equation}
    \label{eq:dilation-thm}
    \begin{split}
        \rho_{\mathrm{in},\mathfrak{S}} \mapsto \Lambda_{t}\left[\rho_{\mathrm{in},\mathfrak{S}}\right] &\coloneqq \mathrm{Tr}_{\mathfrak{F}}\left[\widehat{U}_{t,\mathfrak{SE}}\left(\rho_{\mathrm{in},\mathfrak{S}}\otimes\rho_{\mathfrak{E}}\right)\widehat{U}_{t,\mathfrak{SE}}^{\dagger}\right] \\[4pt]
        &\,= \mathrm{Tr}_{\mathfrak{F}}\left[\mathrm{e}^{-\mathrm{i}\widehat{H}\left(t_\mathrm{f}\:-\:t_{0}\right)}\left(\rho_{\mathrm{in},\mathfrak{S}}\otimes\rho_{\mathfrak{E}}\right)\;\!\mathrm{e}^{\mathrm{i}\widehat{H}\left(t_\mathrm{f}\:-\:t_{0}\right)}\right].
    \end{split}
\end{equation}
As we noted earlier, the final state may not be in the same Hilbert space as the initial state; \emph{i.e.}, $\Lambda_{t}$ may not necessarily map $\mathfrak{S}$ to itself. This is reflected in the fact that we take the final trace over $\mathfrak{F}\subseteq\mathfrak{U}$, where $\mathfrak{F}$ may not necessarily be the same space as $\mathfrak{E}$. However, for our purposes, we will always have $\mathfrak{F} = \mathfrak{E}$.

In \eqref{eq:dilation-thm}, we defined $\widehat{U}_{t,\mathfrak{S^\prime E}} \coloneqq \mathrm{e}^{-\mathrm{i}\widehat{H}\left(t_\mathrm{f}\:-\:t_{0}\right)}$ as the global unitary evolution operator over all of $\mathfrak{U} = \mathfrak{SE}$. Here, $\widehat{H} = \widehat{H}_{\mathfrak{S}} + \widehat{H}_{\mathfrak{E}} + \widehat{H}_{\mathrm{I},\mathfrak{SE}}$ is the global Hamiltonian over all of $\mathfrak{SE}$, divided into the Hamiltonian $\widehat{H}_{\mathfrak{S}}$ over $\mathfrak{S}$ alone, the Hamiltonian $\widehat{H}_{\mathfrak{E}}$ over $\mathfrak{E}$ alone, and the interaction Hamiltonian $\widehat{H}_{\mathrm{I},\mathfrak{SE}}$ between $\mathfrak{S}$ and $\mathfrak{E}$. This representation forms the basis for both the existing NEQT results on Landauer's principle, as well as the GKSL framework for examining open quantum systems. Beyond the rules of quantum mechanics, the only additional assumptions here are that $\mathfrak{S}$ is coupled to some environment $\mathfrak{E}$ which it jointly evolves with unitarily, and that the initial state of $\mathfrak{SE}$ can be factorized \cite{Pechukas94}.

In terms of the dilation theorem, the set of transformations on $\rho_{\mathrm{in},\mathfrak{S}}$ allowed by thermodynamics is simply the set of unitary transformations $\widehat{U}_{t,\mathfrak{SE}}$ that preserves the total energy over all of $\mathfrak{SE}$. These transformations are explicitly described by the \emph{resource theory of quantum thermodynamics} (RTQT) \cite{NW18, Lostalgio19}. In general, quantum resource theories (QRT) provide a information-theoretic framework for describing all possible operations on a given state $\rho$, by describing the information cost of operations and states (in terms of new information we require about the system) \cite{CG19}. In particular, QRTs describe the conditions on operations to act at no additional information cost and provide the conditions on the types of states of new systems that can be prepared at no additional information cost and appended to the overall system. These are respectively known as the \emph{free operations} and \emph{free states}. In addition to these, the quantum resource theory provides the conditions on transformations on $\rho$, known as the \emph{state conversion conditions}. The nature of free operations, free states, and the conversion conditions depends on the specific resource theory.\footnote{As an example, the resource theory of asymmetry tells us the free operations and states of a system with an overall symmetry described by a compact Lie group $G$, with the free operations and conversion conditions given in terms of the unitary representations of $G$. Operations that are covariant with $G$ and states that are invariant under $G$ require no additional information beyond the group already specified; thus, these are respectively the free operations and free states of this resource theory. This example is expanded upon in \cite{CG19}, which also provides the illustrative example of the resource theory of bipartite entanglement.}

In RTQT, we start with the system Hamiltonian $\widehat{H}_{\mathfrak{S}}$ and the (inverse) environment temperature $\beta=1/T$. The thermal (Gibbs) states $\tau \coloneqq \mathrm{e}^{-\beta\widehat{H}}/ \mathrm{Tr}\left[\mathrm{e}^{-\beta\widehat{H}}\right]$ are the maximum-entropy states, which must necessarily be preserved by energy-preserving unitary operations \cite{BHNOW15}. Thus, these are the free states of the environment. As such, if we examine a system using the dilation theorem, it takes no additional information to set the initial state of the environment $\mathfrak{E}$ to be the thermal state $\tau_{\mathfrak{E}}$. (Conversely, selecting any other state \emph{does} involve extra information not specified in the resource theory; namely, information about the distribution of states over $\mathfrak{E}$.) Setting $\mathfrak{F}=\mathfrak{E}$, this gives us a direct expression for the free operations, which are known as the \emph{thermal operations}:
\begin{equation}
    \label{eq:TO}
    \rho_{\mathrm{in},\mathfrak{S}} \mapsto \Xi_{t}\left[\rho_{\mathrm{in},\mathfrak{S}}\right] \coloneqq \mathrm{Tr}_{\mathfrak{E}} \left[ \widehat{U}_{t,\mathfrak{SE}} \left( \rho_{\mathrm{in},\mathfrak{S}} \otimes \tau_{\mathfrak{E}} \right) \widehat{U}_{t,\mathfrak{SE}}^{\dagger} \right].
\end{equation}
The necessary conditions for these transformations to occur are called the \emph{thermomajorization conditions}. When the commutator $\left[\Xi_{t}\left[\rho_{\mathrm{in},\mathfrak{S}}\right],\widehat{H}_{\mathfrak{S}}\right] = 0$ (\emph{i.e.}, when the final state of $\mathfrak{S}$ has a definite energy value, as is the case for all of the systems we will be considering), these conditions are both necessary and sufficient.\footnote{The case of $\left[\Xi_{t}\left[\rho_{\mathrm{in},\mathfrak{S}}\right],\widehat{H}_{\mathfrak{S}}\right] \ne 0$ remains an open question \cite{CG19}; fortunately, that is also beyond the scope of our model.}

These conditions are defined in terms of the \emph{$\beta$-ordering} of a state $\rho_{\mathfrak{S}}$, which has eigenvalues $\{p_{i}\}_{i=1}^{n}$ and corresponds to a Hamiltonian $\widehat{H}_{\mathfrak{S}}$ (with $\left[\rho_{\mathfrak{S}},\widehat{H}_{\mathfrak{S}}\right] = 0$). The $\beta$-ordering $p^{\downarrow} = \left(p^{\downarrow}_{1}, \cdots, p^{\downarrow}_{n}\right)$ is defined \cite{HO13} as an ordering of the $p_{i}$ that satisfies $p^{\downarrow}_{i} \mathrm{e}^{-\beta E_{i}} \geq p^{\downarrow}_{j} \mathrm{e}^{-\beta E_{j}}$ for all $i < j$, where $E_{i}$ is the energy corresponding to $p_{i}$. (Thus, the $\beta$-ordering of the $\{p_{i}\}$s is defined by decreasing values of $p^{\downarrow}_{i} \mathrm{e}^{-\beta E_{i}}$.) From this ordering, we can define the \emph{thermomajorization curve} as the curve defined by the points:\footnote{It's worth noting that the ordering of the $\{p_{i}^{\downarrow}\}$s may \emph{not} be unique when some of the $p^{\downarrow}_{i} \mathrm{e}^{-\beta E_{i}}$ values are equal to each other, but even in this case the thermomajorization curve is \emph{always} unique.}
\begin{equation}
    \label{eq:thermomaj}
    \left\{ \left(0,0\right), \left(\mathrm{e}^{-\beta E_{1}}, p^{\downarrow}_{1} \right), \left(\mathrm{e}^{-\beta E_{1}} + \mathrm{e}^{-\beta E_{2}}, p^{\downarrow}_{1} + p^{\downarrow}_{2} \right), \cdots, \left(\sum_{i=1}^{n} \mathrm{e}^{-\beta E_{i}}, \sum_{i=1}^{n} p^{\downarrow}_{i} \right) \right\}.
\end{equation}
Then, finally, the thermal operation $\rho_{\mathrm{in},\mathfrak{S}} \mapsto \Xi_{t}\left[\rho_{\mathrm{in},\mathfrak{S}}\right]$ can occur if the thermomajorization curve of $\Xi_{t}\left[\rho_{\mathrm{in},\mathfrak{S}}\right]$ is below or equal to the thermomajorization curve of $\rho_{\mathrm{in},\mathfrak{S}}$ everywhere. Collectively, the thermal states, thermal operations, and thermomajorization conditions determine the complete set of states we can generate and transformations we can perform in quantum thermodynamics.

\mypara{Catalytic thermal operations and correlated systems.}
\label{par:ctos}

The concept of a thermal operation can be extended to the case of a \textit{catalytic thermal operation} (CTO), in which a component of the system is a so-called \textit{catalyst} subsystem which cycles back to the initial state. This can be an appropriate model for certain types of subsystems in a computer---for example, a periodic clock signal, such as a resonant clock-power oscillator for an adiabatic circuit (see \S\ref{ssec:tech}). Further, every digital data signal in a typical reversible logic technology (\emph{e.g.}, \cite{Fra+20b}) cycles from a standard ``neutral'' or no-information state to an information-bearing state, and then back to neutral; thus, every node in a typical reversible circuit effectively acts like a catalyst. (This will be discussed in more detail in~\S\ref{ssec:ctos-rc} and \S\ref{ssec:cto-rc-sys}.)

We now present the most general type of CTOs explicitly, following the presentation in \cite{Mueller18}. (Note that this examines a \emph{more} general class of transformations than the ones traditionally examined in the ``second laws of thermodynamics'' framework; the relationship between this presentation and the ``second laws'' is discussed in \S\ref{par:alpha-rre}.) If we divide the overall system $\mathfrak{S}$ into the subsystems $\mathfrak{T}$ and $\mathfrak{K}$, the catalyst $\mathfrak{K}$ is defined as a subsystem which is required within the overall dynamics of the system for the state transition $\rho_{\mathrm{in},\mathfrak{T}} \mapsto \Xi_{t} \left[\rho_{\mathrm{in},\mathfrak{T}}\right]$.\footnote{The reasons the catalyst is required are \emph{not} explored in the CTO framework; rather, the CTO framework merely assumes that a catalyst is required.} If the state of the catalyst is given as $\sigma_{\mathfrak{K}}$, then the transition of the state $\left(\rho_{\mathrm{in},\mathfrak{T}} \otimes \sigma_{\mathfrak{K}} \right)$ is given \cite{Mueller18} by:
\begin{equation}
    \label{eq:Markus-CTO}
    \left(\rho_{\mathrm{in},\mathfrak{T}} \otimes \sigma_{\mathfrak{K}} \right) \mapsto \xi_{\mathfrak{TK}} \coloneqq \mathrm{Tr}_{\mathfrak{E}} \left[ \widehat{U}_{t,\mathfrak{TKE}} \left( \rho_{\mathrm{in},\mathfrak{T}} \otimes \sigma_{\mathfrak{K}} \otimes \tau_{\mathfrak{E}} \right) \widehat{U}_{t,\mathfrak{TKE}}^{\dagger} \right].
\end{equation}
Here, we have $\left[\widehat{U}_{t,\mathfrak{TKE}}, \widehat{H}_{\mathfrak{TKE}}\right] = 0$, and $\mathrm{Tr}_{\mathfrak{KE}}\left[\xi_{\mathfrak{TK}}\right]$ is arbitrarily close to $\Xi_{t} \left[\rho_{\mathrm{in},\mathfrak{T}}\right]$ under the trace norm: for all $\epsilon \in \mathbb{R}^{+}$, there are allowed transformations with:
\begin{equation}
    \label{eq:Markus-CTO-trace-norm}
    \lVert\mathrm{Tr}_{\mathfrak{KE}}\left[\xi_{\mathfrak{TK}}\right] - \Xi_{t} \left[\rho_{\mathrm{in},\mathfrak{T}}\right]\rVert_{1} < \epsilon.
\end{equation}
Meanwhile, the catalytic condition requires $\mathrm{Tr}_{\mathfrak{TE}}\left[\xi_{\mathfrak{TK}}\right] = \sigma_{\mathfrak{K}}$. This is the most general type of CTO, which can be realized if and only if the final Helmholtz free energy $\mathcal{F}$ is less than or equal to the initial Helmholtz free energy. Out of equilibrium, the Helmholtz free energy of a state $\rho$ in a system governed by a Hamiltonian $\widehat{H}$ is given by \cite{BL55,PHS15}:
\begin{equation}
    \label{eq:noneq-F}
    \mathcal{F}\left(\rho\right) \coloneqq \mathrm{Tr}\left[\widehat{H}\rho\right] - k_{\mathrm{B}}T\!\;S\left(\rho\right).
\end{equation}
Here, $S\left(\rho\right)$ is the Rényi-1 entropy (\emph{i.e.}, the von Neumann entropy) of $\rho$. In terms of $\mathcal{F}$, the condition we require for the CTO in \eqref{eq:Markus-CTO} to be realizable is:\footnote{In the case that $\mathcal{F}\left(\rho_{\mathrm{in},\mathfrak{T}}\right) - \mathcal{F}\left(\mathrm{Tr}_{\mathfrak{K}} \left[ \xi_{\mathfrak{TK}} \right] \right) = E$, the condition \eqref{eq:Markus-2nd-law} can be extended by including an uncorrelated source $\mathfrak{W}$ of free energy. In this case, if $\mathfrak{W}$ has the states $\lambda_{\mathfrak{W}}$ and $\kappa_{\mathfrak{W}}$ with $\mathcal{F}\left(\kappa_{\mathfrak{W}}\right) - \mathcal{F}\left(\lambda_{\mathfrak{W}}\right) \geq E$, then we can realize the CTO $\rho_{\mathrm{in},\mathfrak{T}} \otimes \sigma_{\mathfrak{K}} \otimes \kappa_{\mathfrak{W}} \mapsto \xi_{\mathfrak{TK}} \otimes \lambda_{\mathfrak{W}}$ when $\mathcal{F}\left( \rho_{\mathrm{in},\mathfrak{T}} \otimes \sigma_{\mathfrak{K}} \otimes \kappa_{\mathfrak{W}} \right) \geq \mathcal{F}\left( \xi_{\mathfrak{TK}} \otimes \lambda_{\mathfrak{W}} \right)$. Here, the condition \eqref{eq:Markus-QMI-diff} remains. Naturally, if we generalize to the case where correlations between $\mathfrak{TK}$ and $\mathfrak{W}$ are permitted, we return to the CTO given in \eqref{eq:Markus-CTO}.}
\begin{equation}
    \label{eq:Markus-2nd-law}
    \mathcal{F}\left(\mathrm{Tr}_{\mathfrak{KE}} \left[ \xi_{\mathfrak{TK}} \right] \right) \leq \mathcal{F}\left(\rho_{\mathrm{in},\mathfrak{T}}\right).
\end{equation}
Notably, these CTOs do not impose any additional major constraints on the shape of the correlations between $\mathfrak{T}$ and $\mathfrak{K}$: for \emph{any} $\delta \in \mathbb{R}^{+}$, there exists some $\mathfrak{K}$ and $\xi_{\mathfrak{TK}}$ such that $\widehat{H}_{\mathfrak{K}} = 0 $ and the \emph{quantum mutual information} $I\left(\mathfrak{T}:\mathfrak{K}\right)$ between $\mathfrak{T}$ and $\mathfrak{K}$ is bounded by $\delta$:
\begin{equation}
    \label{eq:Markus-QMI-diff}
    I\left(\mathfrak{T} :\mathfrak{K} \right) \coloneqq S\left(\xi_{\mathfrak{TK}} \Vert \;\! \mathrm{Tr}_{\mathfrak{TE}} \left[\xi_{\mathfrak{TK}}\right] \otimes \mathrm{Tr}_{\mathfrak{KE}} \left[\xi_{\mathfrak{TK}}\right] \right) < \delta.
\end{equation}
In practical terms, this means that we can achieve state transitions from $\rho_{\mathrm{in},\mathfrak{T}}$ to $\mathrm{Tr}_{\mathfrak{K}} \left[ \xi_{\mathfrak{TK}} \right]$ by engineering the catalyst and the CPTP map $\Xi_{t}$ to minimize the correlation $I\left(\mathfrak{T} :\mathfrak{K} \right)$. This process of \emph{correlation engineering} \cite{Fra18,Mueller18} lies at the heart of reversible computing: By engineering interacting subsystems bearing computational degrees of freedom and the transformations $\Xi_{t}$ applied on them, we can achieve the CTOs given in \eqref{eq:Markus-CTO}, with the net energy dissipation given by the free energy difference in \eqref{eq:Markus-2nd-law}.

\mypara{Uncorrelated catalytic thermal operations.}
\label{par:alpha-rre}

The expressions in \S\ref{par:ctos} may come as a surprise to those familiar with thermal operations and catalytic thermal operations. Conventionally, CTOs are defined \cite{NW18} by the transformation:
\begin{equation}
    \label{eq:normal-CTO}
    \left(\rho_{\mathrm{in},\mathfrak{T}} \otimes \sigma_{\mathfrak{K}} \right) \mapsto \mathrm{Tr}_{\mathfrak{E}} \left[ \widehat{U}_{t,\mathfrak{TKE}} \left( \rho_{\mathrm{in},\mathfrak{T}} \otimes \sigma_{\mathfrak{K}} \otimes \tau_{\mathfrak{E}} \right) \widehat{U}_{t,\mathfrak{TKE}}^{\dagger} \right] = \Pi_{t} \left[\rho_{\mathrm{in},\mathfrak{T}}\right] \otimes \sigma_{\mathfrak{K}}.
\end{equation}
The data processing inequality (DPI) \cite{Bea12} can give us necessary conditions for these CTOs to be realized, which become necessary and sufficient when $\left[\Pi_{t}\left[\rho_{\mathrm{in},\mathfrak{T}}\right]\otimes\sigma_{\mathfrak{K}},\widehat{H}_{\mathfrak{TK}}\right]=0$ \cite{NW18,RW20}. For any information distance function $f\left(\rho\Vert\sigma\right)$ of the density matrices $\rho$ and $\sigma$, and for any CPTP map $\Lambda_{t}$, the DPI gives:
\begin{equation}
    \label{eq:DPI}
    f\left(\rho\Vert\sigma\right) \geq f\left(\Lambda_{t}\left(\rho\right)\Vert\Lambda_{t}\left[\sigma\right]\right).
\end{equation}
In other words, the DPI is a requirement that must be satisfied for all functions $f$ for $\Lambda_{t}$ to be a valid CPTP map. One family of such functions are the $\alpha$\emph{-relative Rényi entropies} ($\alpha$-RRE), which are defined \cite{BHNOW15, MDSFT13, vEH14} as:
\end{paracol}
\begin{equation}
    \label{eq:alpha-RRE}
    S_{\alpha}\left(\rho\Vert\sigma\right) \coloneqq
    \begin{dcases}
        \frac{\mathrm{sgn}\:\alpha}{\alpha - 1}\:\mathrm{ln}\frac{\mathrm{Tr} \left[\rho^{\alpha}\:\sigma^{1-\alpha}\right]}{\mathrm{Tr}\:\rho} & \alpha \in \left(-1,0\right)\cup\left(0, 1\right); \\[2pt]
        \frac{\mathrm{sgn}\:\alpha}{\alpha - 1}\:\mathrm{ln} \left\{\frac{\mathrm{Tr}\:\left[\left(\sigma^{\left(1-\alpha\right)/2\alpha} \rho\: \sigma^{\left(1-\alpha\right)/2\alpha}\right)^{\alpha}\right]}{\mathrm{Tr}\:\rho}\right\} & \alpha \in \left(-\infty,-1\right)\cup\left(1,\infty\right); \\[2pt]
        \mathrm{Tr}\:\left[\rho\:\!\left(\mathrm{ln}\:\rho - \mathrm{ln}\:\sigma\right)\right] & \mathrm{lim}\:\alpha \rightarrow 1.
    \end{dcases}
\end{equation}
\begin{paracol}{2}
\switchcolumn
The $\alpha \rightarrow 1$ limit provides us with the familiar expression for the \emph{quantum relative divergence} (QRD)\footnote{The $\alpha \rightarrow 1$ limit can be taken either with a monotonically increasing sequence from the $\alpha \in \left(0,1\right)$ case, or with a monotonically decreasing sequence from the $\alpha \in \left(1,\infty\right)$ case. As with the QRDs, other familiar entropies are recovered as limiting cases of the $\alpha$-RREs: the \emph{0-RRE} $S_{0}\left(\rho\Vert\sigma\right) = \mathrm{ln}\:\mathrm{Tr}_{\mathrm{supp}\:\rho}\:\left[\sigma\right]$ is given by the $\alpha \searrow 0$ limit, the \emph{max-RRE} $S_{\infty}\left(\rho\Vert\sigma\right) = \mathrm{inf}\:\!\left\{\lambda\in\mathbb{R}\:\mid\:\rho\leq \mathrm{e}^{\lambda}\sigma\right\}$ is given by the $\alpha \nearrow \infty$ limit, and the $S_{-1}\left(\rho\Vert\sigma\right)$ and $S_{-\infty}\left(\rho\Vert\sigma\right)$ cases are given by interchanging $\rho$ and $\sigma$ in the arguments. It's also worth noting that the expression for $\alpha\in\left(1,\infty\right)$ in \eqref{eq:alpha-RRE} is the conventional form for the $\alpha$-RRE at these values of $\alpha$; this expression is called the \emph{sandwiched RRE}. However, because in general $\rho$ and $\sigma$ do not commute with each other, there are an infinite number of ways to arrange powers of $\rho$ and $\sigma$ that satisfy the Rényi entropy axioms \cite{Renyi55} and retrieve the appropriate limiting cases. These can all be expressed as a single two-parameter family of entropies \cite{AD15}, known as the the $\alpha$\emph{-z-RREs}.}. As such, the DPI imposes the requirement that the CTO \eqref{eq:normal-CTO} must satisfy
\begin{equation}
    \label{eq:alpha-RRE-CTO-condition}
    S_{\alpha}\left(\rho_{\mathrm{in},\mathfrak{T}}\otimes\sigma_{\mathfrak{K}}\Vert\tau_{\mathfrak{TK}}\right) \geq S_{\alpha}\left(\Pi_{t}\left[\rho_{\mathrm{in},\mathfrak{T}}\right]\otimes\sigma_{\mathfrak{K}}\Vert\tau_{\mathfrak{TK}}\right)
\end{equation}
for all $\alpha$ as a necessary condition.\footnote{Indeed, the DPI tells us that a necessary condition for the CTO \eqref{eq:normal-CTO} to be valid is that $f\left(\rho_{\mathrm{in},\mathfrak{T}}\otimes\sigma_{\mathfrak{K}}\Vert\tau_{\mathfrak{TK}}\right)\geq f\left(\Pi_{t}\left[\rho_{\mathrm{in},\mathfrak{T}}\right]\otimes\sigma_{\mathfrak{K}}\Vert\tau_{\mathfrak{TK}}\right)$ for \emph{all} functions $f$ of $\rho_{\mathrm{in},\mathfrak{T}}\otimes\sigma_{\mathfrak{K}}$ and $\tau_{\mathfrak{TK}}$.} In the case that we have transition from the product state $\rho_{\mathrm{in},\mathfrak{T}}\otimes\sigma_{\mathfrak{K}}$ to the product state $\Pi_{t}\left[\rho_{\mathrm{in},\mathfrak{T}}\right]\otimes\sigma_{\mathfrak{K}}$, these are in fact \emph{sufficient} conditions, beyond being simply necessary ones \cite{Klimesh04, Klimesh07, Turgut07}. Thus, \eqref{eq:alpha-RRE-CTO-condition} tells us the constraints we need to satisfy the CTOs \eqref{eq:normal-CTO}. The $S_{\alpha}\left(\rho\Vert\sigma\right)$ in turn define \cite{BHNOW15} the $\alpha$\emph{-Helmholtz ``free energies}:''
\begin{equation}
    \label{eq:alpha-F}
    \mathcal{F}_{\alpha}\left(\rho_{\mathrm{in},\mathfrak{S}}\right) \coloneqq -k_{\mathrm{B}}T\!\;\mathrm{ln}\;\!Z + S_{\alpha}\left(\rho_{\mathrm{in},\mathfrak{S}}\Vert\tau_{\mathfrak{S}}\right).
\end{equation}
We can immediately recognize the $\alpha = 1$ case as equivalent to the expression \eqref{eq:noneq-F}. The CTOs defined in \eqref{eq:normal-CTO} are realized when we have
\begin{equation}
    \label{eq:2nd-laws-thermo}
    \mathcal{F}_{\alpha}\left(\Pi_{t}\left[\rho_{\mathrm{in},\mathfrak{T}}\right]\right) \leq \mathcal{F}_{\alpha}\left(\rho_{\mathrm{in},\mathfrak{T}}\right)
\end{equation}
for all $\alpha$. These conditions are known as the ``second laws of thermodynamics'' \cite{BHNOW15}.


\mypara{Correlated vs.\ uncorrelated CTOs.}
\label{par:markus-vs-2ndlaws}
The expressions for the $\alpha$-RREs in \S\ref{par:alpha-rre} might initially be cause for some concern, since a CTO must satisfy \eqref{eq:2nd-laws-thermo} for \emph{all} alpha to be a viable transition. This concern may escalate to alarm when we consider that the $\alpha$-RREs (and thus the $\alpha$-free energies) are monotone in $\alpha$; \emph{i.e.}, $F_{\beta}\left(\rho\right) \leq F_{\gamma}\left(\rho\right)$ for all $\beta \leq \gamma$. Beyond the standard Helmholtz free energy ($F_{1}$), two notable cases are the \emph{extractable work} $F_{0}$ and the \emph{work of formation} $F_{\infty}$. As their names imply, these are respectively the amount of work we can extract from a given state and the amount of work it takes to form that same state. Since in most cases we have $F_{0} < F_{\infty}$, it would appear that the energy difference $F_{\infty} - F_{0}$ is simply dissipated in the process of creating a state and then extracting work from that state. As a corollary, this would imply that our only hope for a viable reversible computing framework in this formulation is to find sets of states $\left\{\rho_{i}\right\}$ where the equality between $F_\alpha\left(\rho_{i}\right)$ is satisfied for all $\alpha$, which may be a highly restrictive condition.

As discussed in \cite{Mueller18}, however, the ``second laws of thermodynamics'' (and these attendant issues) arise from an additional assumption about the shape of CTOs. Specifically, the CTOs \eqref{eq:normal-CTO} that give rise to the ``second laws of thermodynamics'' assume that the final state of $\mathfrak{S}$ after the thermal operation is a product state of $\mathfrak{T}$ and $\mathfrak{K}$, \emph{i.e.}, that catalytic thermal operations transform the state $\rho_{\mathrm{in},\mathfrak{T}} \otimes \sigma_{\mathfrak{K}}$ to the state $\Pi_{t} \left[\rho_{\mathrm{in},\mathfrak{T}}\right] \otimes \sigma_{\mathfrak{K}}$. However, by \emph{definition} of the catalytic thermal operation, we needed the presence of $\sigma_{\mathfrak{K}}$ to induce the transformation to begin with. Thus, the CTO on $\rho_{\mathrm{in},\mathfrak{T}} \otimes \sigma_{\mathfrak{K}}$ \emph{necessitates} an increase in the QMI between $\mathfrak{T}$ and $\mathfrak{K}$, specifically given by \eqref{eq:Markus-QMI-diff}. Indeed, as proven in \cite{Mueller18}, this mutual information can be made to be infinitesimally small, but \emph{cannot} be zero. Thus, the CTO in \eqref{eq:normal-CTO}, in which we demand that the final state of $\mathfrak{S}$ be in the product state $\Pi_{t} \left[\rho_{\mathrm{in},\mathfrak{T}}\right] \otimes \sigma_{\mathfrak{K}}$, can be thought of as performing the general CTO \eqref{eq:Markus-CTO} and then \emph{ejecting} the QMI \eqref{eq:Markus-QMI-diff}. A direct consequence of this is that, as proven in \cite{Mueller18}, in the general CTO \eqref{eq:Markus-CTO} where we permit correlations to develop between the system and catalyst, the ($\alpha = 1$) Helmholtz free energy \emph{uniquely} specifies the condition required for the transition to take place.\footnote{Thus, \cite{Mueller18} verifies the conjecture first provided in \cite{WGE17}: in the words of \cite{WGE17} directly, ``the [($\alpha = 1$) Helmholtz] free energy is singled out as a measure of athermality''.}

Consequently, if we seek to develop a framework for computing which reduces energy dissipation by avoiding the energy cost of expelling the built up QMI, our computing operations must follow the CTO expression given in \eqref{eq:Markus-CTO}. Since reversible computing is precisely this framework, \eqref{eq:Markus-CTO} provides an explicit expression for the shape of reversible computing operations in terms of CTOs. As a trade-off, we achieve these operations via a buildup of QMI \eqref{eq:Markus-QMI-diff}, which can be made arbitrarily small but cannot be precisely zero. In the framework of reversible computing, this is an acceptable (indeed, \emph{preferred}) trade-off to make\footnote{As pointed out in \cite{Mueller18}, the more general CTOs \eqref{eq:Markus-CTO} aren't necessarily an \emph{improved} form of the CTOs \eqref{eq:normal-CTO}, but rather simply offer a different setup. In lieu of the unavoidable ``free energy'' differences $F_{\infty} - F_{0}$, we've accepted the unavoidable buildup of QMI as a trade-off. For our purposes here, building up QMI and engineering the system and CTO to minimize the QMI and the difference $\lVert\mathrm{Tr}_{\mathfrak{K}}\left[\xi_{\mathfrak{T}\mathfrak{K}}\right] - \Xi_{t}\left[\rho_{\mathrm{in},\mathfrak{T}}\right]\rVert$ is preferred, but the optimal type of CTO will in general be a function of the type of process we're interested in.}.

\subsubsection{Quantum Mechanical Models of the Landauer Bound}
\label{sssec:quant-land}

The general CTOs given in \cite{Mueller18} and discussed in \S\ref{par:ctos} further give us a conceptual framework for understanding the nonequilibrium Landauer bound \cite{Goold15} and the difference between conditional and unconditional Landauer state reset \cite{Anderson19}. First, we briefly review the Landauer principle, following the excellent presentation found in \cite{Anderson19}, before connecting these to the nonequilibrium Landauer bound and the general CTOs found in \cite{Goold15,Mueller18}.

\mypara{Conditional vs.\ unconditional Landauer erasure.}
\label{par:neal}

\end{paracol}

\begin{figure}[t] 
\widefigure
    \centerline{(a) \includegraphics[width=8 cm]{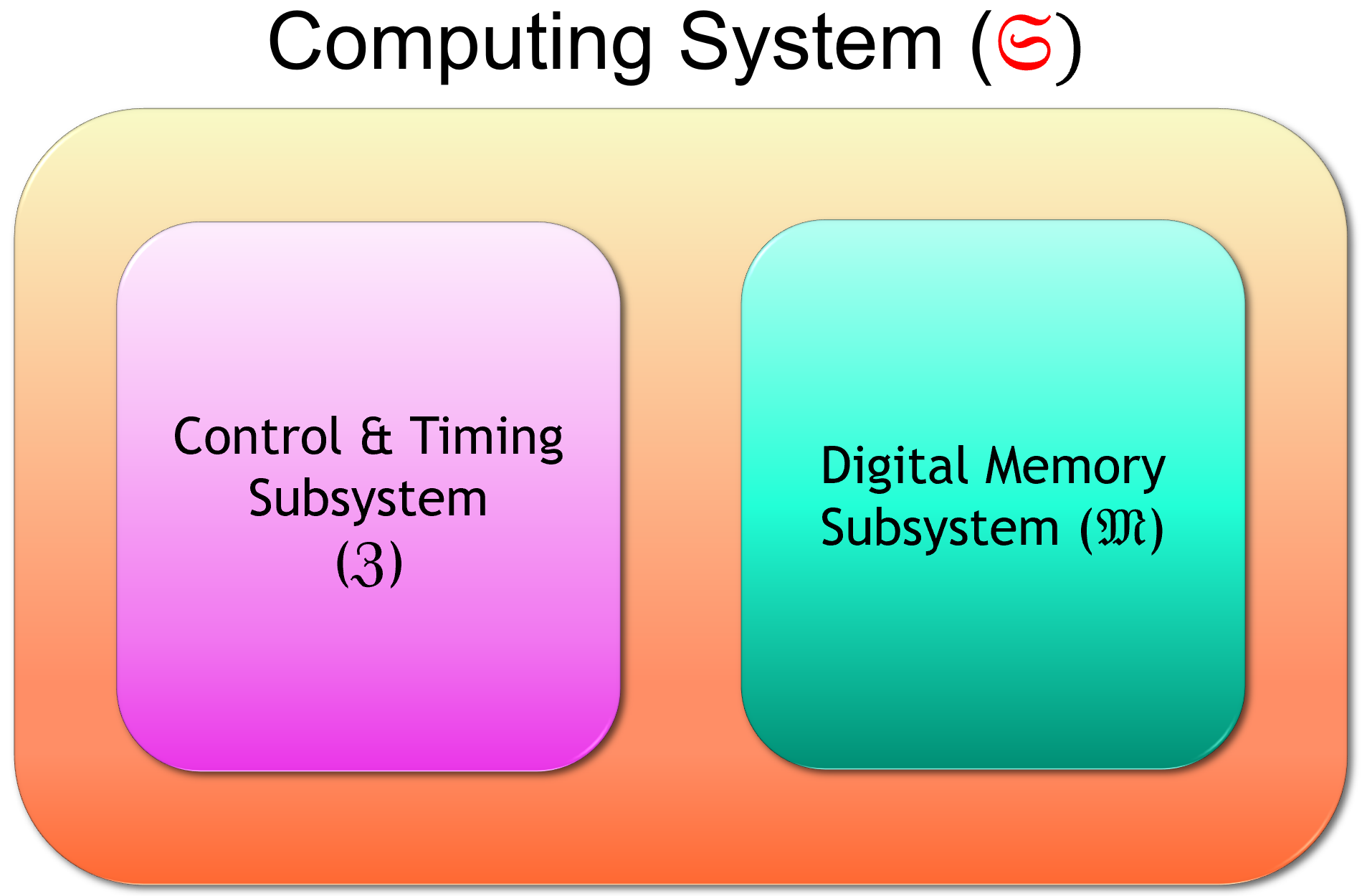} (b) \includegraphics[width=9 cm]{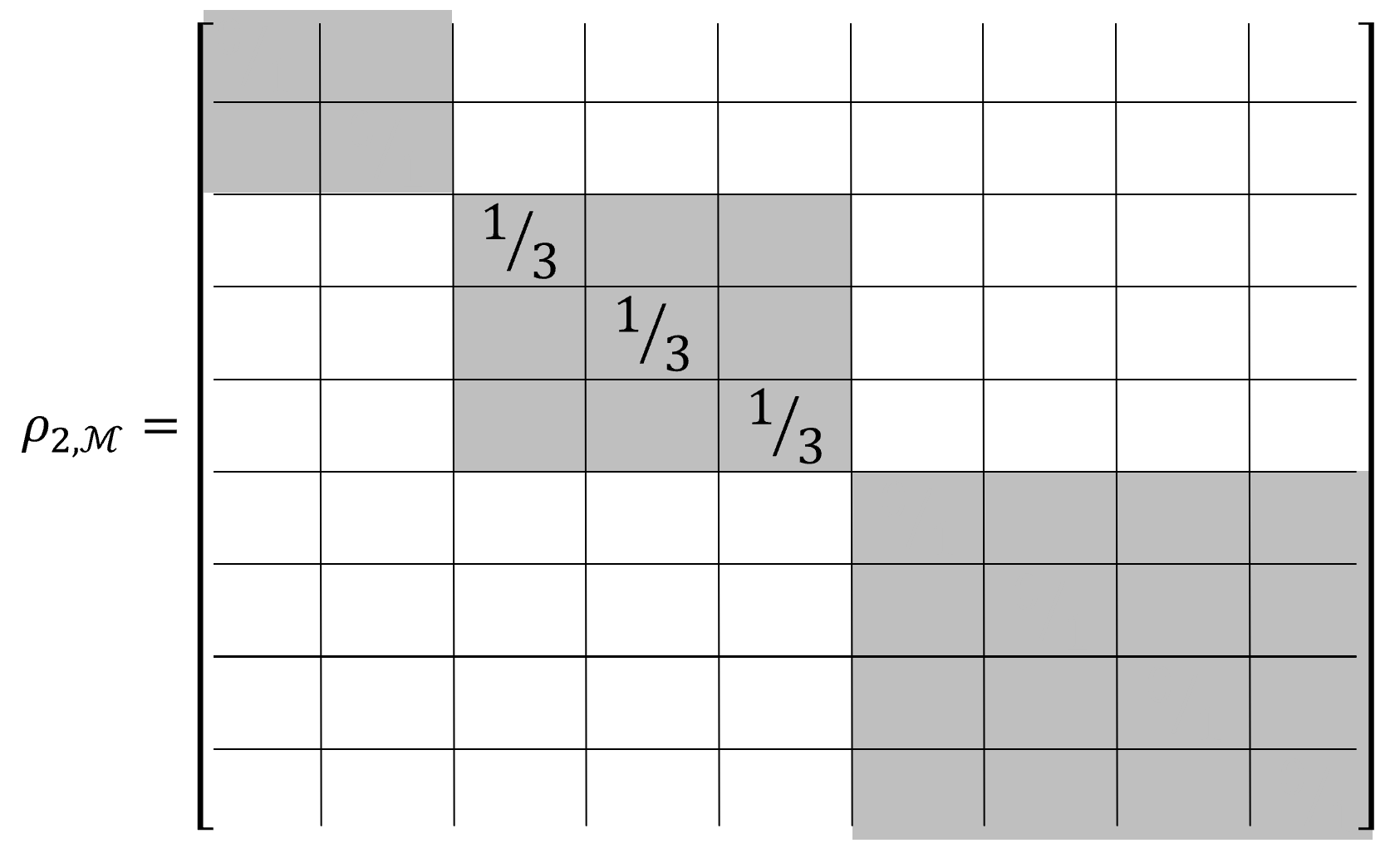}}
    \caption{Breakdown of subsystems for purposes of \S\ref{par:neal}, \textit{etc}\@. (a) For our purposes in this section, and in much of what follows, we focus our attention on a subsystem $\mathfrak{S}^{\prime} = \mathfrak{M}$ of the entire physical computer system (a ``memory'') that exists for the purpose of passively registering some computational data of interest, but does not include any active mechanisms for controlling the timing and performance of state transitions. (b) An example of a density matrix representation of a computational state of $\mathfrak{M}$ in the block-diagonal picture from Fig.~\ref{fig:block-diag}. In this example, the non-computational subsystem of $\mathfrak{M}$ is assumed to be in a maximum-entropy mixed state conditioned on the computational state being $c=c_2$.
    \label{fig:neal}}
\end{figure}
\begin{paracol}{2}
\switchcolumn

Here, and in much of what follows, we restrict our attention to a subsystem $\mathfrak{S}^{\prime} = \mathfrak{M}$ of the entire computer system $\mathfrak{S}$ that plays the role of passively registering data; we can generically call such a subsystem a ``memory,'' without implying any particular architectural structure ({\em i.e.}, it could be any information-bearing set of signals in the machine).  This is to be distinguished from other components that actively manipulate the state of the machine and control the timing of operations, which we will assume are separated into another subsystem $\mathfrak{Z}$ which will usually be left implicit. See Figure~\ref{fig:neal}(a). However, note that even $\mathfrak{M}$, as a physical system, can still be separated into computational and non-computational subsystems as in Fig.~\ref{fig:c-vs-n}, and thus (assuming, as usual, no coherences between digital states) still has states in block-diagonal form as in Fig.~\ref{fig:block-diag}. Each computational state $c_i\in\boldsymbol{C}$ of $\mathfrak{M}$ thus has a unique corresponding representation $\rho_{i,\mathfrak{M}}$ as a density matrix if we assume minimal information about the non-computational part of the state, that is, taking it to have maximum entropy, given the specifications as to what constitutes the set $\boldsymbol{B}_i$ of microstates of $\mathfrak{M}$ validly representing the given computational state $c_i$ in a given technological scenario. See Fig.~\ref{fig:neal}(b). Note that the following discussion blurs the distinction between these density matrices and the abstract states $c_i$ that they represent, and calls them ``computational states'' even though, in the density matrix form, they are also manifestly \textit{physical} entities.

Now, for a subsystem $\mathfrak{S}^{\prime}=\mathfrak{M}$ carrying some computational degrees of freedom, the Landauer state reset process, following \cite{Anderson19}, is the process by which the state $\rho_{\ell,\mathfrak{S}^{\prime}}$ is set to some standard reset state $\rho_{\mathrm{r},\mathfrak{S}^{\prime}}$. In a practical implementation of general computational operations on the system $\mathfrak{S}^{\prime}$ which bears computational degrees of freedom, the reset state is important for describing the operation of typical reversible computing systems such as described in \S\ref{ssec:tech}. The reset state is a standard, known reference state; when we perform operations on the system which correspond to operations, we typically transform the system from $\rho_{\mathrm{r},\mathfrak{S}^{\prime}}$ in a known way such that the operations we perform on $\rho_{\mathrm{r},\mathfrak{S}^{\prime}}$ correspond to sensible computational operations. The end result of this series of operations will be the final computational state, which we then typically need to reset to the standard state in order to perform a new set of operations.

As discussed in \S\ref{ssec:found}, in general, we expect the computer to have a possibly very large, but finite number $N$ of total possible computational states $\left\{\rho_{\ell,\mathfrak{S}^{\prime}}\right\}_{\ell\:=\:1}^{N}$ that it could be in at any time. Then, the reset process we're interested in is the set of the CPTP maps $\left\{\rho_{\ell,\mathfrak{S}^{\prime}}\right\} \mapsto \rho_{\mathrm{r},\mathfrak{S}^{\prime}}$ for all $\ell \in \left\{1,\cdots, N\right\}$. We can start by taking these operations to be thermal operations of the form \eqref{eq:TO}:
\begin{equation}
    \label{eq:cond-Land-TO}
    \begin{split}
        \rho_{\ell,\mathfrak{S}^{\prime}} \mapsto \Xi_{\ell,t}\left[\rho_{\ell,\mathfrak{S}^{\prime}}\right] &\coloneqq \mathrm{Tr}_{\mathfrak{E}} \left[ \widehat{U}_{\ell,t,\mathfrak{S^\prime E}} \left( \rho_{\ell,\mathfrak{S}^{\prime}} \otimes \tau_{\mathfrak{E}} \right) \widehat{U}_{\ell,t,\mathfrak{S^\prime E}}^{\dagger} \right] \\[4pt]
        &\,= \mathrm{Tr}_{\mathfrak{E}} \left[ \rho_{\mathrm{f}\mathrm{c},\mathfrak{U}} \right]= \rho_{\mathrm{r},\mathfrak{S}^{\prime}}.
    \end{split}
\end{equation}
(As before, $\tau_{\mathfrak{X}}$ denotes the thermal / Gibbs state in the system / subsystem $\mathfrak{X}$.) Here, we've defined $\rho_{\mathrm{f}\mathrm{c},\mathfrak{U}}$ as the final \emph{global} state of the entire universe $\mathfrak{U} = \mathfrak{S^\prime E}$ following application of the $\widehat{U}_{\ell,t,\mathfrak{S^\prime E}}$ evolution to $\left(\rho_{\ell,\mathfrak{S}^{\prime}}\otimes\tau_{\mathfrak{E}}\right)$:
\begin{equation}
    \label{eq:cond-Land-final-global-state}
    \rho_{\mathrm{f}\mathrm{c},\mathfrak{U}} \coloneqq \widehat{U}_{\ell,t,\mathfrak{S^\prime E}} \left( \rho_{\ell,\mathfrak{S}^{\prime}} \otimes \tau_{\mathfrak{E}} \right) \widehat{U}_{\ell,t,\mathfrak{S^\prime E}}^{\dagger}.
\end{equation}
Crucially, the effect of the unitary evolution operators $\widehat{U}_{\ell,t,\mathfrak{S^\prime E}}$ over all of $\mathfrak{U}$ is to transform the set of $N$ initial states into the \emph{same} final state over all of $\mathfrak{U}$, given by $\rho_{\mathrm{f}\mathrm{c},\mathfrak{U}}$. The overall unitary evolution operators $\widehat{U}_{\ell,t,\mathfrak{S^\prime E}}$ are given by:\footnote{Note that the exponentials in (\ref{eq:cond-Land-U}) are implicitly required to be time-ordered if the potential $\widehat{V}_{\ell r,\mathfrak{S}^{\prime}}$ is time-dependent, as is the general case.}
\begin{equation}
    \label{eq:cond-Land-U}
    \begin{split}
        \widehat{U}_{\ell,t,\mathfrak{S^\prime E}} &= \mathcal{T} \mathrm{exp} \left\{-\frac{\mathrm{i}}{\hbar}\int \limits_{t_{0}}^{t_\mathrm{f}} \mathrm{d}t'\:\widehat{H} \right\} \\[4pt]
        &= \mathcal{T} \mathrm{exp} \left\{-\frac{\mathrm{i}}{\hbar}\int \limits_{t_{0}}^{t_\mathrm{f}} \mathrm{d}t'\: \left[\widehat{H}_{\mathfrak{S}^{\prime}} + \widehat{H}_{\mathfrak{E}} + \widehat{H}_{\mathrm{I}, \mathfrak{S^\prime E}} + \widehat{V}_{\ell r,\mathfrak{S}^{\prime}}\right] \right\}.
    \end{split}
\end{equation}
Here, in addition to the terms contributing to $\widehat{H}$ we outlined in \eqref{eq:dilation-thm}, we also explicitly pulled out the reset Hamiltonian $\widehat{V}_{\ell r,\mathfrak{S}^{\prime}}$. Note that the reset Hamiltonian is applied \emph{solely} to $\mathfrak{S}^{\prime}$.

An extremely important feature of the set 
$\left\{\widehat{U}_{\ell,t,\mathfrak{S^\prime E}}\right\}$ of unitary operators is that we have a \emph{unique} operator for each state $\rho_{\ell,\mathfrak{S}^{\prime}}$, and that the distinctiveness of each of these operators comes \emph{solely} from the fact that the reset Hamiltonian $\widehat{V}_{\ell r,\mathfrak{S}^{\prime}}$ is individualized for each $\rho_{\ell,\mathfrak{S}^{\prime}}$. Each of these operators gives us a distinct CPTP map $\Xi_{\ell,t}\left(\rho_{\ell,\mathfrak{S}^{\prime}}\right)$. As a result, the expressions \eqref{eq:cond-Land-TO} and \eqref{eq:cond-Land-U} correspond to \emph{conditional Landauer reset}; \emph{i.e.}, the process of resetting the computational state $\rho_{\ell,\mathfrak{S}^{\prime}}$ to the standard state $\rho_{\mathrm{r},\mathfrak{S}^{\prime}}$ where the reset protocol $\widehat{U}_{\ell,t,\mathfrak{S^\prime E}}$ is conditioned on the specific state $\rho_{\ell,\mathfrak{S}^{\prime}}$. In other words, the process of conditional Landauer erasure involves selecting $\widehat{U}_{\ell,t,\mathfrak{S^\prime E}}$ (and, even more specifically, selecting $\widehat{V}_{\ell r,\mathfrak{S}^{\prime}}$) for each $\rho_{\ell,\mathfrak{S}^{\prime}}$ such that the final state $\rho_{\mathrm{f}\mathrm{c},\mathfrak{U}}$ is the same for \emph{any} initial state $\rho_{\ell,\mathfrak{S}^{\prime}} \otimes \tau_{\mathfrak{E}}$ we choose.

A central quantity of interest in the Landauer reset process is the lower bound on the amount of energy transfer (a.k.a. the \emph{dissipation}) from the system to the environment. This can be calculated by examining the change in the environment energy $\Delta\left\langle E_{\ell,\mathfrak{E}}\right\rangle$ during the evolution \eqref{eq:cond-Land-TO}. In terms of this evolution, the final state of the environment is given by:
\begin{equation}
    \label{eq:cond-Land-final-state-env}
    \rho_{\mathrm{f}\mathrm{c},\mathfrak{E}} = \mathrm{Tr}_{\mathfrak{S}^{\prime}} \left[\rho_{\mathrm{f}\mathrm{c},\mathfrak{U}}\right] = \mathrm{Tr}_{\mathfrak{S}^{\prime}} \left[ \widehat{U}_{\ell,t,\mathfrak{S^\prime E}} \left( \rho_{\ell,\mathfrak{S}^{\prime}} \otimes \tau_{\mathfrak{E}} \right) \widehat{U}_{\ell,t,\mathfrak{S^\prime E}}^{\dagger} \right].
\end{equation}
Because $\rho_{\mathrm{f}\mathrm{c},\mathfrak{E}}$ is the same for all $\ell$, we can directly examine the energy increase of the environment as a result of conditional Landauer reset protocol applied to any of the initial states:
\begin{equation}
    \label{eq:cond-Land-Delta-E}
    \begin{split}
        \Delta \left\langle E_{\ell,\mathfrak{E}}\right\rangle_{\mathrm{c}} &= \mathrm{Tr}\left[\rho_{\mathrm{f}\mathrm{c},\mathfrak{E}} \widehat{H}_{\mathfrak{E}} \right] - \mathrm{Tr}\left[\tau_{\mathfrak{E}} \widehat{H}_{\mathfrak{E}} \right]\\[4pt] 
        &= \mathrm{Tr}_{\mathfrak{E}}\left[\mathrm{Tr}_{\mathfrak{S}^{\prime}} \left[ \widehat{U}_{\ell,t,\mathfrak{S^\prime E}} \left( \rho_{\ell,\mathfrak{S}^{\prime}} \otimes \tau_{\mathfrak{E}} \right) \widehat{U}_{\ell,t,\mathfrak{S^\prime E}}^{\dagger} \right] \widehat{H}_{\mathfrak{E}} \right] - \mathrm{Tr}\left[\tau_{\mathfrak{E}} \widehat{H}_{\mathfrak{E}} \right].
    \end{split}
\end{equation}
(Here, the subscript $\mathrm{c}$ indicates that this is specifically for the conditional Landauer reset.) For a pair of interacting systems $a$ and $b$ in which $b$ is initially in a thermal state, we can straightforwardly derive the inequality $\Delta \left(S_{b} - \beta U_{b}\right) \leq 0$ (where $U_{b} \coloneqq \mathrm{Tr}\left[\rho_{b} \widehat{H}_{b}\right]$) from the basic definition of entropy and its convexity property \cite{Partovi89}.\footnote{This is sometimes referred to in the literature as \emph{Partovi's inequality}.} For this system, this gives $\Delta\left\langle E_{\mathfrak{E}} \right\rangle \geq -k_{\mathrm{B}}T\:\mathrm{ln}\:2\,\Delta S_{\mathfrak{E}}$. Combining this inequality with the strong subadditivity of the von Neumann entropy, $\Delta\left\langle E_{\ell,\mathfrak{E}}\right\rangle_{\mathrm{c}}$ has a lower bound given by \cite{Anderson19}:
\begin{equation}
    \label{eq:cond-Land-bound}
    \Delta \left\langle E_{\ell,\mathfrak{E}}\right\rangle_{\mathrm{c}} \geq -k_{\mathrm{B}}T\:\mathrm{ln}\:2\:\Delta S_{\ell,\mathfrak{S}^{\prime}} \coloneqq -k_{\mathrm{B}}T\:\mathrm{ln}\:2\:\left[S\left(\Xi_{\ell,t}\left(\rho_{\ell,\mathfrak{S}^{\prime}}\right)\right) - S\left(\rho_{\ell,\mathfrak{S}^{\prime}}\right)\right].
\end{equation}
Here, $\Delta S_{\ell,\mathfrak{S}^{\prime}}$ is the change in the von Neumann entropy between the initial and final states of $\mathfrak{S}^{\prime}$. Thus, when the reset protocol is given as in \eqref{eq:cond-Land-TO}, the sole contribution to the bound on dissipation into the environment is given by the change in entropy in $\mathfrak{S}^{\prime}$ induced by the overall unitary evolution $\widehat{U}_{\ell,t,\mathfrak{S^\prime E}}$.\footnote{As a reminder, the unitary dynamics over $\mathfrak{S^\prime E}$ by construction cannot increase the entropy over $\mathfrak{S^\prime E}$ as a whole \cite{Zur03,PRGWE16}. Since we don't \emph{a priori} assume that $\widehat{U}_{\ell,t,\mathfrak{S^\prime E}}$ maps $\mathfrak{S^\prime E}$ product states to $\mathfrak{S^\prime E}$ product states, however, this can still increase the subsystem entropy of $\mathfrak{S}^{\prime}$.} Notably, when the initial states and final state have the same von Neumann entropies, the expression $\Delta S_{\ell,\mathfrak{S}^{\prime}}$ is zero, and thus the lower bound on dissipation is zero in this case.

By contrast to the conditional Landauer reset, we can also define the \emph{unconditional} Landauer reset protocol, in which transitions from each of the states $\left\{\rho_{\ell,\mathfrak{S}^{\prime}}\right\}$ to the reset state $\rho_{\mathrm{r},\mathfrak{S}^{\prime}}$ are achieved by applying a single, standard potential $\widehat{V}_{r,\mathfrak{S}^{\prime}}$ to \emph{any} of the states $\rho_{\mathrm{in},\mathfrak{S}^{\prime}}$ that $\mathfrak{S}^{\prime}$ may be in. Thus, in lieu of the set of $N$ unitary operators given in \eqref{eq:cond-Land-TO}, we have a single unitary operator for all of the states defined by:
\begin{equation}
    \label{eq:uncond-Land-U}
    \begin{split}
        \widehat{U}_{\mathrm{u},\mathfrak{S^\prime E}} &= \mathcal{T} \mathrm{exp} \left\{-\frac{\mathrm{i}}{\hbar}\int \limits_{t_{0}}^{t_\mathrm{f}} \mathrm{d}t'\:\widehat{H} \right\} \\[4pt]
        &= \mathcal{T} \mathrm{exp} \left\{-\frac{\mathrm{i}}{\hbar}\int \limits_{t_{0}}^{t_\mathrm{f}} \mathrm{d}t'\: \left[\widehat{H}_{\mathfrak{S}^{\prime}} + \widehat{H}_{\mathfrak{E}} + \widehat{H}_{\mathrm{I}, \mathfrak{S^\prime E}} + \widehat{V}_{r,\mathfrak{S}^{\prime}}\right] \right\}.
    \end{split}
\end{equation}
The corresponding set of thermal operations in this case are given by:
\begin{equation}
    \label{eq:uncond-Land-TO}
    \rho_{\ell,\mathfrak{S}^{\prime}} \mapsto \Xi\left[\rho_{\ell,\mathfrak{S}^{\prime}}\right] \coloneqq \mathrm{Tr}_{\mathfrak{E}} \left[ \widehat{U}_{\mathrm{u},\mathfrak{S^\prime E}} \left( \rho_{\ell,\mathfrak{S}^{\prime}} \otimes \tau_{\mathfrak{E}} \right) \widehat{U}_{\mathrm{u},\mathfrak{S^\prime E}}^{\dagger} \right] = \rho_{\mathrm{r},\mathfrak{S}^{\prime}}.
\end{equation}

The set of evolutions given in \eqref{eq:uncond-Land-TO} provide a sharp contrast with those given in \eqref{eq:cond-Land-TO}. In \eqref{eq:cond-Land-TO}), we chose $\widehat{V}_{\ell r,\mathfrak{S}^{\prime}}$ such that $\widehat{U}_{\ell,t,\mathfrak{S^\prime E}}$ mapped every $\rho_{\ell,\mathfrak{S}^{\prime}} \otimes \tau_{\mathfrak{E}}$ to the same final state $\rho_{\mathrm{f},\mathfrak{U}}$. By contrast, in \eqref{eq:uncond-Land-TO} we have only a single unitary operator for every possible state under consideration. As a consequence, $\widehat{U}_{\mathrm{u},\mathfrak{S^\prime E}}$ maps each $\rho_{\ell,\mathfrak{S}^{\prime}} \otimes \tau_{\mathfrak{E}}$ to a \emph{different} global final state:
\begin{equation}
    \label{eq:uncond-Land-final-global-states}
    \left\{\rho_{\ell,f,\mathfrak{U}}\right\}_{\ell\:=\:1}^{N} \coloneqq \left\{\widehat{U}_{\mathrm{u},\mathfrak{S^\prime E}} \left(\rho_{\ell,\mathfrak{S}^{\prime}}\otimes\tau_{\mathfrak{E}}\right)\widehat{U}_{\mathrm{u},\mathfrak{S^\prime E}}^{\dagger}\right\}_{\ell\:=\:1}^{N}.
\end{equation}
In other words, for each $\rho_{\ell,\mathfrak{S}^{\prime}}$, the result of the evolution given by $\widehat{U}_{\mathrm{u},\mathfrak{S^\prime E}}$ is to produce a \emph{distinct} final state over all of $\mathfrak{U}$. The only constraints on the evolutions in \eqref{eq:uncond-Land-TO} (and thus, on the states $\rho_{\ell,f,\mathfrak{U}}$) beyond the laws of quantum mechanics and quantum thermodynamics is that the final \emph{subsystem} state of $\mathfrak{S}^{\prime}$ must be the reset state: we require $\mathrm{Tr}_{\mathfrak{E}}\left[\rho_{\ell,f,\mathfrak{U}}\right] = \rho_{\mathrm{r},\mathfrak{S}^{\prime}}$ for all $\ell$.

Because each of the final global states is different, the energy increase of the environment (calculated as in \eqref{eq:cond-Land-Delta-E} and \eqref{eq:cond-Land-bound}) will be a distinct expression for each initial state. However, we can collect these expressions together by examining the \emph{average} energy increase of the conditional and unconditional Landauer resets, over a collection of reset operations performed over a set of individual states. If the states $\rho_{\ell,\mathfrak{S}^{\prime}}$ appear in our collection a fraction of $p_{\ell}$ times, then the average energy increase of the environment will be given by:\footnote{We can alternately think \cite{Anderson19} of this as a series of conditional or unconditional Landauer resets, respectively, over a single joint system $\mathfrak{S^\prime E}$, with the population fractions representing the number of times $\mathfrak{S^\prime E}$ is set to that state.}
\begin{gather}
    \label{cond-Land-avg-E}
    \left\langle \Delta \left\langle E_{\ell,\mathfrak{E}} \right\rangle_{\mathrm{c}} \right\rangle = \mathrm{Tr}_{\mathfrak{E}}\left[\mathrm{Tr}_{\mathfrak{S}^{\prime}} \left[ \sum_{\ell\:=\:1}^{N} p_{\ell}\:\widehat{U}_{\ell,t,\mathfrak{S^\prime E}} \left( \rho_{\ell,\mathfrak{S}^{\prime}} \otimes \tau_{\mathfrak{E}} \right) \widehat{U}_{\ell,t,\mathfrak{S^\prime E}}^{\dagger} \right] \widehat{H}_{\mathfrak{E}} \right] - \mathrm{Tr}\left[\tau_{\mathfrak{E}} \widehat{H}_{\mathfrak{E}} \right]; \\[4pt]
    \left\langle \Delta \left\langle E_{\ell,\mathfrak{E}}\right\rangle_{\mathrm{c}}\right\rangle \geq  -k_{\mathrm{B}}T\:\mathrm{ln}\:2\sum_{\ell\:=\:1}^{N}p_{\ell}\:\Delta S_{\ell,\mathfrak{S}^{\prime}}.
\end{gather}
We can compare these expressions to the average energy increase and average energy bound of the unconditional Landauer reset protocol across all of the $\left\{\rho_{\ell,\mathfrak{S}^{\prime}}\right\}$s. Since convex linear combinations of density matrices form another density matrix, we can express the weighted sum of the $\left\{\rho_{\ell,\mathfrak{S}^{\prime}}\right\}$s as a new fiducial density matrix $\rho_{\mathrm{in},\mathfrak{S}^{\prime}}$:
\begin{equation}
    \label{uncond-Land-fiducial-DM}
    \rho_{\mathrm{in},\mathfrak{S}^{\prime}} \coloneqq \sum_{\ell\:=\:1}^{N}p_{\ell}\:\rho_{\ell,\mathfrak{S}^{\prime}}.
\end{equation}
Then, the average energy increase of the unconditional Landauer reset protocol corresponds \cite{Anderson19} to the average energy increase of $\rho_{\mathrm{in},\mathfrak{S}^{\prime}}$:
\begin{equation}
    \label{eq:uncond-Land-Delta-E}
    \begin{split}
        \left\langle \Delta \left\langle E_{\ell,\mathfrak{E}}\right\rangle_{\mathrm{u}} \right\rangle &=  \mathrm{Tr}_{\mathfrak{E}}\left[\mathrm{Tr}_{\mathfrak{S}^{\prime}} \left[ \widehat{U}_{t,\mathfrak{S^\prime E}} \left( \sum_{\ell\:=\:1}^{N} p_{\ell}\:\rho_{\ell,\mathfrak{S}^{\prime}} \otimes \tau_{\mathfrak{E}} \right) \widehat{U}_{t,\mathfrak{S^\prime E}}^{\dagger} \right] \widehat{H}_{\mathfrak{E}} \right] - \mathrm{Tr}\left[\tau_{\mathfrak{E}} \widehat{H}_{\mathfrak{E}} \right]\\[4pt]
        &=  \mathrm{Tr}_{\mathfrak{E}}\left[\mathrm{Tr}_{\mathfrak{S}^{\prime}} \left[ \widehat{U}_{t,\mathfrak{S^\prime E}} \left( \rho_{\mathrm{in},\mathfrak{S}^{\prime}} \otimes \tau_{\mathfrak{E}} \right) \widehat{U}_{t,\mathfrak{S^\prime E}}^{\dagger} \right] \widehat{H}_{\mathfrak{E}} \right] - \mathrm{Tr}\left[\tau_{\mathfrak{E}} \widehat{H}_{\mathfrak{E}} \right].
    \end{split}
\end{equation}
(In this expression, since $\widehat{U}_{t,\mathfrak{S^\prime E}}$ is independent of $\ell$, we were able to move the sum inside the expression.)

As with \eqref{eq:cond-Land-bound}, the strong subadditivity of the von Neumann entropy and Partovi's inequality gives \cite{Anderson19} a lower bound on $\left\langle \Delta \left\langle E_{\ell,\mathfrak{E}}\right\rangle_{\mathrm{u}} \right\rangle$ in terms of the entropy:
\begin{equation}
    \label{eq:uncond-Land-bound}
    \begin{split}
        \left\langle \Delta \left\langle E_{\ell,\mathfrak{E}}\right\rangle_{\mathrm{u}} \right\rangle &\geq -k_{\mathrm{B}}T\:\mathrm{ln}\:2 \sum_{\ell\:=\:1}^{N}p_{\ell}\:\left(\Delta S_{\ell,\mathfrak{S}^{\prime}} + \:\mathrm{log}_{2}\:p_\ell\right);\\[4pt]
        \left\langle \Delta \left\langle E_{\ell,\mathfrak{E}}\right\rangle_{\mathrm{u}} \right\rangle &\geq -k_{\mathrm{B}}T\:\mathrm{ln}\:2 \,\left(\sum_{\ell\:=\:1}^{N}p_{\ell}\:\Delta S_{\ell,\mathfrak{S}^{\prime}} - \Delta I_{\mathrm{er},\mathfrak{S}^{\prime}} \right).
    \end{split}
\end{equation}
Here, we recognize $\sum_{\ell\:=\:1}^{N} p_{\ell}\:\mathrm{log}_{2}\:p_{\ell} \eqqcolon -H\left(\left\{p_{\ell}\right\}\right) = \Delta I_{\mathrm{er},\mathfrak{S}^{\prime}}$ as the Shannon entropy of the distribution $\left\{p_{\ell}\right\}$, which is equivalent to the information quantity transferred from $\mathfrak{S}^{\prime}$ to $\mathfrak{E}$. As before, when the initial states and final state have the same von Neumann entropies, the expression $\Delta S_{\ell,\mathfrak{S}^{\prime}}$ is zero. In this case, the lower bound on the unconditional Landauer reset protocol is given \emph{entirely} by the amount of information $\Delta I_{\mathrm{er},\mathfrak{S}^{\prime}}$ transferred from $\mathfrak{S}^{\prime}$ to $\mathfrak{E}$.

The only difference between the evolutions \eqref{eq:cond-Land-U} and \eqref{eq:uncond-Land-U} is whether or not the reset potential $\widehat{V}$ is conditioned on the initial state of $\mathfrak{S}^{\prime}$. This seems to indicate that $\Delta I_{\mathrm{er},\mathfrak{S}^{\prime}}$ contains within it some correlated information (\emph{i.e.}, QMI) between $\mathfrak{S}^{\prime}$ and whatever implemented the potential $\widehat{V}$. In fact, as rigorously proven in \cite{Anderson18}, the \emph{entirely} of $\Delta I_{\mathrm{er},\mathfrak{S}^{\prime}}$ is the QMI that arises specifically from the process of conditioning (or not conditioning) $\widehat{V}$ on the initial state of $\mathfrak{S}^{\prime}$. The correlated nature of $\Delta I_{\mathrm{er},\mathfrak{S}^{\prime}}$ plays a central role in understanding the conditional and unconditional Landauer bounds. Likewise, the details of where the reset potentials $\widehat{V}_{\ell r,\mathfrak{S}^{\prime}}$ and $\widehat{V}_{r,\mathfrak{S}^{\prime}}$ come from will play a key role in understanding the distinction between these two. These issues will be discussed in detail in \S\ref{ssec:ctos-rc}, as we tie in this model to the CTO framework.

\mypara{Nonequilibrium Landauer bound.}
\label{par:john}

We can understand the expressions in \S\ref{par:neal} very straightforwardly from the NEQT point of view, both from the point of view of quantum thermodynamic fluctuation relations \cite{Goold15, GCGPVP17} and from the point of view of the general CTOs discussed in the previous section. We start with a general CPTP map given in terms of the dilation theorem \eqref{eq:dilation-thm} with $\mathfrak{F} = \mathfrak{E}$ and a general environment state $\rho_{\mathfrak{E}}$; \emph{i.e.}, a CPTP map given by (henceforth just writing $\mathfrak{S}$ for $\mathfrak{S}^{\prime}$): \begin{equation}
    \label{eq:general-CPTP-map}
    \rho_{\mathrm{in},\mathfrak{S}} \mapsto \Lambda_{t}\left[\rho_{\mathrm{in},\mathfrak{S}}\right] = \mathrm{Tr}_{\mathfrak{E}}\left[\widehat{U}_{t,\mathfrak{SE}}\left(\rho_{\mathrm{in},\mathfrak{S}}\otimes\rho_{\mathfrak{E}}\right)\widehat{U}_{t,\mathfrak{SE}}^{\dagger}\right].
\end{equation}
If we label the eigenstates of $\rho_{\mathfrak{E}}$ as $\ket{e_{a}}$, the eigenvalues of $\rho_{\mathfrak{E}}$ as $e_{a}$, and perform the partial trace over the basis $\ket{v_{b}}$ of $\mathfrak{E}$, then the expression of $\Lambda_{t}\left[\rho_{\mathrm{in},\mathfrak{S}}\right]$ expands to give:
\begin{equation}
    \label{eq:Kraus-CPTP-map}
    \Lambda_{t}\left[\rho_{\mathrm{in},\mathfrak{S}}\right] = \left(\mathbbm{1}_{\mathfrak{S}} \otimes \sum_{b} \bra{v_{b}}\right) \widehat{U}_{t,\mathfrak{SE}}\left(\rho_{\mathrm{in},\mathfrak{S}}\otimes\sum_{a}e_{a}\ket{e_{a}}\bra{e_{a}}\right)\widehat{U}_{t,\mathfrak{SE}}^{\dagger}\left(\mathbbm{1}_{\mathfrak{S}} \otimes \sum_{b} \ket{v_{b}}\right).
\end{equation}
The distributivity of the tensor product allows us to write this expression solely in terms of operators on $\mathfrak{S}$. This defines the \emph{system Kraus operators} (usually simply the \emph{Kraus operators} \cite{Kraus71}) as:
\begin{equation}
    \label{eq:syst-Kraus-op}
    \widehat{M}_{ab}\coloneqq \sum_{a,\:b}
    \sqrt{e_{a}}\:\Braket{v_{b}|\!\widehat{U}_{t,\mathfrak{SE}}\!|e_{a}}.
\end{equation}
It's worth noting that the Kraus operators are \emph{dependent} on the global operator $\widehat{U}_{t,\mathfrak{SE}}$ and the environment expressions $\ket{e_{a}}$, $e_{a}$, and $\ket{v_{b}}$, but as operators themselves solely map density matrices over $\mathcal{H}_{\mathfrak{S}}$ to $\mathcal{H}_{\mathfrak{S}}$. In other words, even though we have $\widehat{M}_{ab}$ dependent on quantities outside of $\mathfrak{S}$, nevertheless we have $\widehat{M}_{ab}\in\mathrm{Aut}\left(\mathcal{D}\left(\mathcal{H}_{\mathfrak{S}}\right)\right)$ when considering it as an operator. Also note that a given set $\left\{\widehat{M}_{ab}\right\}$ of Kraus operators is emphatically not unique: any unitary rotation of the basis $\ket{v_{b}}$ defines a new set of Kraus operators.

Any given set of Kraus operators satisfies the completeness relation:
\begin{equation}
    \label{eq:Kraus-completeness}
    \sum_{a,\:b}\widehat{M}_{ab}^{\dagger}\widehat{M}_{ab} = \mathbbm{1}.
\end{equation}
The Kraus operators in turn give the \emph{operator-sum representation} of the CPTP map $\Lambda_{t}\left[\rho_{\mathrm{in},\mathfrak{S}}\right]$:
\begin{equation}
    \label{eq:op-sum-rep}
    \Lambda_{t}\left[\rho_{\mathrm{in},\mathfrak{S}}\right] = \sum_{a,\:b}\widehat{M}_{ab}^{\dagger}\,\rho_{\mathrm{in},\mathfrak{S}}\,\widehat{M}_{ab} = \mathbbm{1}.
\end{equation}
From the Kraus operator completeness relation \eqref{eq:Kraus-completeness}, the (Hölder) dual of any CPTP map is always \emph{unital}; \emph{i.e.}, for any CPTP map $\Lambda_{t}$, we always have $\Lambda_{t}^{\dagger}\left(\mathbbm{1}_{\mathfrak{S}}\right) = \mathbbm{1}_{\mathfrak{S}}$. The same may not necessarily be true for $\Lambda_{t}$ itself; instead, the unitality condition for $\Lambda_{t}$ is given in terms of the Kraus operators by the condition:
\begin{equation}
    \label{eq:Kraus-unitality-requirement}
    \sum_{a,\:b}\widehat{M}_{ab}\widehat{M}_{ab}^{\dagger} = \mathbbm{1}.
\end{equation}
Unital channels are notable since they map the identity $\mathbbm{1}_{\mathfrak{S}}$ to itself (and thus the maximally mixed state $\mathbbm{1}_{\mathfrak{S}}/\mathrm{Tr}\left[\mathbbm{1}\right]$ to itself): we have  $\Lambda_{t}\left[\mathbbm{1}_{\mathfrak{S}}\right] = \mathbbm{1}_{\mathfrak{S}}$ only when $\Lambda_{t}$ is unital (by definition). It's worth noting that even though the Kraus operators themselves can be arbitrarily changed by a unitary transform, this sum is invariant under such a transform, so the unitality condition is independent of the specific basis we evaluate the Kraus operators in.

We can very straightforwardly understand the difference between the conditional and unconditional Landauer reset, and in particular the terms in the unconditional Landauer bound \eqref{eq:uncond-Land-bound}, in terms of the Kraus operators and the unitality condition. In the same way as we defined the system Kraus operators, we can define the \emph{environment Kraus operators} $\widehat{N}_{cd}\in\mathrm{Aut}\left(\mathcal{D}\left(\mathcal{H}_{\mathfrak{E}}\right)\right)$ as Kraus operators on $\mathfrak{E}$. Labelling the eigenstates of $\rho_{\mathrm{in},\mathfrak{S}}$ as $\ket{s_{c}}$, the eigenvalues of $\rho_{\mathrm{in},\mathfrak{S}}$ as $s_{c}$, and the basis of $\mathfrak{S}$ as $\ket{w_{d}}$, we can define $\widehat{N}_{cd}$ as:
\begin{equation}
    \label{eq:env-Kraus-op}
    \widehat{N}_{cd}\coloneqq \sum_{c,\:d}
    \sqrt{s_{c}}\,\Braket{w_{d}|\!\widehat{U}_{t,\mathfrak{SE}}\!|s_{c}}.
\end{equation}
Then, as with \eqref{eq:TO}, \eqref{eq:cond-Land-TO}, and \eqref{eq:uncond-Land-TO}, we examine the evolution of $\mathfrak{SE}$ when we couple $\mathfrak{S}$ initially in the state $\rho_{\mathrm{in},\mathfrak{S}}$ to the environment $\mathfrak{E}$, initially in the thermal state:
\begin{equation}
    \label{eq:noneq-Land-TO}
    \rho_{\mathrm{in},\mathfrak{S}}\otimes\tau_{\mathfrak{E}} \mapsto \widehat{U}_{t,\mathfrak{SE}} \left(\rho_{\mathrm{in},\mathfrak{S}}\otimes\tau_{\mathfrak{E}}\right)\widehat{U}_{t,\mathfrak{SE}}^{\dagger}.
\end{equation}
As with \eqref{eq:cond-Land-final-state-env}, we're interested in the final environment state of $\mathfrak{E}$, which can tell us the bound on the energy increase of $\mathfrak{E}$. The final state of the environment as a result of the transformation \eqref{eq:noneq-Land-TO} is given by:
\begin{equation}
    \label{eq:noneq-Land-final-state-env}
    \rho_{\mathrm{f},\mathfrak{E}} = \mathrm{Tr}_{\mathfrak{S}}\left[\widehat{U}_{t,\mathfrak{SE}}\left(\rho_{\mathrm{in},\mathfrak{S}}\otimes\tau_{\mathfrak{E}}\right)\widehat{U}_{t,\mathfrak{SE}}^{\dagger}\right] = \sum_{c,\:d}\widehat{N}_{cd}\,\tau_{\mathfrak{E}}\,\widehat{N}_{cd}^{\dagger}.
\end{equation}
From this, and using the two-time measurement formalism \cite{TLH07}, the probability distribution $P\left(Q\right)$ of the environment heat $Q$ in the eigenbasis $\ket{e_{a}}$ of $\tau_{\mathfrak{E}}$ is given \cite{Goold15} by:
\begin{equation}
    \label{eq:noneq-Land-heat-dist}
    \begin{split}
        &P\left(Q\right) = \sum_{c,\:d;\:g,\:h}\Braket{e_{g}|\!\widehat{N}_{cd}\!|e_{h}}\Braket{e_{h}|\tau_{\mathfrak{E}}|e_{h}}\Braket{e_{h}|\!\widehat{N}_{cd}^{\dagger}\!|e_{g}}\,\delta\left(Q-\left(E_{h}-E_{g}\right)\right)\\
        &= \sum_{c,\:d;\:g,\:h}\Braket{e_{g}|\!\widehat{N}_{cd}\!|e_{h}}\Braket{e_{h}|\!\frac{\mathrm{e}^{-\beta\widehat{H}_{\mathfrak{E}}}}{\mathrm{Tr}_{\mathfrak{E}}\left[\mathrm{e}^{-\beta\widehat{H}_{\mathfrak{E}}}\right]}\!|e_{h}}\Braket{e_{h}|\!\widehat{N}_{cd}^{\dagger}\!|e_{g}}\,\delta\left(Q-\left(E_{h}-E_{g}\right)\right).
    \end{split}
\end{equation}
This gives the moment-generating function of the dissipated heat given by: 
\begin{equation}
    \label{eq:noneq-Land-moment-generating-fn-env}
    \left\langle \mathrm{e}^{-\beta Q}\right\rangle = \sum_{c,\:d}\mathrm{Tr}\left[\widehat{N}_{cd}^{\dagger}\,\tau_{\mathfrak{E}}\,\widehat{N}_{cd}\right] = \sum_{c,\:d}\mathrm{Tr}\left[\widehat{N}_{cd}\,\widehat{N}_{cd}^{\dagger}\,\tau_{\mathfrak{E}}\right].
\end{equation}
Then, a direct consequence of Jensen's inequality is that the energy increase in this process is given in terms of the Kraus operators:
\begin{equation}
    \label{eq:noneq-Land-bound-env}
    \Delta\left\langle E_{\mathfrak{E}}\right\rangle \geq -k_{\mathrm{B}}T\:\mathrm{ln}\:\mathrm{Tr}\left[\sum_{c,\:d}\widehat{N}_{cd}\,\widehat{N}_{cd}^{\dagger}\,\tau_{\mathfrak{E}}\right].
\end{equation}

The expression \eqref{eq:noneq-Land-bound-env} immediately helps us understand the conditional Landauer bound \eqref{eq:cond-Land-bound} and the unconditional Landauer bound \eqref{eq:uncond-Land-bound}: the overall evolution \eqref{eq:noneq-Land-TO} corresponds to a CPTP map (a.k.a. quantum channel) over $\mathfrak{E}$\footnote{It's worth noting that although \eqref{eq:noneq-Land-bound-env} is slightly unclear compared to \eqref{eq:cond-Land-bound} and \eqref{eq:uncond-Land-bound}, in the sense that Kraus operator expression mixes both the entropy increase contribution and the correlated information ejection contribution, it serves as the most clear expression from a quantum information theory point of view, and is also the tightest bound available \cite{Goold15}.}. This channel may or may not be unital over $\mathfrak{E}$, and the degree to which this channel fails to be unital is exactly the degree to which the channel increases the overall entropy of $\mathfrak{S}$ and expels the information quantity $\Delta I_{\mathrm{er},\mathfrak{S}}$ to $\mathfrak{E}$. The fact that unital quantum channels map maximally mixed states to maximally mixed states is essential: the degree to which this channel fails to be unital tells us the extent to which the channel perturbs the maximally mixed state $\tau_{\mathfrak{E}}$ of the environment. Indeed, we see that for a perfectly unital channel, the sum of the Kraus operators retrives $\mathbbm{1}_{\mathfrak{E}}$, and the energy bound is zero.

The degree of unitality stands out as a key quantity of interest in examining the nonequilibrium Landauer bound in a given system. Using the technique of full counting statistics \cite{EHM09}, the expressions \eqref{eq:noneq-Land-moment-generating-fn-env}--\eqref{eq:noneq-Land-bound-env} can be extended \cite{GCGPVP17} to a one-parameter family of expressions (replacing $\beta$ with a more general parameter). This technique gives an explicit way to quantify the non-unitality of $\widehat{N}_{cd}\widehat{N}_{cd}^{\dagger}$ in the above expressions:
\begin{equation}
    \label{eq:nonunitality}
    \mathcal{N}_{\mathfrak{E}} \coloneqq \left\Vert\sum_{c,d} \widehat{N}_{cd}\,\widehat{N}_{cd}^{\dagger}-\mathbbm{1}_{\mathfrak{E}}\right\Vert_{2}.
\end{equation}
Here, $\left\Vert\widehat{A}\right\Vert_{2}$ represents the Hilbert-Schmidt norm. Finally, from \eqref{eq:noneq-Land-moment-generating-fn-env}, we have the average energy dissipated into the environment given \cite{Goold15,GCGPVP17,EHM09,Reeb14,TD21} by:
\begin{equation}
    \label{eq:noneq-Land-env-heat-inc}
    \begin{split}
        \beta\left\langle Q\right\rangle\left(t\right) &= -\Delta S_{\mathfrak{S}} - I(\mathfrak{S}:\mathfrak{E}) - S\left(\rho_{\mathfrak{E}}\left(t\right)\Vert\tau_{\mathfrak{E}}\right) \\[4pt]
        & = - S\left(\Lambda_{t}\left[\rho_{\mathrm{in},\mathfrak{S}}\right]\right) + S\left(\rho_{\mathrm{in},\mathfrak{S}}\right) - \Delta I_{\mathrm{er},\mathfrak{S}} - I(\mathfrak{S}:\mathfrak{E}) - S\left(\rho_{\mathrm{f},\mathfrak{E}}\left(t\right)\Big\Vert\tau_{\mathfrak{E}}\right).
    \end{split}
\end{equation}
We can immediately recognize this expression as simply the extension of the expressions \eqref{eq:cond-Land-bound} and \eqref{eq:uncond-Land-bound} to include the possibility of initial correlations between $\mathfrak{S}$ and $\mathfrak{E}$ and the possibility that the environment may not start out in the thermal state. For our setup, neither of these conditions are applicable, and thus the last two terms vanish.

We would expect that the environment is not a ``special'' subsystem in terms of these derivations, and that an equivalent expression can be derived by considering the system. From each subsystem's point of view, the other serves as the ancillary system in the dilation theorem sense. Indeed, expanding the Kraus operators in terms of $\ket{s_{c}}$ and $\ket{w_{d}}$, and rearranging terms in the overall trace, provides us with an equivalent expression to \eqref{eq:noneq-Land-moment-generating-fn-env}:
\begin{equation}
    \label{eq:noneq-Land-moment-generating-fn-sys}
    \left\langle \mathrm{e}^{-\beta Q}\right\rangle = \mathrm{Tr}_{\mathfrak{S}}\bigg[\mathrm{Tr}_{\mathfrak{E}}\left[\widehat{U}_{t,\mathfrak{SE}}^{\dagger}\left(\mathbbm{1}_{\mathfrak{S}}\otimes\tau_{\mathfrak{E}}\right)\widehat{U}_{t,\mathfrak{SE}}\right]\rho_{\mathrm{in},\mathfrak{S}}\bigg].
\end{equation}
As with the Kraus operators, the expression $\mathrm{Tr}_{\mathfrak{E}}\left[\widehat{U}_{t,\mathfrak{SE}}^{\dagger}\left(\mathbbm{1}_{\mathfrak{S}}\otimes\tau_{\mathfrak{E}}\right)\widehat{U}_{t,\mathfrak{SE}}\right]$ is an operator which depends on properties outside of $\mathfrak{S}$, but as an operator lives in $\mathrm{Aut}\left(\mathcal{D}\left(\mathcal{H}_{\mathfrak{S}}\right)\right)$; \emph{i.e.}, it maps density matrices in $\mathfrak{S}$ to density matrices in $\mathfrak{S}$. The connection between these expressions to the conditional and unconditional Landauer reset protocols is apparent, but the connection between both of these to the CTO framework is slightly more subtle. The connection between all three is discussed in \S\ref{ssec:ctos-rc}.

As mentioned, the expectation value \eqref{eq:noneq-Land-moment-generating-fn-env} derived in \cite{Goold15}, and its extension derived in \cite{GCGPVP17}, rely on the two-time measurement formalism \cite{TLH07}. This might cause some trepidation---when considering the final energy, we generally must also consider the impact on the system of performing the measurement itself \cite{NC00}. Conventionally, measuring the system in the state $\ket{k}$ corresponds \cite{Born26,vN32} to a projection $\widehat{\Pi}_{k}$ upon the pre-measurement state $\rho$ onto $\ket{k}$. This is given by Born's rule, which in terms of density matrices we can express as:
\begin{equation}
    \label{eq:born-rule}
    \rho \mapsto \frac{\widehat{\Pi}_{k}\,\rho\,\widehat{\Pi}_{k}}{\mathrm{Tr}\left[\widehat{\Pi}_{k}\,\rho\right]}.
\end{equation}
This corresponds to a change in the von Neumann entropy given by:
\begin{equation}
    \label{eq:measurement-vN-entropy-change}
    \Delta S_{\mathrm{m}} = -\sum\limits_{k} \mathrm{Tr}\left[\widehat{\Pi}_{k}\,\rho\,\widehat{\Pi}_{k}\:\mathrm{ln}\: \widehat{\Pi}_{k}\,\rho\,\widehat{\Pi}_{k} \right] - \mathrm{Tr}\left[\rho\:\mathrm{ln}\:\rho\right].
\end{equation}
As we've just seen in \eqref{eq:noneq-Land-env-heat-inc}, this corresponds to a change in energy. In fact, ``ideal'' projective measurements (\emph{i.e.}, those in which the measurements reproduce the measured statistics of the system, those which exhibit a one-to-one correspondence between measurement states and measured states, and which do not change the measurement statistics after measurement) cost an \emph{infinite} amount of energy \cite{GFH20}.

Quite fortuitously, an alternate formulation of quantum work can be developed \cite{DPZ16} which completely avoids this issue, by focusing on the change in the expectation values of the energy eigenstates. For a time-evolving Hamiltonian starting at $\widehat{H}_{0}$ with an initial eigenstate $\ket{k_{0}}$, this formulation is defined by:
\begin{equation}
    \label{eq:one-time-work}
    \Tilde{W}_{k} \coloneqq \Braket{k_{0}|\widehat{U}_{t}^{\dagger}\widehat{H}\left(t\right)\widehat{U}_{t}|k_{0}}
\end{equation}
Notably, this formulation retrieves the \emph{same} average work as the two-time measurement formalism:
\begin{equation}
    \label{eq:one-time-vs-two-time-avg-work}
    \begin{split}
        \left\langle\Tilde{W}\right\rangle &= \sum\limits_{k_{0}}\,\Braket{k_{0}|\widehat{U}_{t}^{\dagger}\widehat{H}\left(t\right)\widehat{U}_{t}|k_{0}}\frac{ \mathrm{e}^{-\beta E_{k_{0}}}}{\mathrm{Tr}\left[\mathrm{e}^{-\beta\widehat{H}_{0}}\right]} - \mathrm{Tr}\left[\tau_{0}\,\widehat{H}_{0}\right]\\[4pt]
        & = \mathrm{Tr}\left[\rho\left(t\right)\widehat{H}\left(t\right)\right] - \mathrm{Tr}\left[\tau_{0}\,\widehat{H}_{0}\right] = \left\langle W \right\rangle
    \end{split}
\end{equation}
(Here, $E_{k_{0}}$ is the energy eigenvalue of $\ket{k_{0}}$, $\tau_{0}$ is the thermal (Gibbs) state corresponding to $\widehat{H}_{0}$, and $\rho\left(t\right)$ is the system state at time $t$.) As a direct consequence of the $\left\langle\Tilde{W}\right\rangle =  \left\langle W \right\rangle$ equality, we see that although \eqref{eq:noneq-Land-moment-generating-fn-env}--\eqref{eq:noneq-Land-moment-generating-fn-sys} are derived using the two-time measurement formalism, they're compatible with any alternate definitions of quantum work which provide the same expectation values.



\color{black}

\subsubsection{Gorini-Kossakowski-Sudarshan-Lindblad (GKSL) Dynamics}
\label{sssec:gksl}

Beyond providing NEQT justifications for Landauer's principle and providing a new explanation for the difference between the conditional and unconditional Landauer recent protocols, another central aim of this work is to lay out the foundations for representing classical RC operations explicitly in terms of open quantum systems. Here, we discuss the framework of \emph{GKSL equations with multiple asymptotic states} \cite{Albert14, ABFJ16, Albert18}, which we apply in \S\ref{ssec:rc-op-rep} to model reversible computing operations.

\mypara{Markov assumption.}
\label{par:gksl-intro}

In a closed system governed by a Hamiltonian $\widehat{H}$ and whose state at time $t$ is given by $\rho\left(t\right)$, the dynamics are given by the \emph{Liouville-von Neumann (LvN) equation}:
\begin{equation}
    \label{eq:closed-LvN-eq}
    \frac{\mathrm{d}\rho\left(t\right)}{\mathrm{d}t} = -\mathrm{i}\left[\widehat{H},\rho\right].
\end{equation}
By analogy with the Liouville theorem of classical statistical mechanics and symplectic geometry, we can define \cite{Breuer07,Banerjee18} the \emph{Liouville superoperator} as $\hat{\hat{\mathcal{L}}}\left[\rho\left(t\right)\right] \coloneqq -\mathrm{i}\left[\widehat{H},\rho\right]$. This gives the superoperator version of the LvN equation as:
\begin{equation}
    \label{eq:LvN-eq-superop}
    \frac{\mathrm{d}\rho\left(t\right)}{\mathrm{d}t} = \hat{\hat{\mathcal{L}}}\left[\rho\left(t\right)\right] \coloneqq -\mathrm{i}\left[\widehat{H},\rho\right].
\end{equation}
(This is also known in the literature as the \emph{quantum master equation} or the \emph{Liouvillian}.) As with the unitary evolution of states, the formal solution to this is given by a Volterra integral equation:
\begin{equation}
    \label{eq:LvN-formal-sol}
    \rho\left(t\right) = \mathcal{T}\:\mathrm{exp}\left\{\int\limits_{t_{0}}^{t}\mathrm{d}t'\:\hat{\hat{\mathcal{L}}}\left(t'\right)\right\}\,\rho\left(t_{0}\right).
\end{equation}
In the specific case that $\hat{\hat{\mathcal{L}}}$ is independent of time, this simplifies to $\rho\left(t\right) = \mathrm{e}^{t\hat{\hat{\mathcal{L}}}}\rho\left(t_{0}\right)$. In general, this may not be guaranteed to converge, let alone have a closed-form solution \cite{BBP04,BGM06,Ivanov15}.\footnote{This is of course true for the unitary evolution of states as well; convergence is only guaranteed when the algebra of the argument of the integral has a commuting structure (\emph{i.e.}, when the Volterra integral equation is over c-numbers).} Nevertheless, this is the formal solution to the LvN equation \eqref{eq:LvN-eq-superop}. Using the dilation theorem, we expect the time evolution of $\rho_{\mathfrak{S}}\left(t\right)$ to follow the same principle; \emph{i.e.}, that we can determine the dynamics of $\rho_{\mathfrak{S}}\left(t\right)$ by examining the time evolution of the closed system $\mathfrak{U} = \mathfrak{SE}$ and taking the partial trace over $\mathfrak{E}$. Thus, we have the LvN equation for $\rho_{\mathfrak{S}}\left(t\right)$ given by:
\begin{equation}
    \label{eq:open-LvN-eq}
    \frac{\mathrm{d}\rho_{\mathfrak{S}}\left(t\right)}{\mathrm{d}t} = \mathrm{Tr}_{\mathfrak{E}}\bigg[-\mathrm{i}\left[\widehat{H}_{\mathfrak{SE}},\rho_{\mathfrak{SE}}\left(t\right)\right]\bigg] = -\mathrm{i}\:\mathrm{Tr}_{\mathfrak{E}}\left[\widehat{H}_{\mathfrak{SE}},\widehat{U}_{t,\mathfrak{SE}}\left(\rho_{\mathfrak{S}}\left(t_{0}\right)\otimes\rho_{\mathfrak{E}}\right)\widehat{U}_{t,\mathfrak{SE}}^{\dagger}\right].
\end{equation}
(Just to be clear about the notation, in the last expression we're taking the partial trace over $\mathfrak{E}$ of the commutator of $\widehat{H}_{\mathfrak{SE}}$ with the state given by the unitary time evolution of $\rho_{\mathfrak{S}}\left(t_i\right)\otimes \rho_{\mathfrak{E}}$.)

To find a solution to this equation for $\rho_{\mathfrak{S}}$, we would need to evaluate the Volterra integral equation \eqref{eq:LvN-formal-sol} for the global evolution over $\mathfrak{SE}$ and then trace over $\mathfrak{E}$. This has the exact same problems of convergence and closed form as before, since we haven't changed the problem itself. Instead, as a first step to determining the dynamics of $\rho_{\mathfrak{S}}$, we can make the simplifying assumption of \emph{Markovian dynamics}; \emph{i.e.}, that over a differential time evolution $t\mapsto t + \mathrm{d}t$, the properties of $\rho\left(t + \mathrm{d}t\right)$ are determined \emph{entirely} by the properties of $\rho\left(t\right)$. Since this assumption explicitly states that $\rho_{\mathfrak{S}}\left(t+\mathrm{d}t\right)$ depends only on $\rho_{\mathfrak{S}}\left(t\right)$, we \emph{must} make the Markov approximation in order to write down a differential evolution equation $\rho_{\mathfrak{S}}$ that's first-order in time. We might be concerned that this is an overly restrictive assumption for a sensible model of reversible computing; fortunately, this assumption is in fact entirely in line with some of the key assumptions we make in our generalized models of reversible computing. The relation between these assumptions, and their suitability, is discussed in \S\ref{ssec:markov-rc}.

The map $\rho_{\mathfrak{S}}\left(t_{0}\right) \mapsto \rho_{\mathfrak{S}}\left(t\right)$ is a quantum channel $\rho_{\mathrm{in},\mathfrak{S}} \mapsto \Lambda_{t}\left[\rho_{\mathrm{in},\mathfrak{S}}\right]$; thus, we can express $\rho\left(t+\mathrm{d}t\right)$ in terms of the operator-sum representation \eqref{eq:op-sum-rep} of the CPTP map. In this representation, the Markov approximation appears as:
\begin{equation}
    \label{eq:Markov-dynamics-Kraus-rep-Schr}
    \rho_{\mathfrak{S}}\left(t+\mathrm{d}t\right) = \sum_{c,\:d}\widehat{N}_{cd}\left(\mathrm{d}t\right)\,\rho_{\mathfrak{S}}\left(t\right)\,\widehat{N}_{cd}^{\dagger}\left(\mathrm{d}t\right).
\end{equation}
We can retrieve a Liouville-type superoperator in the Markov approximation by examining the differential evolution of a quantum channel:
\begin{equation}
    \label{eq:GKSL-difference-quotient-def}
    \Lambda_{\mathrm{d}t}\left[\rho_{\mathfrak{S}}\right] = \hat{\hat{\mathcal{I}}} + \mathrm{d}t\:\underset{\mathrm{d}t\:\rightarrow\:0}{\mathrm{lim}}\frac{\Lambda_{\mathrm{d}t}-\hat{\hat{\mathcal{I}}}}{\mathrm{d}t} + \cdots \coloneqq \hat{\hat{\mathcal{I}}} + \mathrm{d}t\:\hat{\hat{\mathcal{L}}} + \cdots.
\end{equation}
By expanding the Kraus operators and keeping the terms up to order $\mathcal{O}\left(\mathrm{d}t\right)$, we get the \emph{Gorini-Kossakowski-Sudarshan-Lindblad (GKSL) superoperator / equation} \cite{Lindblad76,GKS76}:
\end{paracol}
\begin{equation}
    \label{eq:GKSL-def}
    \frac{\mathrm{d}\rho_{\mathfrak{S}}}{\mathrm{d}t} = \hat{\hat{\mathcal{L}}}\left[\rho_{\mathfrak{S}}\left(t\right)\right] \coloneqq -\mathrm{i}\left[\widehat{H}_{\mathfrak{S}},\rho_{\mathfrak{S}}\right] + \frac{1}{2}\sum_{a,\:b\:>\:0}\kappa_{ab}\left(2\widehat{F}_{ab}\,\rho_{\mathfrak{S}}\:\!\widehat{F}_{ab}^{\dagger} + \widehat{F}_{ab}^{\dagger}\:\!\widehat{F}_{ab}\,\rho_{\mathfrak{S}} + \rho_{\mathfrak{S}}\:\!\widehat{F}_{ab}^{\dagger}\:\!\widehat{F}_{ab}\right).
\end{equation}
\begin{paracol}{2}
\switchcolumn
\noindent (These are also referred to in the literature as \emph{Lindbladians} or \emph{quantum Markov equation}.) Here, $\widehat{F}_{cd}$ are the so-called \emph{jump operators}. These induce ``quantum jumps''; \emph{i.e.}, the quantum state transitions that are distinct from the (closed-system) evolution of $\rho_{\mathfrak{S}}$ under $\widehat{H}_{\mathfrak{S}}$.\footnote{More precisely, $\widehat{F}_{ab}\,\rho_{\mathfrak{S}}\:\!\widehat{F}_{ab}^{\dagger}$ induces the jumps, and $\widehat{F}_{ab}^{\dagger}\:\!\widehat{F}_{ab}\,\rho_{\mathfrak{S}} + \rho_{\mathfrak{S}}\:\!\widehat{F}_{ab}^{\dagger}\:\!\widehat{F}_{ab}$ normalizes the evolution in the case that there are no jumps.} $\kappa_{ab}$ are the rates corresponding to the $ab$th jump. 

The Markov approximation made in \eqref{eq:Markov-dynamics-Kraus-rep-Schr} has important consequences for the time scales we consider. By definition, the Markov approximation assumes that the state $\rho_{\mathfrak{S}}\left(t + \mathrm{d}t\right)$ at time $t + \mathrm{d}t$ depends \emph{only} on the state $\rho_{\mathfrak{S}}\left(t\right)$ at time $t$. In doing so, we explicitly preclude \cite{Breuer07,Banerjee18,Preskill19} the possibility of \emph{fluctuations} that can take information from $\mathfrak{S}$ to $\mathfrak{E}$ during an intermediate time period and then have that information return to $\mathfrak{S}$ at time $t + \mathrm{d}t$. Instead, this assumption is equivalent to saying that any information that is ejected from the system to the environment cannot be returned to the system. As a result, this approximation requires a separation between the time scales of these fluctuations, the time scales available at our resolution (and the dynamics of interest), and the relaxation time of the system. If we denote $\tau_{\mathsf{F}}$ as the time scale of these fluctuations, $\tau_{\mathsf{S}}$ as the time scale available to us at the resolution we're capable of (and thus, the time scale of the dynamics we're interested in), and $\tau_{\mathsf{R}}$ as the relaxation time of the system, the Markov approximation requires a clean separation between all three time scales, corresponding to:
\begin{equation}
    \label{eq:Markov-time-scales}
    \tau_{\mathsf{F}} \ll \tau_{\mathsf{S}} \ll \tau_{\mathsf{R}}.
\end{equation}
Before continuing, it's worth mentioning that a change in the relationship between $\tau_{\mathsf{S}}$ and $\tau_{\mathsf{F}}$ leads to a substantial change in the dynamics of the system. By contrast with the $\tau_{\mathsf{F}} \ll \tau_{\mathsf{S}}$ condition, the condition $\tau_{\mathsf{S}} \ll \tau_{\mathsf{F}} \ll \tau_{\mathsf{R}}$ leads to \emph{quantum Brownian motion (QBM)} \cite{Breuer07,Banerjee18,Erdos12}. The most famous example of QBM is the \emph{Caldeira-Leggett model} \cite{CL83}, which examines the quantum Brownian dynamics of a particle coupled to a bath described by a set of harmonic oscillators. At high temperatures, this gives rise to a different Markovian master equation than the GKSL equation \eqref{eq:GKSL-def}; at low temperatures, this gives rise to substantial non-Markovianities.

The GKSL evolution equations \eqref{eq:GKSL-def} provide the evolution of the system in the Schrödinger picture; \emph{i.e.}, when the operators $\widehat{A}\in\mathcal{B}\left(\mathcal{H}_{\mathfrak{S}}\right)$ on $\mathcal{H}_{\mathfrak{S}}$ are stationary and the states (and thus the density matrices) evolve with time.\footnote{The notation $\mathcal{B}(\mathcal{H})$ represents the set of bounded operators on $\mathcal{H}$ with finite trace norm.} We can examine the Heisenberg picture (where the states are stationary and the operators evolve with time) using the adjoint differential evolution expression:
\begin{equation}
    \label{eq:Markov-dynamics-Kraus-rep-Heis}
    \widehat{A}\left(t+\mathrm{d}t\right) = \sum_{c,\:d}\widehat{N}_{cd}^{\dagger}\left(\mathrm{d}t\right)\widehat{A}\left(t\right)\widehat{N}_{cd}\left(\mathrm{d}t\right).
\end{equation}
As with the GKSL equation \eqref{eq:GKSL-def}, we can expand the Kraus operators (or alternately take the adjoint of the GKSL equation directly) to get the \emph{adjoint GKSL equation} governing the time evolution of operators in the Heisenberg picture:
\end{paracol}
\begin{equation}
    \label{eq:GKSL-adj-def}
    \frac{\mathrm{d}\widehat{A}}{\mathrm{d}t} = \hat{\hat{\mathcal{L}}}^{\ddagger}\left[\widehat{A}\left(t\right)\right] \coloneqq \mathrm{i}\left[\widehat{H}_{\mathfrak{S}},\widehat{A}\right] + \frac{1}{2}\sum_{a,\:b\:>\:0}\kappa_{ab}\left(2\widehat{F}_{ab}^{\dagger}\,\widehat{A}\:\!\widehat{F}_{ab} + \widehat{F}_{ab}^{\dagger}\:\!\widehat{F}_{ab}\,\widehat{A} + \widehat{A}\:\!\widehat{F}_{ab}^{\dagger}\:\!\widehat{F}_{ab}\right).
\end{equation}
\begin{paracol}{2}
\switchcolumn
\noindent 
The adjoint GKSL superoperator provides us with the time evolution of operators, including the conserved quantities of the system; these are discussed in more detail in \S\ref{par:gksl-mas}. 
The formal solution to this evolution equation is, as we'd expect:
\begin{equation}
    \label{eq:adj-LvN-formal-sol}
    \widehat{A}\left(t\right) = \mathcal{T}\:\mathrm{exp}\left\{\int\limits_{t_{0}}^{t}\mathrm{d}t'\:\hat{\hat{\mathcal{L}}}\left(t'\right)\right\}\,\widehat{A}\left(t_{0}\right).
\end{equation}
As always, this simplifies to $\widehat{A}\left(t\right) = \mathrm{e}^{t\hat{\hat{\mathcal{L}}}^{\ddagger}}\widehat{A}\left(t_{0}\right)$ when $\hat{\hat{\mathcal{L}}}^{\ddagger}$ is independent of time. Unfortunately, the expressions of $\hat{\hat{\mathcal{L}}}$ and $\hat{\hat{\mathcal{L}}}^{\ddagger}$ defined in \eqref{eq:GKSL-def} and \eqref{eq:GKSL-adj-def} are not unique: we can have unitary transformations which redefine the jump operators or mix the jump operators and Hamiltonian (or that do both) while leaving the overall forms of $\hat{\hat{\mathcal{L}}}$ and $\hat{\hat{\mathcal{L}}}^{\ddagger}$ invariant.\footnote{This is unfortunate but not surprising; we saw the same kind of ambiguity in the definition of the Kraus operators. Indeed, the first kind of unitary transform is due to precisely that ambiguity from earlier.}

The GKSL equation \eqref{eq:GKSL-def} also serves as the generator of the \emph{quantum dynamical semigroups} \cite{Alicki07,Kossakowski72,Ingarden75}. Starting with the dilation theorem \eqref{eq:dilation-thm}, we as always select $\mathfrak{F} = \mathfrak{E}$. As discussed before, seeking a first-order differential equation of the form \eqref{eq:GKSL-def} automatically imposes the Markov property. If we consider the set of all of the CPTP maps which start at the same starting environment state $\rho_{\mathfrak{E}}$ and are evolved to various times $t$, the Markov property is equivalent to the semigroup property $\Lambda_{t_{1}}\Lambda_{t_{2}} = \Lambda_{t_{1}\,+\,t_{2}}$. Thus, we define the quantum dynamical semigroup as the family $\left\{\Lambda_{t}\mid t\geq0\right\}$ that satisfies the Markov property.\footnote{According to Lindblad's theorem \cite{Lindblad76}, \emph{any} quantum operation that satisfies the semigroup property will satisfy the GKSL equation and vice versa. However, note that not every CPTP map satsifies the GKSL equation \cite{Wolf08, WECC08}; it just so happens that we're only concerned with the ones that do.

Oftentimes, the requirement that $\mathrm{Tr}\left[\left(\Lambda_{t}\left[\rho\right]\right)\widehat{A}\right]$ is a continuous function of $t$ for all trace-norm-bounded operators $\widehat{A}\in\mathcal{B}\left(\mathcal{H}_{\mathfrak{S}}\right)$ is also specified. However, because we already specified that $\Lambda_{t}$ is a CPTP map for all $t$, that $\widehat{A}\in\mathcal{B}\left(\mathcal{H}_{\mathfrak{S}}\right)$ is a bounded function, and that the semigroup property is specified where $t\in\mathbb{R}$ is a continuous parameter, this requirement is a direct consequence of what we already have. Finally, we note that the notation $\mathsf{Op}\left(\mathsf{H}\right)$ is sometimes used for $\mathcal{B}\left(\mathcal{H}_{\mathfrak{S}}\right)$, \emph{e.g.}, as in \cite{Albert14, ABFJ16, Albert18}.}

As always, the differential equation \eqref{eq:GKSL-def} is solved by \eqref{eq:LvN-formal-sol}, which simplifies to $\Lambda_{t}\left[\rho_{\mathfrak{S}}\left(t_{0}\right)\right] = \rho_{\mathfrak{S}}\left(t\right) = \mathrm{e}^{t\hat{\hat{\mathcal{L}}}}$ when $\hat{\hat{\mathcal{L}}}$ is independent of $t$. Since the family $\left\{\Lambda_{t}\mid t\geq0\right\}$ is a one-parameter semigroup, we can recognize $\hat{\hat{\mathcal{L}}}$ as the generator \cite{Alicki07,Woit17} of this semigroup. As usual, we can then define $\mathrm{e}^{t\hat{\hat{\mathcal{L}}}}$ in terms of its Taylor-Madhava series expansion:
\begin{equation}
    \label{eq:exp-Madhava}
    \mathrm{e}^{t\hat{\hat{\mathcal{L}}}} \coloneqq \underset{t\:\rightarrow\:\infty}{\mathrm{lim}}\left(1 - \frac{t\hat{\hat{\mathcal{L}}}}{n}\right)^{n}.
\end{equation}
We see that the Markovian approximation gives a vital tool for explicitly calculating the dynamics of the evolution of $\rho_{\mathfrak{S}}$: it gives us a first-order differential equation in time for the evolution of $\rho_{\mathfrak{S}}$, entirely in terms of the known quantities $\widehat{H}_{\mathfrak{S}}$ and $\widehat{N}_{ab}$. Furthermore, by cleverly engineering $\mathfrak{S}$, these may even be controllable expressions in experiments.

\mypara{GKSL dynamics with multiple asymptotic states.}
\label{par:gksl-mas}

Central quantities of interest in GKSL evolution are the \emph{asymptotic states}, which are the set of states that $\rho_{\mathfrak{S}}\left(t\right)$ evolves into in the infinite time limit under the $\mathrm{e}^{t\hat{\hat{\mathcal{L}}}}$ evolution:
\begin{equation}
    \label{eq:as-states-def}
    \left. \ket{\rho_{\infty}}\!\right\rangle \coloneqq \underset{t\:\rightarrow\:\infty}{\mathrm{lim}}\mathrm{e}^{t\hat{\hat{\mathcal{L}}}}\left.\ket{\rho_{\mathrm{in},\mathfrak{S}}}\!\right\rangle.
\end{equation}
(Here, we've employed the ``double-ket'' notation that appears in the vectorization of $\mathcal{B}\left(\mathcal{H}_{\mathfrak{S}}\right)$; this is discussed in more detail in Appendix \ref{app:vectorization}. This notation will be used for the remainder of the text.)

We can determine the asymptotic states by examining the spectral decomposition of $\hat{\hat{\mathcal{L}}}$. 
Here, we follow the excellent presentation in \cite{Albert18}, which is the thesis corresponding to the original papers \cite{Albert14, ABFJ16} which develop this framework. (This is simply an abbreviated version; the reader interested in more details of the framework is highly encouraged to read these references.) 
In general, $\hat{\hat{\mathcal{L}}}$ is not necessarily Hermitian. If it's still unitarily diagonalizable, we denote the eigenvalues by $\lambda_{a}$, the (right) eigenvectors by $\left.\ket{\mathsf{p}_{a}}\!\right\rangle$, and the left eigenvectors by $\left\langle\!\bra{\mathsf{q}_{a}}\right.$. Denoting $\hat{\hat{\mathcal{D}}} = \mathrm{diag}\left(\lambda_{a}\right)$, $\hat{\hat{\mathcal{P}}}$ as the matrix formed by setting $\left\{\!\left.\ket{\mathsf{p}_{a}}\!\right\rangle\!\right\}$ as rows, and $\hat{\hat{\mathcal{P}}}^{-1}$ as the matrix formed by setting $\left\{\!\left\langle\!\bra{\mathsf{q}_{a}}\right.\!\right\}$ as columns, we have $\left.\ket{\rho_{\mathfrak{S}}\left(t\right)}\!\right\rangle$ in terms of the spectral decomposition of $\hat{\hat{\mathcal{L}}}$ given by:
\begin{equation}
    \label{eq:GKSL-time-evol-spec-decomp-normal}
    \begin{split}
        \left.\ket{\rho_{\mathfrak{S}}\left(t\right)}\!\right\rangle &= \mathrm{e}^{t\hat{\hat{\mathcal{L}}}}\left.\ket{\rho_{\mathrm{in},\mathfrak{S}}}\!\right\rangle = \hat{\hat{\mathcal{P}}}\mathrm{e}^{t\hat{\hat{\mathcal{D}}}}\hat{\hat{\mathcal{P}}}^{-1}\left.\ket{\rho_{\mathrm{in},\mathfrak{S}}}\!\right\rangle \\[4pt]
        &= \sum_{a}\left.\ket{\mathsf{p}_{a}}\!\right\rangle\, \mathrm{e}^{t\lambda_{a}}\,\left\langle\!\Braket{\mathsf{q}_{a}\!|\!\rho_{\mathrm{in},\mathfrak{S}}}\!\right\rangle = \sum_{a} c_{a}\,\mathrm{e}^{t\lambda_{a}}\!\:\left.\ket{\mathsf{p}_{a}}\!\right\rangle.
    \end{split}
\end{equation}
Here, $c_{a} = \left\langle\!\Braket{\mathsf{q}_{a}\!|\!\rho_{\mathrm{in},\mathfrak{S}}}\!\right\rangle$ are simply the coefficients of the (Hilbert-Schmidt) inner product $\left\langle\!\Braket{\mathsf{q}_{a}\!|\!\rho_{\mathrm{in},\mathfrak{S}}}\!\right\rangle = \mathrm{Tr}_{\mathfrak{S}}\left[\mathsf{q}_{a}^{\dagger}\,\rho_{\mathrm{in},\mathfrak{S}}\right].$ Since $t\in\mathbb{R}$, the eigenvalues can only satisfy $\mathfrak{Re}\left\{\lambda_{a}\right\} < 0$ or $\mathfrak{Re}\left\{\lambda_{a}\right\} = 0$ to have finite $\left.\ket{\rho_{\infty}}\!\right\rangle$. The eigenstates for $\mathfrak{Re}\left\{\lambda_{a}\right\} < 0$ are the \emph{damped} or \emph{decaying states}, while the eigenstates for $\mathfrak{Re}\left\{\lambda_{a}\right\} = 0$ are the asymptotic states.\footnote{Most examinations of GKSL dynamics examine the case of a single asymptotic state, given by the right eigenvector corresponding to $\lambda = 0$. The corresponding left eigenvector is $\mathbbm{1}$. GKSL systems with multiple asymptotic states form a set of measure zero \cite{Albert18} over the set of all possible GKSL systems. However, we're representing an actual engineered system: we're not interested in the set of \emph{all} possible GKSL systems, we're interested in the one that actually represents our system, so this isn't a problem.}

In the case that $\hat{\hat{\mathcal{L}}}$ isn’t diagonalizable, it still has a decomposition in Jordan normal form. Each Jordan block has a specific eigenvalue, but eigenvalues could be spread over multiple Jordan blocks. If we remove the assumption of unitary diagonalizability, the decomposition given by \eqref{eq:GKSL-time-evol-spec-decomp-normal} still holds for diagonal Jordan blocks, where $\lambda_{a}$ is now specifically the eigenvalue of $\left.\ket{\mathsf{p}_{a}}\!\right\rangle$. (Since the left eigenvectors are the same as the right eigenvectors of the adjoint operator, the eigenvalue of $\left\langle\!\bra{\mathsf{q}_{a}}\right.$ is just $\lambda_{a}^{\ast}$. The overall spectral decomposition of the operators now use the right eigenvalue specifically.) For non-diagonal Jordan blocks, the expression $\mathrm{e}^{t\hat{\hat{\mathcal{L}}}}$ pulls down a factor of $t^{n}/n!$ for each instance of a nonzero superdiagonal entry. Thus, for a given non-diagonal Jordan block with right eigenvalue $\lambda_{a}$ and $N$ generalized eigenvectors, if we index the generalized eigenvectors by $\mu,\nu\in\mathbb{N}_{N}^{+}$, we have the decomposition of $\mathrm{e}^{t\hat{\hat{\mathcal{L}}}}\left.\ket{\rho_{\mathrm{in},\mathfrak{S}}}\!\right\rangle$ for that specific block given by \cite{Albert14, Albert18}:
\begin{equation}
    \label{eq:GKSL-time-evol-spec-decomp}
    \begin{split}
        \left.\ket{\rho_{\mathfrak{S}}\left(t\right)}\!\right\rangle = \mathrm{e}^{t\hat{\hat{\mathcal{L}}}}\left.\ket{\rho_{\mathrm{in}, \mathfrak{S}}}\!\right\rangle &= \sum_{\nu\:\leq\:\mu}\frac{t^{\left(\nu\:-\:\mu\right)}}{\left(\nu-\mu\right)!} \left.\ket{\mathsf{p}_{a}}\!\right\rangle\, \mathrm{e}^{t\lambda_{a}}\,\left\langle\!\Braket{\mathsf{q}_{a}\!|\!\rho_{\mathrm{in},\mathfrak{S}}}\!\right\rangle \\[4pt]
        &= \sum_{\nu\:\leq\:\mu}\frac{c_{a}\:\mathrm{e}^{t\lambda_{a}}\:t^{\left(\nu\:-\:\mu\right)}}{\left(\nu-\mu\right)!} \left.\ket{\mathsf{p}_{a}}\!\right\rangle.
    \end{split}
\end{equation}
A direct consequence is that the Jordan normal form of the asymptotic state eigenvalue blocks (\emph{i.e.}, the eigenvalue blocks with pure imaginary eigenvalues) are all diagonal, since the factor of $t^{\left(\nu\:-\:\mu\right)}$ would blow up otherwise.

As mentioned earlier, the factor of $\mathrm{e}^{t\lambda_{a}}$ in the spectral decompositions  \eqref{eq:GKSL-time-evol-spec-decomp-normal} or \eqref{eq:GKSL-time-evol-spec-decomp} of $\left.\ket{\mathsf{p}_{a}}\!\right\rangle$ tells us that the set of pure imaginary eigenvalues correspond to the asymptotic states; we can denote these as $\Lambda_{a} = \mathfrak{Im}\left\{\lambda_{a}\right\}$. The corresponding right eigenvectors form a subspace of $\mathcal{B}\left(\mathcal{H}_{\mathfrak{S}}\right)$, called the \emph{asymptotic subspace} and denoted $\mathsf{As}\left(\mathcal{H}_{\mathfrak{S}}\right)$. The asymptotic left and right eigenvectors are respectively the asymptotic states and the (asymptotic) conserved quantities of the GKSL system we're examining.\footnote{These conserved quantities are slightly different from some of the more familiar conserved currents in classical and quantum mechanics; for instance, the conserved quantity given by the $\left\langle\!\bra{\mathbbm{1}}\right.$ eigenvector is the trace of $\rho_{\infty}$.} Indexing these by their value of $\Lambda$ and their corresponding degeneracy $\mu$, we'll denote the right and left eigenvectors as $\left.\ket{\mathsf{s}_{\Lambda\mu}}\!\!\right\rangle$ and $\left\langle\!\!\bra{\mathsf{J}_{\Lambda\mu}}\right.$ respectively. (For clarity, this means that we have $\hat{\hat{\mathcal{L}}}\left.\ket{\mathsf{s}_{\Lambda\mu}}\!\!\right\rangle = \mathrm{i}\Lambda\left.\ket{\mathsf{s}_{\Lambda\mu}}\!\!\right\rangle$ and $\hat{\hat{\mathcal{L}}}^{\ddagger}\left.\ket{\mathsf{J}_{\Lambda\mu}}\!\!\right\rangle = -\mathrm{i}\Lambda\left.\ket{\mathsf{J}_{\Lambda\mu}}\!\!\right\rangle$.)

\begin{figure}[h]
    \centerline{\includegraphics[width=5 cm]{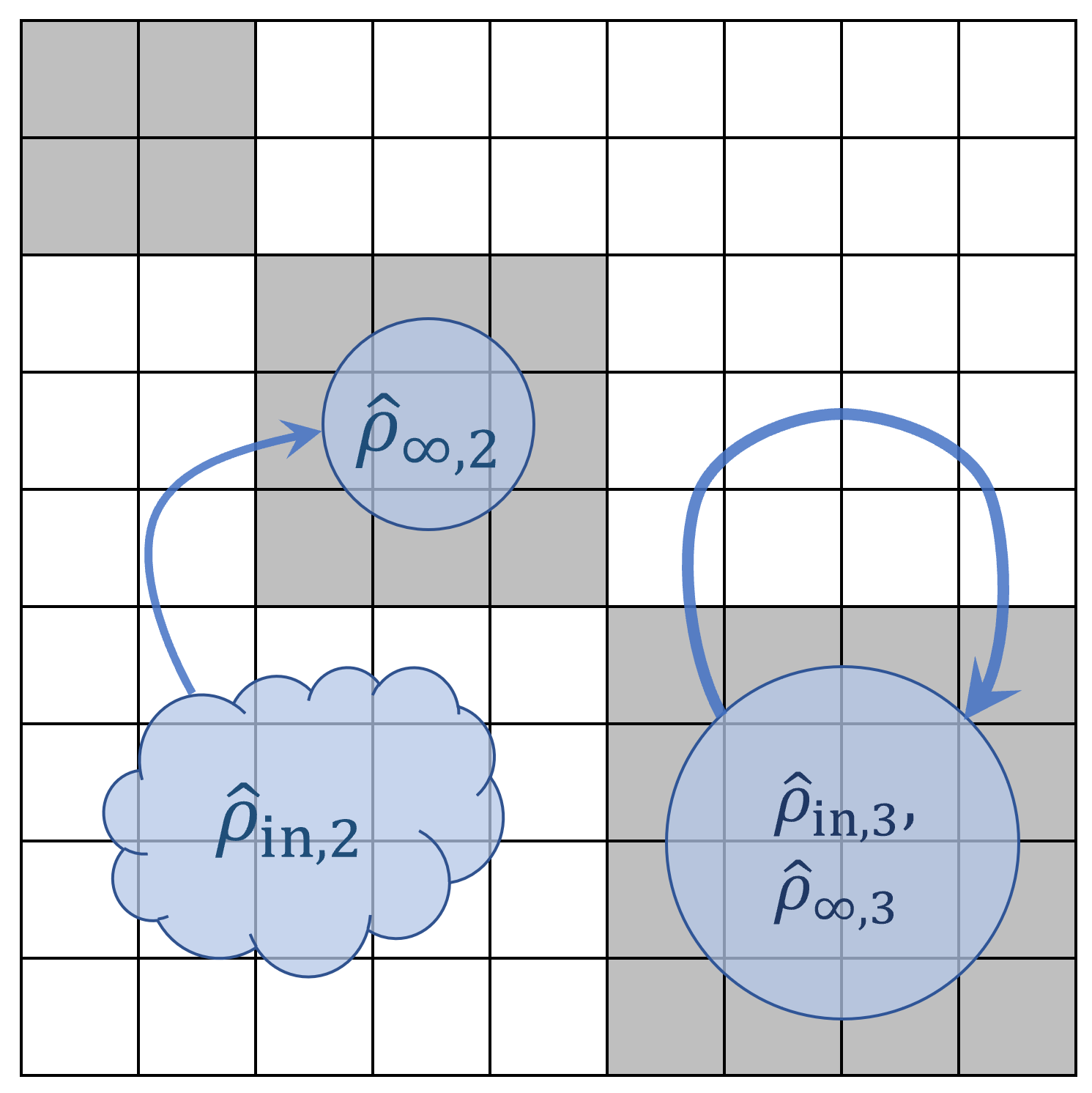}}
    \caption{Two examples of evolution of states under GKSL dynamics to $\mathsf{As}\left(\mathcal{H}_{\mathfrak{S}}\right)$ in the $t\rightarrow\infty$ limit. Here, the gray blocks denote further subspaces of $\mathsf{As}\left(\mathcal{H}_{\mathfrak{S}}\right)$. $\left.\ket{\rho_{\mathrm{in},2}}\!\right\rangle$ starts outside of $\mathsf{As}\left(\mathcal{H}_{\mathfrak{S}}\right)$ and settles into one such subspace of $\mathsf{As}\left(\mathcal{H}_{\mathfrak{S}}\right)$ as $t \rightarrow \infty$, whereas $\left.\ket{\rho_{\mathrm{in},3}}\!\right\rangle$ starts inside one subspace of $\mathsf{As}\left(\mathcal{H}_{\mathfrak{S}}\right)$, leaves it but stays within $\mathsf{As}\left(\mathcal{H}_{\mathfrak{S}}\right)$, and then returns to the same subspace as as $t \rightarrow \infty$. Many further dynamics are possible: for instance, a state in one of these subspaces could leave $\mathsf{As}\left(\mathcal{H}_{\mathfrak{S}}\right)$ altogether, and then settle into the same or even a different subspace of $\mathsf{As}\left(\mathcal{H}_{\mathfrak{S}}\right)$ as $t \rightarrow \infty$; meanwhile, a state could start in a subspace of $\mathsf{As}\left(\mathcal{H}_{\mathfrak{S}}\right)$, leave it but stay within $\mathsf{As}\left(\mathcal{H}_{\mathfrak{S}}\right)$ overall, and settle into a different subspace as $t \rightarrow \infty$. There are, indeed, no restrictions on where the initial state can start from, as long as it's somewhere in the universe and settles to $\mathsf{As}\left(\mathcal{H}_{\mathfrak{S}}\right)$ as $t \rightarrow \infty$: as with $\left.\ket{\rho_{\mathrm{in},2}}\!\right\rangle$, a state could start outside of $\mathsf{As}\left(\mathcal{H}_{\mathfrak{S}}\right)$ altogether, or a state could alternately start inside $\mathsf{As}\left(\mathcal{H}_{\mathfrak{S}}\right)$ but outside one of the subspaces. These subspaces are defined in the infinite time dynamics; thus, they only have relevance \emph{after} the GKSL dynamics has finished. \label{fig:settle}}
\end{figure}

Under the action of $\mathrm{e}^{t\hat{\hat{\mathcal{L}}}}$, \emph{all} initial states $\left.\ket{\rho_{\mathrm{in}}}\!\right\rangle$ converge to $\mathsf{As}\left(\mathcal{H}_{\mathfrak{S}}\right)$ in the $t\rightarrow\infty$ limit. Naturally, the GKSL dynamics maps states outside of $\mathsf{As}\left(\mathcal{H}_{\mathfrak{S}}\right)$---the states that don't survive in the infinite time limit---to some linear combination of states in $\mathsf{As}\left(\mathcal{H}_{\mathfrak{S}}\right)$. However, at a \emph{finite} time $t$, there's no requirement that if
$\left.\ket{\rho_{\mathfrak{S}}\left(t\right)}\!\right\rangle = \mathrm{e}^{t\hat{\hat{\mathcal{L}}}}$ is in $\mathsf{As}\left(\mathcal{H}_{\mathfrak{S}}\right)$, then it must stay in that state for all future times. Indeed, the GKSL dynamics permits the time evolution of a state in $\mathsf{As}\left(\mathcal{H}_{\mathfrak{S}}\right)$ to leave that subspace altogether at some other time, as long as the state returns to $\mathsf{As}\left(\mathcal{H}_{\mathfrak{S}}\right)$ at $t\rightarrow\infty$. Two examples of GKSL state evolution are provided in Figure \ref{fig:settle}.

From $\left.\ket{\mathsf{s}_{\Lambda\mu}}\!\!\right\rangle$ and $\left.\ket{\mathsf{J}_{\Lambda\mu}}\!\!\right\rangle$, we can construct a superoperator projector $\hat{\hat{\mathcal{P}}}_{\infty}$, known as the \emph{asymptotic projection}, which projects solely onto $\mathsf{As}\left(\mathcal{H}\right)$:\footnote{The last two equalities come from recognizing $\hat{\hat{\mathcal{P}}}_{\infty}$ as the projector onto the peripheral spectrum of $\mathrm{e}^{t\hat{\hat{\mathcal{L}}}}$. The second-to-last equality expresses $\hat{\hat{\mathcal{P}}}_{\infty}$ as the Cesàro mean of $\mathrm{e}^{t\hat{\hat{\mathcal{L}}}}$, and the last equality expresses $\hat{\hat{\mathcal{P}}}_{\infty}$ as the Dunford-Taylor integral of $\mathrm{e}^{t\hat{\hat{\mathcal{L}}}}$ via the resolvent $\left(z\hat{\hat{\mathcal{I}}}-\hat{\hat{\mathcal{L}}}\right)^{-1}$.}
\begin{equation}
    \label{eq:asymp-proj}
    \hat{\hat{\mathcal{P}}}_{\infty} \coloneqq \sum_{\Lambda,\:\mu} \left.\ket{\mathsf{s}_{\Lambda\mu}}\!\!\right\rangle\left\langle\!\!\bra{\mathsf{J}_{\Lambda\mu}}\right. = \underset{T\:\rightarrow\:\infty}{\mathrm{lim}}\:\!\frac{1}{T}\:\!\sum_{\Lambda}\:\!\int\limits_{0}^{T}\mathrm{d}t\:\mathrm{e}^{t\left(\hat{\hat{\mathcal{L}}}\:-\:\mathrm{i}\Lambda\hat{\hat{\mathcal{I}}}\right)} = \frac{1}{2\pi \mathrm{i}}\oint\limits_{C} \frac{\mathrm{d}z\:\mathrm{e}^{t\hat{\hat{\mathcal{L}}}}}{z\hat{\hat{\mathcal{I}}}-\hat{\hat{\mathcal{L}}}}.
\end{equation}
(Here, $C$ is the contour which encloses only the $\left\{\lambda\right\}$s.) Using $\hat{\hat{\mathcal{P}}}_{\infty}$, we can write $\left.\ket{\rho_{\infty}}\!\right\rangle$ as:
\begin{equation}
    \label{eq:infinite-time-dynamics}
    \left.\ket{\rho_{\infty}}\!\right\rangle = \mathrm{e}^{-\mathrm{i}\widehat{H}_{\infty}s}\left(\hat{\hat{\mathcal{P}}}_{\infty}\left.\ket{\rho_{\mathrm{in}}}\!\right\rangle\right)\mathrm{e}^{\mathrm{i}\widehat{H}_{\infty}s}
\end{equation}
Here, $\widehat{H}_{\infty}$ is the Hamiltonian that governs the asymptotic dynamics, parametrized by the time parameter $s \gg t$; \emph{i.e.}, $s$ \emph{solely} describes the dynamics of the system after it has already equilibrated. From this expression, we can directly see that $\mathsf{As}\left(\mathcal{H}_{\mathfrak{S}}\right) = \hat{\hat{\mathcal{P}}}_{\infty}\left[\mathcal{B}\left(\mathcal{H}_{\mathfrak{S}}\right)\right]$. 

At this point, an important subtlety about time scales must be addressed. As mentioned in \S\ref{par:gksl-intro}, the Markov assumption involves the separation of multiple time scales. Three time scales in particular are relevant: the time scale $\tau_{\mathsf{F}}$ of the fluctuations between $\mathfrak{S}$ and $\mathfrak{E}$, the time scale $\tau_{\mathsf{S}}$ of the GKSL dynamics, and the relaxation time scale $\tau_{\mathsf{R}}$; all of which are separated according to \eqref{eq:Markov-time-scales}. The introduction of multiple asymptotic states gives rise to the possibility of dynamics within the asymptotic space \emph{after} the relaxation. By contrast, the GKSL equation \eqref{eq:GKSL-def} examines the process by which a system initially in a state $\left.\ket{\rho_{\mathrm{in}}}\!\right\rangle$ relaxes to one or more of the asymptotic state(s) $\left.\ket{\rho_{\infty}}\!\right\rangle$. This assumes that the relaxation occurs on a timescale much faster than the timescale involved in the dynamics of the asymptotic state(s). Thus, implicit in this expression is the notion that $t$ is a parameter that \emph{only} sees timescales on the order of the relaxation timescale. In other words, the GKSL dynamics is entirely \emph{before} the dynamics of the asymptotic states, and the limit as $t \rightarrow \infty$ is then still \emph{before} the asymptotic state dynamics. Labelling the asymptotic dynamics timescale $s_{\mathsf{As}}$, we now modify \eqref{eq:Markov-time-scales} to include $s_{\mathsf{As}}$, giving:
\begin{equation}
    \label{eq:multiplte-As-Markov-time-scales}
    \tau_{\mathsf{F}} \ll \tau_{\mathsf{S}} \ll \tau_{\mathsf{R}} \ll s_{\mathsf{As}}.
\end{equation}
Intuitively, the GKSL dynamics is interested in the limit as $t \rightarrow \tau_{\mathsf{R}}$, but due to the parametrization of the dynamics, we take the limit as $t \rightarrow \infty.$ In order to describe the asymptotic dynamics, we use a separate parameter $s$.

The spectral properties of $\hat{\hat{\mathcal{L}}}$ play a central role in understanding the GKSL dynamics of the system in question. As such, the projector decomposition of these dynamics which separates the asymptotic dynamics from the dissipative dynamics serves as an essential tool to understand the overall structure of GKSL evolution. This is given by the \emph{four-corners decomposition} 
, first developed in \cite{ABFJ16}. (As before, we follow the presentation in \cite{Albert18}.) 
We can define the operator $\widehat{P}_{\mathsf{A}} \in \mathcal{H}_{\mathfrak{S}}$ as the projector onto the asymptotic states. Explicitly, $\widehat{P}_{\mathsf{A}}$ is defined in terms of $\rho_{\infty}$ by:
\begin{equation}
    \label{eq:asymptotic-state-projector}
    \begin{split}
        \widehat{P}_{\mathsf{A}}\,\rho_{\infty}\,\widehat{P}_{\mathsf{A}} &= \rho_{\infty};\\[4pt]
        \mathrm{Tr}_{\mathfrak{S}}\left[\widehat{P}_{\mathsf{A}}\right] &= \underset{\rho_{\infty}}{\mathrm{max}}\left[\mathrm{rank}\left(\rho_{\infty}\right)\right].
    \end{split}
\end{equation}
(The second expression helps us ensure that $\widehat{P}_{\mathsf{A}}$ is defined so that it \emph{only} projects onto the asymptotic states.) Meanwhile, the complement of $\widehat{P}_{\mathsf{A}}$ is given by: $\widehat{Q} \coloneqq \mathbbm{1}_{\mathfrak{S}} - \widehat{P}_{\mathsf{A}}$, with $\widehat{Q}\,\rho\left(t\right)\,\widehat{Q}\rightarrow 0$ as $t \rightarrow \infty$. Together, $\widehat{P}_{\mathsf{A}}$ and $\widehat{Q}$ provide the four-corners projections of operators $\widehat{A}\in\mathcal{B}\left(\mathcal{H}_{\mathfrak{S}}\right)$:
\begin{equation}
    \label{eq:four-corners-decomp}
    \begin{split}
        \widehat{A}_{\ytableausetup{boxsize = 1mm}\begin{ytableau}*(black)&*(white)\\*(white)&*(white)\\\end{ytableau}} \coloneqq  \hat{\hat{\mathcal{P}}}_{\ytableausetup{boxsize = 1mm}\begin{ytableau}*(black)&*(white)\\*(white)&*(white)\\\end{ytableau}} \widehat{A} \coloneqq \widehat{P}_{\mathsf{A}}\widehat{A}\widehat{P}_{\mathsf{A}}; \quad & \quad \widehat{A}_{\ytableausetup{boxsize = 1mm}\begin{ytableau}*(white)&*(white)\\*(black)&*(white)\\\end{ytableau}} \coloneqq  \hat{\hat{\mathcal{P}}}_{\ytableausetup{boxsize = 1mm}\begin{ytableau}*(white)&*(white)\\*(black)&*(white)\\\end{ytableau}} \widehat{A} \coloneqq \widehat{Q}\widehat{A}\widehat{P}_{\mathsf{A}}; \\[4pt]
        \widehat{A}_{\ytableausetup{boxsize = 1mm}\begin{ytableau}*(white)&*(black)\\*(white)&*(white)\\\end{ytableau}} \coloneqq  \hat{\hat{\mathcal{P}}}_{\ytableausetup{boxsize = 1mm}\begin{ytableau}*(white)&*(black)\\*(white)&*(white)\\\end{ytableau}} \widehat{A} \coloneqq \widehat{P}_{\mathsf{A}}\widehat{A}\widehat{Q}; \quad & \quad \widehat{A}_{\ytableausetup{boxsize = 1mm}\begin{ytableau}*(white)&*(white)\\*(white)&*(black)\\\end{ytableau}} \coloneqq  \hat{\hat{\mathcal{P}}}_{\ytableausetup{boxsize = 1mm}\begin{ytableau}*(white)&*(white)\\*(white)&*(black)\\\end{ytableau}} \widehat{A} \coloneqq \widehat{Q}\widehat{A}\widehat{Q}. \\[4pt]
        \widehat{A} = &\begin{pmatrix}
                       \widehat{A}_{\ytableausetup{boxsize = 1mm}\begin{ytableau}*(black)&*(white)\\*(white)&*(white)\\\end{ytableau}} & \widehat{A}_{\ytableausetup{boxsize = 1mm}\begin{ytableau}*(white)&*(black)\\*(white)&*(white)\\\end{ytableau}} \\
                       \widehat{A}_{\ytableausetup{boxsize = 1mm}\begin{ytableau}*(white)&*(white)\\*(black)&*(white)\\\end{ytableau}} & \widehat{A}_{\ytableausetup{boxsize = 1mm}\begin{ytableau}*(white)&*(white)\\*(white)&*(black)\\\end{ytableau}}
                   \end{pmatrix}
   \end{split}
\end{equation}
Thus, $\widehat{P}_{\mathsf{A}}$ and $\widehat{Q}$ provide a decomposition of every $\widehat{A}\in\mathcal{B}\left(\mathcal{H}_{\mathfrak{S}}\right)$. \eqref{eq:four-corners-decomp} also provides a definition of the \emph{four corners projection superoperators} $\left\{\hat{\hat{\mathcal{P}}}_{\ytableausetup{boxsize = 1mm}\begin{ytableau}*(black)&*(white)\\*(white)&*(white)\\\end{ytableau}}, \hat{\hat{\mathcal{P}}}_{\ytableausetup{boxsize = 1mm}\begin{ytableau}*(white)&*(white)\\*(black)&*(white)\\\end{ytableau}}, \hat{\hat{\mathcal{P}}}_{\ytableausetup{boxsize = 1mm}\begin{ytableau}*(white)&*(black)\\*(white)&*(white)\\\end{ytableau}}, \hat{\hat{\mathcal{P}}}_{\ytableausetup{boxsize = 1mm}\begin{ytableau}*(white)&*(white)\\*(white)&*(black)\\\end{ytableau}}\right\}$, which act on $\widehat{A}$ as indicated.

\eqref{eq:asymptotic-state-projector} and \eqref{eq:four-corners-decomp} serve as the foundation for examining the properties of GKSL systems with multiple asymptotic states, as well as the geometric properties of their quantum state spaces. These in turn are fundamental for examining the properties of classical (including reversible) computing operations in GKSL systems. However, a detailed discussion of the properties of the four-corners decomposition is far beyond the scope of this paper; the interested reader is highly encouraged to examine \cite{Albert14, ABFJ16, Albert18}. For our purposes here, what's relevant is that $\mathsf{As}\left(\mathcal{H}_{\mathfrak{S}}\right)$ forms an identifiable subspace, which we can project onto using the four corners projection superoperators.

As an important note, $\hat{\hat{\mathcal{P}}}_{\ytableausetup{boxsize = 1mm}\begin{ytableau}*(black)&*(white)\\*(white)&*(white)\\\end{ytableau}}$ does \emph{not} project onto $\mathsf{As}\left(\mathcal{H}_{\mathfrak{S}}\right)$ directly. Rather, the $\!\;\ytableausetup{boxsize = 1.35mm, aligntableaux = bottom}\begin{ytableau}*(black)&*(white)\\*(white)&*(white)\\\end{ytableau}\!\;$ subspace contains $\mathsf{As}\left(\mathcal{H}_{\mathfrak{S}}\right)$ in its entirety: $\mathsf{As}\left(\mathcal{H}_{\mathfrak{S}}\right) \subseteq \!\; \ytableausetup{boxsize = 1.35mm, aligntableaux = center}\begin{ytableau}*(black)&*(white)\\*(white)&*(white)\\\end{ytableau}$. The difference between these lies in the asymptotic dynamics governed by $\widehat{H}_{\infty}$: $\mathsf{As}\left(\mathcal{H}_{\mathfrak{S}}\right)$ describes the states that survive in the infinite-time limit (as given in \eqref{eq:asymp-proj} and \eqref{eq:infinite-time-dynamics}). If there are no further dephasing dynamics within $\!\;\ytableausetup{boxsize = 1.35mm, aligntableaux = bottom}\begin{ytableau}*(black)&*(white)\\*(white)&*(white)\\\end{ytableau}$, then $\mathsf{As}\left(\mathcal{H}_{\mathfrak{S}}\right) = \ytableausetup{boxsize = 1.35mm, aligntableaux = center}\begin{ytableau}*(black)&*(white)\\*(white)&*(white)\\\end{ytableau}$; conversely, if there are, then $\mathsf{As}\left(\mathcal{H}_{\mathfrak{S}}\right) \subsetneq \, \ytableausetup{boxsize = 1.35mm, aligntableaux = center}\begin{ytableau}*(black)&*(white)\\*(white)&*(white)\\\end{ytableau}$. This notation also serves as a visual indication for the framework: each operator in $\mathcal{B}\left(\mathcal{H}_{\mathfrak{S}}\right)$ can be subdivided into four regions, corresponding to these projections. We can freely gather $\mathsf{As}\left(\mathcal{H}_{\mathfrak{S}}\right)$ into the top-left corner. Then, $\widehat{P}_{\mathsf{A}}\widehat{A}\widehat{P}_{\mathsf{A}}$ projects into the top-left corner; \emph{i.e.}, $\widehat{P}_{\mathsf{A}}\widehat{A}\widehat{P}_{\mathsf{A}}$ projects into $\!\;\ytableausetup{boxsize = 1.35mm, aligntableaux = bottom}\begin{ytableau}*(black)&*(white)\\*(white)&*(white)\\\end{ytableau}$.

The subspace $\!\:\ytableausetup{boxsize = 1.35mm, aligntableaux = bottom}\begin{ytableau}*(black)&*(white)\\*(white)&*(white)\\\end{ytableau}\!\:$ is a full Hilbert space in its own right, supporting quantum mechanical evolution under $\widehat{H}_{\infty}$ in the $t\rightarrow\infty$ limit. Thus, it supports \emph{any} possible dynamics that can be governed by a Hamiltonian. Indeed, this framework provides a way to describe an open system extension of \emph{any} finite-dimensional system that can be governed by the laws of nonrelativistic quantum mechanics, as long as the open system relaxation is governed by Markov dynamics.\footnote{In principle, infinite-dimensional Hilbert spaces, relativistic quantum mechanical systems, and quantum field theories should be describable as well; however, the details of these descriptions are still in progress.} Thus, we can directly model a system of computational states as discussed in \S\ref{par:qsoc}: each computational state corresponds to a DFS within an overall Hilbert space. The overall Hilbert space $\!\;\ytableausetup{boxsize = 1.35mm, aligntableaux = bottom}\begin{ytableau}*(black)&*(white)\\*(white)&*(white)\\\end{ytableau}\!\;$ is then the direct sum of the individual DFS spaces, known as the \emph{von Neumann algebra} \cite{BN08,TV08,BKNPV10,DFSU16,PP17}: 
\begin{equation}
    \label{eq:vN-algebra}
     \ytableausetup{boxsize = 1.6mm, aligntableaux = center}\begin{ytableau}*(black)&*(white)\\*(white)&*(white)\\\end{ytableau} = \bigoplus\limits_{i}\mathcal{H}_{\mathsf{DFS},i}.
\end{equation}
The fact that this is identical to \eqref{eq:subspace-sum} (at the level of the vectorized space) shows us that the von Neumann algebra is a natural framework to represent reversible computing operations. This representation is discussed in detail in \S\ref{ssec:rc-op-rep}. The embedding of a von Neumann algebra within the four corners representation of an operator evolving under GKSL dynamics is given in Figure~\ref{fig:asymp}.
\begin{figure}[H]
    \centerline{\includegraphics[width=5 cm]{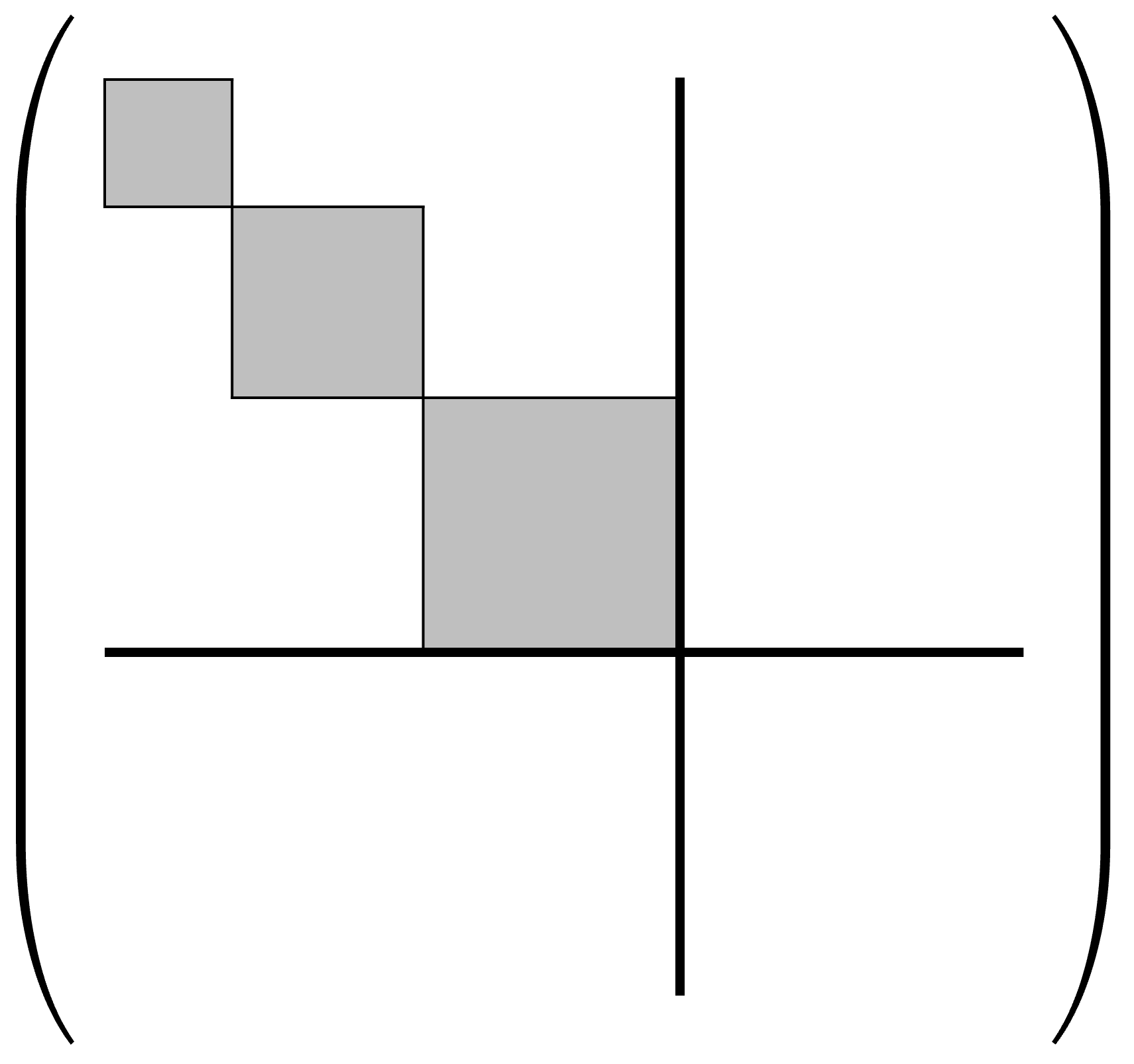}}
    \caption{An example of the type of overall operator algebra $\mathcal{B}\left(\mathcal{H}_{\mathfrak{S}}\right)$ that can support classical (including reversible) computing operations. (Note this matrix operates on the space of vectorized density matrices.) Here, the upper left corner subspace is a von Neumann algebra corresponding to the direct sum of decoherence-free subspaces, with the gray regions representing the individual decoherence-free subspaces. 
    \label{fig:asymp}}
\end{figure}

A striking feature of GKSL dynamics with multiple asymptotic states is that the shape of $\!\;\ytableausetup{boxsize = 1.35mm, aligntableaux = bottom}\begin{ytableau}*(black)&*(white)\\*(white)&*(white)\\\end{ytableau}\!\;$ corresponds to substantially different expressions for the quantum geometric tensor (QGT) over $\!\;\ytableausetup{boxsize = 1.35mm, aligntableaux = bottom}\begin{ytableau}*(black)&*(white)\\*(white)&*(white)\\\end{ytableau}\!\;$. This is a key aspect of GKSL dynamics with multiple asymptotic states, and will also serve as an essential feature of understanding the properties of RC operations in open quantum systems. The dependence of the dynamics on the QGT of $\!\;\ytableausetup{boxsize = 1.35mm, aligntableaux = bottom}\begin{ytableau}*(black)&*(white)\\*(white)&*(white)\\\end{ytableau}\!\;$ is central to the framework developed in \cite{ABFJ16}, and is discussed in detail there and in \cite{Albert18}. Unsurprisingly, because the framework of classical reversible computing operations in open quantum systems relies at its core on GKSL dynamics with multiple asymptotic states, the dependence of GKSL dynamics on the QGT over $\!\;\ytableausetup{boxsize = 1.35mm, aligntableaux = bottom}\begin{ytableau}*(black)&*(white)\\*(white)&*(white)\\\end{ytableau}\!\;$ is an indispensable part of classical RC operations in open quantum systems as well. The quantum geometric properties of RC operations are briefly mentioned in \S\ref{ssec:rc-op-rep}. A more detailed analysis of these properties, and conclusions regarding RC operations, are the central theme of a forthcoming work which follows up on these discussions.

\subsection{Existing and Proposed Implementation Technologies}
\label{ssec:tech}

In this section, we briefly survey a number of conceptual examples of concrete physical mechanisms of operation that may be suitable, to varying degrees, for performing reversible computations. The detailed performance characteristics for these example technologies (\emph{e.g.}, the exact energy dissipation per operation as a function of speed) depend on a great many design details, so we will not attempt to derive those characteristics here. Rather, this survey is just to give the reader an idea regarding the range of physical mechanisms for reversible computing that may be possible. It is likely that many other, much more efficient mechanisms can be invented with further research.

First, here is a concise list of the technologies we will survey, with abbreviations noted (some of which are coined here):

\begin{enumerate}
    \item Reversible adiabatic CMOS (RA-CMOS).
    \item Reversible quantum flux parametron (RQFP).
    \item Reversible quantum-dot cellular automaton (R-QCA).
    \item Reversible nanomechanical rod logics (RNRL).
    \item Ballistic asynchronous reversible computing in superconductors (BARCS).
\end{enumerate}

These particular examples will be described in a bit more detail in the following subsections. These are not the \emph{only} physical mechanisms for reversible computing to have been proposed, but (except for BARCS) are some of the most well-developed implementation concepts so far.

\subsubsection{Reversible Adiabatic CMOS}
\label{sssec:ra-cmos}

This class of implementation technologies for reversible computing refers to a logic design discipline based on ordinary CMOS (complementary metal-oxide-semiconductor) field-effect transistors \cite{YK93,You94,Fra99,Ven+03,ZFW19,Fra+20a,Fra+20b}. To approach physical reversibility in these types of circuits requires several conditions to be met (see Appendix~\ref{app:adia-minE} for some key derivations):

\begin{enumerate}
    \item The \emph{on/off conductance ratio} $r_{\mathrm{on/off}} = G_{\mathrm{on}}/G_{\mathrm{off}}$ of the device channel (at the specified operating points) should diverge, as the technology is improved. The quantity $G_{\mathrm{on}}$ refers to the typical effective peak source-drain conductance through the channel of a device (transistor) when it is in the ``on'' state (with gate voltage set accordingly, \emph{e.g.}, $V_\mathrm{g} = V_\mathrm{dd} = $ logic HIGH for an n-type FET). Meanwhile, $G_{\mathrm{off}}$ refers to the maximum conductance through the device for off-state ``leakage'' current (including both gate leakage and subthreshold current) when the device is nominally turned off (\emph{e.g.}, $V_\mathrm{g} = 0 = $ logic LOW for an nFET). Roughly speaking, $1/G_\mathrm{on}$ ends up being proportional to the characteristic \emph{relaxation timescale} $\tau_{r} = R_\mathrm{on}C$ of the circuit, while $1/G_\mathrm{off}$ ends up being proportional to the characteristic \emph{equilibration timescale} $\tau_e = R_\mathrm{off}C$ of the circuit when its non-equilibrium state is not being actively maintained. One of the classic results of physical reversible computing theory, the roots of which can be traced back to Feynman's lectures on computation, delivered in the early 80s \cite{Fey00}, is that in general, at least for any classic ``adiabatic'' reversible computing technology, the maximum energy recovery efficiency for a reversible device is ultimately limited as a function of the ratio of these two timescales, \emph{e.g.}, as $\eta_\mathrm{er} \leq 1 - c\cdot\sqrt{\tau_\mathrm{r}/\tau_\mathrm{e}}$ for $\tau_\mathrm{e} \gg \tau_\mathrm{r}$. (See App.~\ref{app:adia-minE}) That is, the minimum fraction of signal energy dissipated per operation cycle scales like $\sqrt{\tau_\mathrm{r}/\tau_\mathrm{e}}$, quite generally. For CMOS, this means that, to attain high energy efficiency, we want to make the leakage conductance $G_\mathrm{off}$ as small as possible to extend the equilibration timescale $\tau_\mathrm{e}$, and doing this well in practice requires some combination of various engineering refinements (\emph{e.g.}, higher threshold voltages, thicker gate oxides, lower operating temperatures, higher materials purity). Identifying the most economical manufacturing process to minimize $G_\mathrm{off}$ in practice is not a simple optimization problem by any means. However, there appears to be no fundamental reason why the ratio $r_\mathrm{on/off}$ cannot be made as large as desired with further refinement of the technology over time. Thus, it seems that this class of circuits can approach ideal reversibility with continued development.
    
    \item Since in CMOS, the relaxation timescale $\tau_\mathrm{r}$ is subject to lower bounds, the \emph{transition time} $t_\mathrm{tr}$ for the adiabatic logic transitions should also diverge. For a given technology, the minimum dissipation per cycle will be found when the transition time $t_\mathrm{tr}$ is (within a small constant factor) roughly at the geometric mean $\tau_\mathrm{m} = \sqrt{\tau_\mathrm{r}\tau_\mathrm{e}}$ between the relaxation and equilibration timescales (App.~\ref{app:adia-minE}). 
    But, as long as we can arrange to keep extending the equilibration timescale $\tau_e$, the useful transition time $t_\mathrm{tr} \approx \tau_\mathrm{m}$ can continue increasing as well.
    
    \item The \emph{effective quality factor} $Q_\mathrm{eff}$ of any external resonant oscillatory element serving as the clock-power supply driving the adiabatic circuit should also diverge. For our purposes, $Q_\mathrm{eff}$ can be defined as the ratio between the peak electrostatic energy $E_\mathrm{load}=\frac{1}{2}C_\mathrm{load}V_\mathrm{dd}^2$ stored transiently on the logic nodes, and the energy dissipated by the resonant oscillator per cycle, $Q_\mathrm{eff} = E_\mathrm{load}/E_\mathrm{odiss}$.
    
\end{enumerate}

In addition to the above, two design rules that must always be obeyed in (conditionally) reversible adiabatic CMOS circuits in order for them to be able to approach physically-reversible operation in the above limits are the following:

\begin{enumerate}
    \item Never turn on a transistor when there is a nonzero source-drain voltage across it.
    \item Never turn off a transistor when there is a nonzero source-drain current through it.
\end{enumerate}

If either of these rules is ever broken in the design, this can lead to substantial non-adiabatic dissipation, and the physical computational process as a whole no longer qualifies as being asymptotically physically reversible. This is discussed further in \cite{Fra+20b,Fra03,Fra99}.

A brief description of the overall normal mechanism of reversible operation for these kinds of circuits is as follows.  Periodic voltage waveforms are supplied by a resonant oscillatory circuit that is customized to provide quasi-trapezoidal wave shapes (that is, with roughly flat waveform tops and bottoms).  The flat regions are needed in order to avoid pushing current through devices while they are being switched on or off.  The provided waveforms exist in several different (mutually-offset) phases.  Each phase drives a corresponding section (subset) of adiabatic logic circuits.  The choice of which circuit nodes to charge up in a given section is determined using series-parallel networks of devices, whose gate (control) electrodes are connected to the (quiescent) nodes controlled by a neighboring phase.  After the supplied waveforms have finished causing the desired transitions between valid logic voltage levels for a given section of the circuit, now those circuit nodes can be used to control the adiabatic transitions for the neighboring sections in adjacent clock phases.  The correct architectural design of these kinds of circuits can become somewhat involved, but is conceptually straightforward. See~\cite{Fra+20b} for an example.

\mypara{Description of RA-CMOS in terms of our general framework.}

In reversible adiabatic CMOS technology, a given (time-dependent) computational state $c(\tau_\ell)$ has a very simple description in terms of physical microstates. Essentially, in a given computational state, at a given time, each circuit node exhibits a given well-defined, relatively uniform voltage level, within some tolerances. Of course, there will be local fluctuations about that average level. Physical states in which voltages depart substantially (over a broad region) from any of the computationally-meaningful levels can be relegated to the catch-all ``invalid'' computational state $c_\bot$, but during normal operation of a well-engineered circuit, such states should have an astronomically close to zero probability of arising in any case.

In terms of computational operations, a certain computational operation $O_s^t$ is carried out each time one of the supply waveforms executes a voltage-level transition between two distinct valid logic levels, \emph{i.e.},\ from an initial level $V_\mathrm{i}$ to a final level $V_\mathrm{f} \neq V_\mathrm{i}$, over the time interval between the two time points $s$ (start time) and $t=s+t_\mathrm{tr}$ (end time). During this transition, the voltage levels on the set of circuit nodes that are connected to that particular supply line themselves undergo (with a slight delay, and modulo voltage offsets due to leakage and other non-idealities) the same transition between voltage levels.  In this process, some transistors (\emph{e.g.}, ones whose gates are controlled by the transitioning nodes) may be turned on or turned off, causing the source and drain nodes of those transistors to become connected to or disconnected from each other.  These connection and disconnection events result in the set of accessible computational states changing over time (since even the \emph{number} of independent connected components will be changing, thus, so will the number of available computational states).

In any case, as long as the two rules of adiabatic design are respected throughout a given transition, the operation $O_s^t$ that is performed will be both (conditionally) logically reversible (under the condition that the rules are respected), as well as asymptotically thermodynamically reversible, in the limit described above where $G_\mathrm{off}\rightarrow 0$ and $t_\mathrm{tr}\rightarrow\infty$, while keeping $t_\mathrm{tr}\ll\tau_\mathrm{e}$. Thus, we can allow both $r_\mathrm{on/off}$ and $t_\mathrm{tr}$ to continue increasing as the technology develops, and approach perfect reversibility over time given continued development of this family of technologies. 

The above discussion glosses over the important issue of the effective quality factor $Q_\mathrm{eff}$ of the driving power-clock resonator, which will also limit the overall degree of reversibility, but, as far as we know at present, there is no reason why this quantity cannot diverge as well, with continued engineering refinements. (The development of high-quality trapezoidal resonators suitable for driving adiabatic circuits is in the scope of engineering R\&D work being performed at Sandia.)

\subsubsection{Reversible Quantum Flux Parametron}
\label{sssec:rqfp}

The Reversible Quantum Flux Parametron (RQFP) logic family~\cite{TYY14,TYY18,YTY19} is a logically reversible variant of the well-developed superconducting logic family AQFP (Adiabatic Quantum Flux Parametron), which has been being developed primarily at Yokohama National University in Japan. RQFP (and its not-necessarily-reversible generalization AQFP) rely on adiabatic transformation of the abstract potential energy surface (PES) that obtains within Josephson-junction-based superconducting circuits. The independent variables for the PES describe the current distribution in the circuit, and the phase (order parameter) differences across the junctions. The PES is manipulated in such a way that the occupied potential energy valley of the system is transformed adiabatically to configurations representing different computational states. 

RQFP circuits are controlled by externally supplied waveforms, similarly to the case in RA-CMOS, except that the supplies are providing current signals, not voltage signals (since voltages, except for inductive transients, are normally zero in superconducting circuits). Except for the fact that the state of the circuit and the driving signals at a given time is described in terms of currents instead of voltages, and that the physics of superconductivity dominates the charge transport, the representation of RQFP in terms of computational and physical states is, roughly speaking, qualitatively similar to the case in RA-CMOS. That is to say, the higher-level principles of pipelined reversible logic are roughly comparable between the two technologies. 

One advantage, however, of RQFP compared to CMOS is that, due to Meissner-effect trapping of flux quanta, the natural equilibration timescale $\tau_\mathrm{e}$ is extremely large (effectively infinite) in RQFP, and as a result, scaling to extreme ultra-low levels of dissipation may ultimately prove far easier to do in RQFP than in adiabatic CMOS. The primary disadvantages of RQFP, compared to CMOS, are its lower density and accordingly higher manufacturing cost per-device, together with its requirement for low-temperature operation.

\subsubsection{Reversible Quantum-dot Cellular Automaton}
\label{sssec:r-qca}

The Quantum-dot Cellular Automata (QCA or QDCA) \cite{Lent+93,Amlani+99,LI03} family of technologies operate using single electrons confined to quantum dots, dipole configurations of two such electrons confined to four such dots in a square layout separated by tunnel barriers (a.k.a.\ a ``cell''), and linear/branching arrays of such cells interacting through dipole-dipole Coulombic interactions. As in RA-CMOS and RQFP, externally supplied signals are used to adiabatically raise and lower potential energy barriers that separate neighboring regions of the physical state space, in patterns that (in the technology's reversible variant, here dubbed R-QCA) allow the overall computational state to evolve reversibly---which, as usual, means in a (conditionally) logically reversible and asymptotically physically reversible way. 

An interesting note about QDCA is that it was recently shown \cite{PL18} that exponential scaling of adiabaticity with speed (as in Landau-Zener transitions with a missed level crossing) exists in this system, apparently implying that there is no fundamental lower bound on dissipation-delay product. In \cite{PL21}, Pidaparthi and Lent investigate this phenomenon in more detail using a Lindbladian analysis, finding that when there is weak thermal coupling to the environment, these systems can exhibit substantially suppressed dissipation within a certain regime of speeds. This is a promising result, and we expect that this type of behavior likely generalizes to a wider variety of quantum systems.

\subsubsection{Reversible Nanomechanical Rod Logics}
\label{sssec:rnrl}

This is a concept that goes back to K. Eric Drexler's work in the 1980s leading up to his dissertation on molecular nanotechnology at MIT \cite{Drexler81,Drexler91,Drexler92}. The original idea was that logical bits are encoded in the linear displacements of atomically-precise nano-rods that move within sleeves at the ends of nano-springs. The nods are pushed back and forth (by externally supplied mechanical signals, following the same kind of quasi-trapezoidal waveforms we've talked about previously) to adiabatically transform them between computational states, using nano-scale bumps on the rods to sterically hinder each other's motion in ways that allow them to perform (conditionally reversible) Boolean logic. The whole scheme is closely analogous to RA-CMOS, except that it uses mechanical rather than electrical state variables.

Drexler's rod logic concept was updated more recently \cite{Merkle+16,Merkle+18} by a group led by Ralph Merkle (a pioneering cryptographer and early nanotechnologist). The new concept eliminated the sleeve bearings, whose friction had dominated the dissipation in Drexler's earlier concept. In the new scheme, the only bearings are \emph{rotary} bearings implemented by single carbyne bonds, whose orbitals are circularly symmetrical. Frictional losses in this system were assessed in simulations \cite{HMA17} to be so low that individual joints (operated reversibly) would dissipate $\sim$70,000$\times$ less than Landauer's $k_{\mathrm{B}}T\ln 2$ limit even when operating at frequencies as high as 100 MHz. This example illustrates that in principle, dissipation-delay products for reversible operations can be far smaller than is the case in RA-CMOS. (This particular dissipation-delay value is roughly $10^6\times$ improved versus projected \emph{end-of-roadmap} CMOS---and at room temperature!)

The main problem with the Drexler-Merkle family of nanomechanical rod logic concepts for reversible computing is simply that building them would seemingly require a very general, sophisticated, atomically-precise and fast technology for nano-fabrication and assembly, which does not yet exist, and may continue to not exist for some decades.

\subsubsection{Ballistic Asynchronous Reversible Computing in Superconductors}
\label{sssec:barcs}

Ballistic Asynchronous Reversible Computing (BARC, previously called ABRC) \cite{Fra17b} is a fundamentally new physical model of reversible computing in which the computational degrees of freedom evolve \emph{ballistically} (\emph{i.e.}, under their own inertia) rather than being dragged along adiabatically as a side effect of the oscillatory evolution of an external resonator. This change may provide certain advantages in terms of, \emph{e.g.}, allowing us to avoid having to worry about accidentally exciting undesired modes of the resonator and of the distribution network for the driving signal. The BARC model is required to be asynchronous as a means to prevent the nonlinear interactions between subsystems from chaotically amplifying uncertainties in the subsystem trajectories.

In a current project at Sandia, we are attempting to implement the BARC model in superconducting electronic circuits \cite{Fra+19a}. The computational subsystems are individual polarized flux solitons (or \emph{fluxons}) propagating near-ballistically along long Josephson junction (LJJ) transmission lines. In our circuits, fluxons are conserved and interact asynchronously with stored flux quanta at (stateful) interaction sites or \emph{circuit elements}, transforming the local digital state reversibly, in a deterministic sort of elastic ``scattering'' interaction.

BARC is an extremely novel concept, and has only been developed to a very preliminary level to date.  So far, we have a single (very simple) ``working'' BARC circuit element (\emph{i.e.}, it simulates correctly in SPICE) \cite{Fra+19b}; a test chip for it has been fabricated, and experimental tests of it are in progress. However, a wider variety of useful elements, leading up to a complete logic family, still need to be developed and optimized.

We are also collaborating with a group at the University of Maryland which has been working on a similar ballistic approach which they call Reversible Fluxon Logic (RFL) \cite{WO17,OW18,WO18,YWO19,OW19,OW20,WO20}.  The original RFL concept envisioned synchronous ballistic logic, but the Maryland group is also now also developing asynchronous elements which fall into the BARCS paradigm \cite{WO20b}.



\section{Results}
\label{sec:res}

Much work remains to be done, in terms of fleshing out a complete physical theory of reversible computing informed by NEQT, but in this section, we review some important preliminary results in this area that can be, or have already been obtained.

First, we view it as important, for resolving some of the long-standing controversies in the thermodynamics of computation, to distinguish a couple of different results that have historically been associated with Landauer's Principle:

\begin{enumerate}

    \item First is a simple result regarding the interchangeability of entropy between computational and non-computational forms. This one follows directly from the association of computational states to sets of microstates discussed in \S\ref{sssec:comp-states}. However, it is such an important result that we call it \emph{The Fundamental Theorem of the Thermodynamics of Computation}. We review it in \S\ref{ssec:fun-thm} below. This result implies that non-computational or ``physical'' entropy must be increased when computational (``information'') entropy is reduced, but {\em does not require that total entropy be increased}. 

    \item Second is a result (\S\ref{ssec:lan-prin}) showing that a strict entropy \emph{increase} is required whenever there is a loss of known information (which by itself is not surprising, since entropy increase effectively \emph{means} that known information is reduced), and furthermore, that an example of this necessarily occurs when one of two mutually-correlated subsystems is \emph{obliviously erased}, meaning that, in isolation, its reduced subsystem entropy is ejected to its local thermal environment without regards to its existing correlations. To the extent that the ejected information is then thermalized, with its correlations to the other subsystem being lost, this then corresponds to a strict increase in total entropy. This result follows directly from unitarity, information theory, and the definitions in \S\ref{ssec:found}.

\end{enumerate}

We argue that it is the second result, and not the first one, that is most properly understood as being \textit{Landauer's Principle}, because Landauer's Principle is most properly taken to concern the consequences of information loss \emph{in a computer}---since that was the subject of Landauer's original paper \cite{Lan61}. And, at least in an ordinary, deterministic computer, it is normally the case that \emph{computed bits are correlated}---meaning, there is mutual information between them (and/or, between them and the inputs that they were computed from)---since in fact, one can say that it is the generation of specific desired patterns of correlation between different computational subsystems that is exactly the entire point of what computation, \textit{per se}, is all about.

In particular, we show in \S\ref{ssec:lan-prin} that, for any deterministically computed bit (or larger computational subsystem), the amount of \emph{new entropy that is generated} when that bit is obliviously erased is strictly lower-bounded by the prior reduced subsystem entropy of that bit (or subsystem).\footnote{The form of this argument that we present was previously made explicit in the preprint \href{https://arxiv.org/abs/1901.10327}{arXiv:1901.10327}, but we reprise it here.} Note that, in that statement, we are talking about an absolute \emph{increase} in the \emph{total} entropy of the model universe (including computational \emph{and} non-computational entropy, defined below), and \emph{not} just a \emph{transfer} of entropy from computational to non-computational form.
Thus, Landauer's Principle, when it is properly understood in this way, really does provide a lower bound on new entropy \emph{generation}, and not simply on entropy transfer, as has sometimes been claimed.

In addition to the above clarification of Landauer's Principle, we also (in \S\ref{ssec:rc-thms}) review two fundamental theorems of physical reversible computing theory. (These were previously presented in \cite{Fra17a}, but we reprise them here.)

\begin{enumerate}
    \item \emph{The Fundamental Theorem of Traditional Reversible Computing}, whose proof is summarized in \S\ref{sssec:trad-rc} below, states that the only deterministic computational operations that \emph{always} avoid ejecting computational entropy to non-computational form (and thus, can avoid Landauer's lower limit on entropy increase when operating in isolation on computed bits) are the \emph{unconditionally logically reversible operations} (\textit{e.g.}, Toffoli gate operations) traditionally studied in reversible computing theory.
    
    \item \emph{The Fundamental Theorem of Generalized Reversible Computing}, whose proof is summarized in \S\ref{sssec:gen-rc} below, states that, in a statistically \emph{contextualized} computation, it can suffice (in a properly designed mechanism) to avoid entropy ejection (and the resultant entropy increase due to Landauer's limit) if a computational operation is simply reversible on the \emph{subset} of initial states having nonzero probability in the given statistical operating context \cite{Fra17a}. 
    
\end{enumerate}

Taking the latter observation (the generalized theorem) into account is essential in order for the scope of the theory to adequately encompass the state of the art of the existing best practices (\S\ref{ssec:tech}) in the engineering of reversible computing hardware. The generalized theorem significantly expands the class of computational mechanisms that can be seen to be capable of approaching thermodynamic reversibility when appropriate constraints are met. In particular, all of the actual implementation technologies for reversible computing described in \S\ref{ssec:tech} can \emph{only} be understood properly in the light of the generalized form of the theorem. In other words, all of the real reversible computing technologies that have been implemented to date are only \emph{conditionally} reversible, and so they rely, for their ability to achieve asymptotic reversibility in practice, on the fact that their preconditions for reversibility have been met by design, within the architectures of those machines, and (implicitly) on the \emph{Generalized} Theorem of RC.

Next, the framework of NEQT provides us with several new perspectives from which to understand Landauer's principle, and to begin to characterize the properties of RC operations in open quantum systems. In \S\ref{ssec:ctos-rc}, we relate Landauer's principle and the structure of reversible computing to the CTOs discussed in \S\ref{par:ctos}. By examining the difference between conditional and unconditional Landauer reset using the structure of these CTOs, we find a general motivation for and rationale for the structure of RC directly from RTQT. Quite spectacularly, we see that by treating elements of the physical computer system as the ``catalyst'' in the CTO sense, we can directly represent repeated cycles of computation and Landauer reset. In particular, we see that using the structure of CTOs outlined in \S\ref{par:ctos}, we can cycle through the operation of computational systems as long as we wish, with minimal buildup of QMI. Finally, in \ref{ssec:rc-op-rep}, we begin the foundational work of representing RC operations in terms of open quantum systems from a first-principles level, using the properties of GKSL dynamics with multiple asymptotic states. In particular, we note specific properties of computational and noncomputational operations, and briefly discuss implications in terms of their quantum geometric signatures.

Let us now review these results in a bit more detail.

\subsection{The Fundamental Theorem of the Thermodynamics of Computation}
\label{ssec:fun-thm}

First, starting from the basic conception of (classical, digital) computational states presented in \S\ref{sssec:comp-states}, we can easily derive what we call \emph{The Fundamental Theorem of the Thermodynamics of Computation}. This theorem formalizes the necessary relationship between so-called ``information entropy'' (that is, entropy of the computational state) and physical entropy.

To start, let $\phi$ be a variable representing the (complete, micro-) physical state of the computer system $\mathfrak{S}$, specified by a choice of one of the protocomputational basis vectors $\vec{b}\in\boldsymbol{\mathcal{B}}$. Assume that, at a given point in time $\tau$, the probability mass over the different possible physical states $\phi$ is distributed according to a probability distribution $p(\phi)$, as given in the usual way using the Born rule, or equivalently, by the diagonal elements of the system's instantaneous density matrix $\rho(\tau)$ in the $\boldsymbol{\mathcal{B}}$ basis.

We can then derive an implied probability distribution $P(c)$ over the computational states $c_j\in\boldsymbol{C}$, by simply summing $p$ over the various physical states $\phi=\phi_i\in\boldsymbol{B}_j$, where $\boldsymbol{B}_j$ denotes the specific basis set $\boldsymbol{B}_c\subset\boldsymbol{\mathcal{B}}$ corresponding to computational state $c=c_j$:
\begin{equation}
    \label{eq-28}
    P(c_j)=\sum_{\phi_i\in\boldsymbol{B}_j} p(\phi_i).
\end{equation}
It is then trivial to show that the system's total (von Neumann/Shannon) entropy $S(\Phi)$ (where $\Phi$ is a random variable ranging over values $\phi$) can always be partitioned as:
\begin{equation}
    \label{eq-29}
    S(\Phi) = H(C)+S(\Phi | C),
\end{equation}
where $H(C)$ (with $C$ a random variable ranging over values $c$) refers to the computational entropy or ``information entropy,'' meaning the entropy of the computational state variable $C$ according to the above-derived probability distribution $P(c)$, and meanwhile $S(\Phi|C)$ refers to the conditional entropy of the physical state variable $\Phi$ if the value of the computational state variable $C$ is given. If we then define \emph{non-computational entropy} as $S_\mathrm{nc} = S(\Phi|C)$, then we can just say, ``total entropy equals computational entropy plus non-computational entropy.'' Note that this is true always---no matter how the protocomputational basis $\boldsymbol{\mathcal{B}}$ is defined and partitioned into basis sets corresponding to computational states. See Figure~\ref{fig:fun-thm}. (And for more details, see \cite{Fra18}.)

\begin{figure}[h] 
\centerline{\includegraphics[width=6 cm]{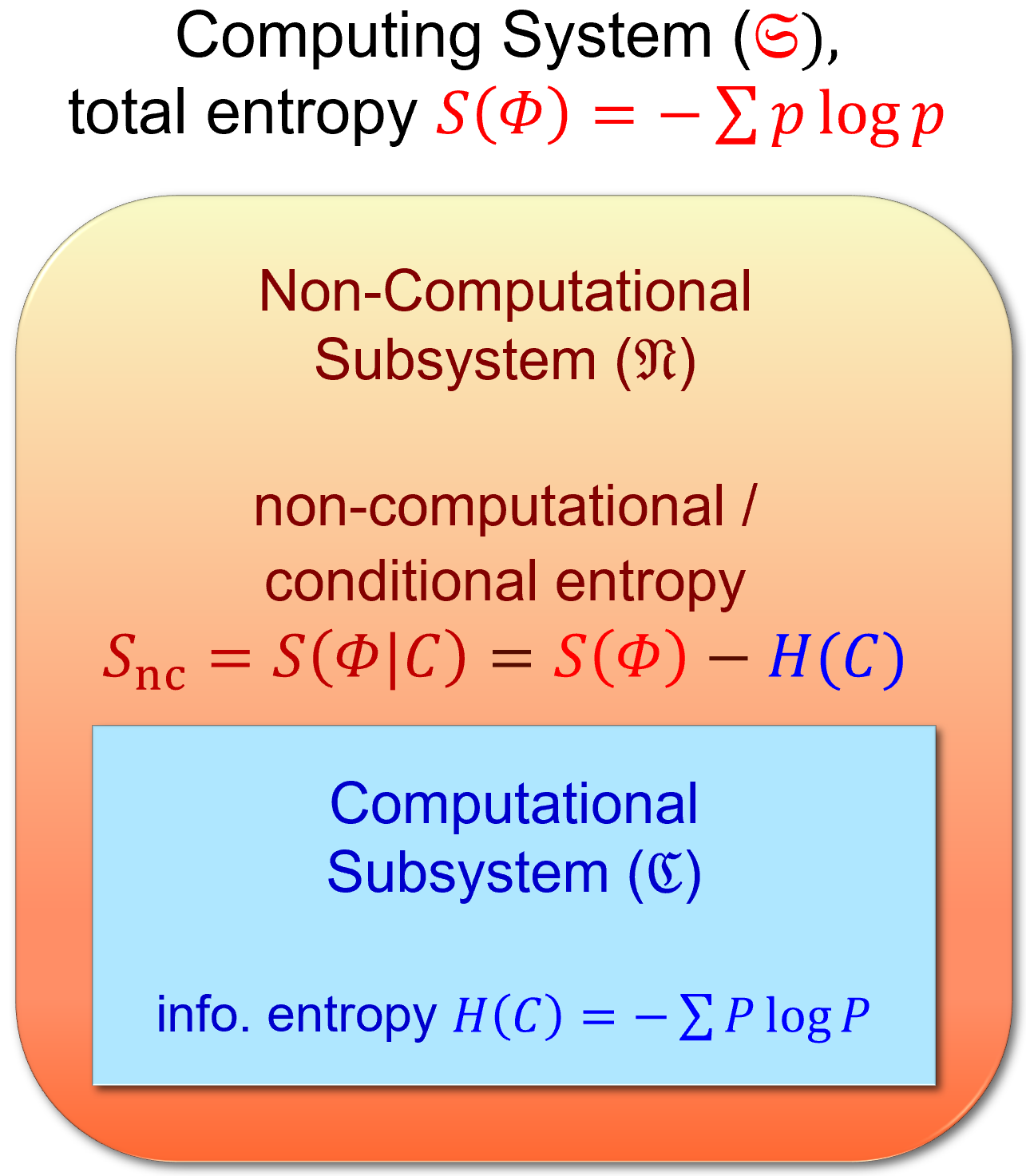}}
\caption{Fundamental Theorem of the Thermodynamics of Computing, illustrated using the picture of Fig.~\ref{fig:c-vs-n}. No matter how we choose the protocomputational basis $\boldsymbol{\mathcal{B}}$ of the computing system $\mathfrak{S}$ and partition it into subsets $\boldsymbol{B}_c$ for distinct computational states $c\in\boldsymbol{C}$, we can always express the total physical entropy $S(\Phi)$ of the system as the sum of the information entropy $H(C)$ of the computational state (state of the computational subsystem $\mathfrak{C}$), and the non-computational entropy $S_\mathrm{nc}(\Phi)$ of $\mathfrak{S}$, which is equal to the conditional entropy $S(\Phi|C)$ of the physical state when the computational state is given.\label{fig:fun-thm}}
\end{figure}   

This fact, together with the Second Law of Thermodynamics (\emph{i.e.}, $\partial S/\partial t \geq 0$ globally\footnote{It's worth noting that the entropy production rate can be negative in local systems due to non-Markovian dynamics \cite{SE19} or systems with delayed response \cite{BND21}.}), implies that one can't ever reduce computational entropy (for example, by merging two computational states, like in Figure~\ref{fig:transitions}(b)) without also increasing non-computational entropy by (at least) a corresponding amount. Of course, this works both ways---meaning, if you \emph{increase} computational entropy (\emph{e.g.}, by \emph{splitting} a computational state, in a stochastic computational operation, like in Figure~\ref{fig:transitions}(c)), you can thereby also \emph{reduce} non-computational entropy accordingly. This is done in practice, for example, in paramagnetic cooling \cite{DWK34,KWG60,PG99}, if we think of the (relatively stable) randomized magnetic domains that form during the cooling process as constituting ``computational'' bits.

A widespread perception\footnote{As an example, \cite{Sagawa14} discusses Landauer's Principle and logical reversibility in a context that \textit{only} considers independent systems, rather than subsystems of correlated systems, and concludes that logical reversibility is not required for thermodynamic reversibility, an observation which was already made explicitly in earlier work such as \cite{Ben03}. Although formally correct, such analyses completely miss the key point that, when obliviously erasing parts of \textit{correlated} systems, such as deterministically computed bits, there is necessarily a loss of mutual information that does indeed result in a required thermodynamic irreversibility. Thus, these analyses have been misinterpreted by some (\textit{e.g.} \cite{Wol19a,Wol19b}) as evidence by the fundamental rationale for reversible computing is incorrect, but this in not in fact the case, since, when the correlated case which is relevant to computing is considered, the connection between logical and thermodynamic reversibility is recovered. We discuss this in more detail in \S\ref{ssec:lan-prin} and \S\ref{ssec:wolpert}.\label{foot:bad-Sagawa}} is that the theorem corresponding to the above observations constitutes a form of Landauer's Principle, but we argue that, although it is indeed a very important basic theorem of the thermodynamics of computation, calling it ``Landauer's Principle'' creates confusion, because it entirely misses what we argue is the central, most important point of Landauer's Principle proper, which has to do more specifically with the loss of computed, \emph{correlated} information, such as typically exists within a computer. This viewpoint is discussed at great length in \cite{Fra18}, and more concisely in the following subsection.

\subsection{Landauer's Principle Proper}
\label{ssec:lan-prin}

Since we wish to argue that Landauer's Principle is, most centrally, a theorem about the consequences of information loss in \emph{computation}, specifically, it behooves us to say a little bit more about what we mean by that. 

A quite general picture of computation involves the concept of function evaluation, \emph{e.g.}, computing $y=f(x)$, given $x$. In fact, historically, the very first formal model of universal computation, due to Alonzo Church, defined general computations in terms of recursive function evaluation \cite{Chu36}. And it is of course well-known today that arbitrarily complex computations can be composed out of simple function evaluations (\emph{e.g.}, Boolean logic operations).

Let us then consider, as an example, two subsystems $\mathfrak{X},\mathfrak{Y}$ of a computational system $\mathfrak{C}$, that exist for purposes of holding the input value $x$ and output value $y$ of some function $f(\cdot)$ to be evaluated. Let us assume that subsystems $\mathfrak{X}$ and $\mathfrak{Y}$ have separable corresponding computational state spaces $\boldsymbol{C}_\mathfrak{X}, \boldsymbol{C}_\mathfrak{Y}$, which is to say, the computational states of subsystems $\mathfrak{X}$ and $\mathfrak{Y}$ are independently measurable. There is then a joint computational state space $\boldsymbol{C}_{\mathfrak{XY}} = \boldsymbol{C}_\mathfrak{X}\times\boldsymbol{C}_\mathfrak{Y}$. Suppose initially we have some distribution $P(C_\mathfrak{X})$ over the initial state of $\mathfrak{X}$. Suppose, then, a deterministic computational $O_{\mathfrak{XY}}$ is performed on the joint system $\mathfrak{XY}$ which leaves $C_\mathfrak{X}$ unchanged, but results in $c_\mathfrak{Y}=f(c_\mathfrak{X})$, which is to say, the computational state $c_\mathfrak{Y}$ of $\mathfrak{Y}$ becomes a state representing the value $y=f(x)$, where $x$ is the value represented by the state $c_\mathfrak{X}$ of $\mathfrak{X}$. Note that this operation $O_{\mathfrak{XY}}$ could also be reversible, \emph{e.g.}\ if $C_\mathfrak{Y}$ contained a known value initially (\emph{e.g.}, is ``cleared memory'').

It then follows from the above setup that:

\begin{enumerate}
    \item First, the \emph{reduced} computational entropy of $\mathfrak{Y}$, written $H(C_\mathfrak{Y})$, after performing $O_{\mathfrak{XY}}$, is \emph{entirely} accounted for by the mutual information between $\mathfrak{Y}$ and $\mathfrak{X}$; that is, $H(C_\mathfrak{Y}) = I(C_\mathfrak{X};C_\mathfrak{Y})$. In other words, $\mathfrak{Y}$ contains an exactly \emph{zero} amount of \emph{independent} entropy, relative to $\mathfrak{X}$, since $H(C_\mathfrak{Y}|C_\mathfrak{X}) = 0$. (\emph{I.e.}, $\mathfrak{Y}$ is completely determined by $\mathfrak{X}$.) This just follows from the fact that, as is typically the case in traditional digital computation, function evaluation is a deterministic operation.
    
    \item Second, now suppose that, next, an \emph{irreversible} computational operation $O_{\mathrm{erase}}$ is performed \emph{locally} on $\mathfrak{Y}$ \emph{in complete isolation from} $\mathfrak{X}$, that is, without any \emph{influence} from the state of $\mathfrak{X}$, or even any applied \emph{knowledge} about the state of $\mathfrak{X}$ (beyond our prior distribution $P(C_\mathfrak{X})$), and suppose, further, that the overall output-state distribution $P(C_\mathfrak{Y})$ resulting from $O_\mathrm{erase}$ has \emph{zero} entropy. This resultant distribution $P(C_\mathfrak{Y})$ is found by computing a weighted sum of $O_\mathrm{erase}(c_\mathfrak{Y})$ over the set of all input computational states of $\mathfrak{Y}$ with probability $P(c_\mathfrak{Y})>0$. For this distribution to have zero entropy implies that all such states of $\mathfrak{Y}$ map to the same value, $c_\mathfrak{Y} = c_0$, which is why $O_{\mathrm{erase}}$ can be considered an ``erasure'' operation. 
    
    \item If we now simply assume that \emph{the non-computational entropy in $\mathfrak{S}$ will shortly be thermalized}---which is to say, the entropy ejected from $\mathfrak{Y}$ is not being preserved in a stable or predictable form elsewhere in the physical state of the system---then it follows that the correlation previously embodied by the mutual information $I(C_\mathfrak{X};C_\mathfrak{Y})$ has now been lost, and therefore, \emph{the total entropy of the model universe $\mathfrak{U}$ is immediately \emph{(\emph{i.e.}, after a thermalization timescale)} increased by (at minimum) the prior value of} the reduced (marginal) subsystem entropy $H(C_\mathfrak{Y})$, just before the erasure. An example is illustrated in Figure~\ref{fig:land}.
\end{enumerate}

We argue that the resulting theorem constitutes what is the \emph{most appropriate} statement of Landauer's principle: Namely, that to erase \emph{any} deterministically-computed information in an isolated computational subsystem \textit{obliviously}, that is, without regards to its correlations with other information that may exist, and if this is followed by allowing the reduced subsystem entropy that was thereby ejected to subsequently thermalize, results in turning that subsystem's previous mutual information (which was \emph{not} independent entropy) into \emph{true} entropy (real, new uncertainty), and thereby results in a permanent increase in the total entropy of the universe by the corresponding amount. 

Finally, perhaps the easiest way to see that a loss of mutual information generally implies entropy increase is simply to note that, for any random variables $X,Y$,
\begin{equation}
    H(X,Y) = H(X) + H(Y) - I(X;Y),
\end{equation}
and thus, if we hold the marginal subsystem entropies $H(X),H(Y)$ steady while their mutual information $I(X;Y)$ is reduced, the total entropy $H(X,Y)$ of the joint distribution of the $XY$ system must increase accordingly. More broadly, one can say that mutual information is a part of the total \textit{known information} $K=C-H$ in a system, where $C$ is the \textit{information capacity} or maximum entropy of the system, and $H$ is its present entropy (which we can consider \textit{unknown information}). Thus, a loss of mutual information is really just a special case of the more general process of the transformation of physical information from a `known' to an `unknown' status, which \textit{is} entropy increase.

Additional details of the argument in this subsection can be found in the extended postprint version \href{https://arxiv.org/abs/1901.10327}{arXiv:1901.10327} of \cite{Fra18}.

\begin{figure}[t]
\includegraphics[width=13.5 cm]{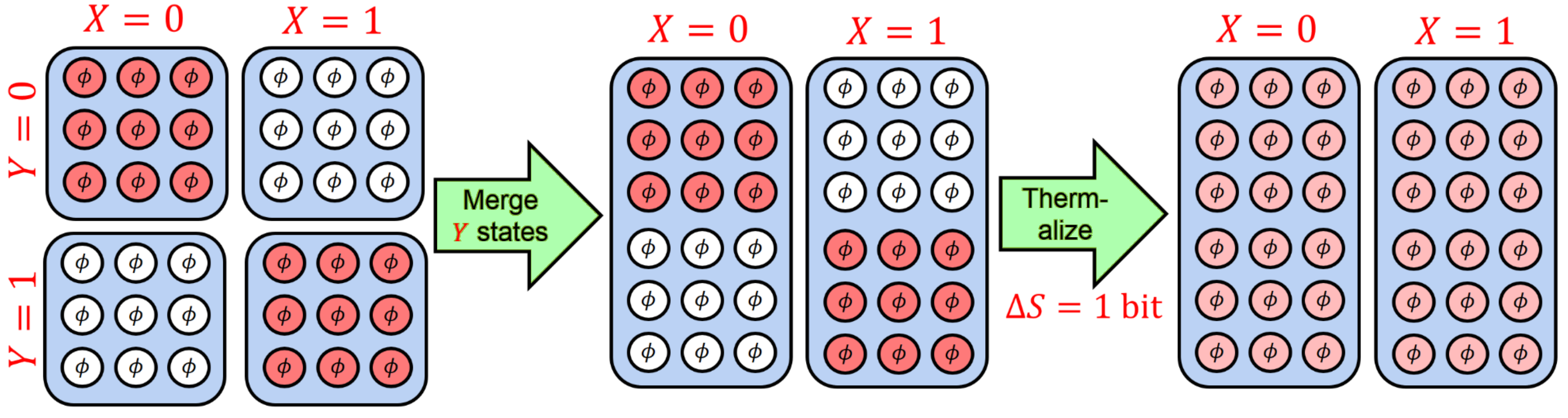}
\caption{\textbf{Landauer's Principle as entropy increase from thermalization of mutual information.} Red shading denotes probability density. (\textbf{Left}) Two perfectly-correlated computational bit-systems $\mathfrak{X}$ and $\mathfrak{Y}$; their states could have been prepared by computing $\mathfrak{Y}$'s value $y$ deterministically from $x$, \emph{e.g.}, using $y:=x$. (\textbf{Middle}) When the variable $Y$ (for the computational state of $\mathfrak{Y}$) is obliviously erased, this amounts to merging the two computational states in each column; we can say that now $Y=0$ (say) in each merged state. Note that now, there briefly exists a correlation between $X$ and the non-computational part of the physical state. (\textbf{Right}) Very quickly (over a thermalization timescale), we lose track of the probabilities of the different physical states making up each computational state, thus losing this correlation information. This is where the absolute increase of total entropy from Landauer’s Principle necessarily occurs. We cannot, of course, then undo this entropy increase by simply reversing the first step (un-merging the $Y$ states), because the correlation information between $X$ and $Y$ has already been irrevocably lost by this point.\label{fig:land}}
\end{figure}   

\subsection{Fundamental Theorems of Reversible Computing}
\label{ssec:rc-thms}

In this section, we review what we call \emph{The Fundamental Theorems of Reversible Computing (RC)} \cite{Fra17a}, which show that, in order for a deterministic computing system to avoid entropy increases due to Landauer's Principle (when understood properly, as in \S\ref{ssec:lan-prin} above), logically reversible computational operations must be utilized. The Fundamental Theorem of RC comes in two versions: The traditional version, which shows that traditional \emph{unconditionally} logically reversible operations are required in order to avoid entropy increase from Landauer's Principle in \emph{all} possible input circumstances, and the (less often recognized) \emph{generalized} version, which shows that a broader class of \emph{conditionally} reversible operations suffice, for use in systems that are properly designed to ensure that the preconditions for reversibility of those operations are met. 

\subsubsection{Fundamental Theorem of Traditional Reversible Computing}
\label{sssec:trad-rc}

Before we review the traditional RC theorem, we first present a simple definition, based on the discussion of \S\S\ref{ssec:fun-thm}--\ref{ssec:lan-prin} above, which will be helpful for stating it. (The following is presented in time-independent terms, but can easily be made time-dependent.)

\mypara{Entropy-ejecting operations.}\label{par:ent-ej-ops}

A computational operation $O$ on a computational state set $\mathbf{C}$ is called \emph{(potentially) entropy-ejecting} if and only if where exists some possible prior distribution $P(C)\in\mathcal{P}(\mathbf{C})$ such that, when the operation $O$ is applied within that context, the increase $\Delta S_\mathrm{nc}$ in the non-computational entropy required by the Fundamental Theorem of the Thermodynamics of Computing (\S\ref{ssec:fun-thm}) is greater than zero. If an operation $O$ is not (even potentially) entropy-ejecting, we call it \emph{non-entropy-ejecting}.

Note that if an operation is entropy-ejecting, and it is performed \emph{in isolation} (by which we implicitly also mean \textit{obliviously}, without external knowledge of the state being applied) on a subsystem that contains mutual information with other subsystems (and if we assume that any non-computational entropy will not be preserved in a predictable form, but will be thermalized), then this entropy ejection will furthermore result in a \emph{global entropy increase}, by a straightforward generalization of Landauer's Principle (in its proper form, stated above in \S\ref{ssec:lan-prin}).

\paragraph{\textbf{Fundamental Theorem of Traditional Reversible Computing.} If a deterministic computational operation $O$ is non-entropy-ejecting (by the above definition), then it follows that $O$ must be unconditionally logically reversible.} The proof of this statement is trivial, but can be found in \href{https://arxiv.org/abs/1806.10183}{arXiv:1806.10183}, the extended postprint version of \cite{Fra17a}.

Note, also, that an immediate corollary of this theorem is that, if we wish to perform computational operations \emph{in isolation} (\emph{i.e.}, obliviously) on subsystems that contain any mutual information with other systems (such as subsystems whose state was computed deterministically from those other systems), then we can \emph{only} avoid a global entropy increase from Landauer's Principle in the \emph{general} case (\emph{i.e.}, for \textit{any} distribution $P(C)$ over initial computational states, and when the non-computational state doesn't preserve information in a predictable form) if those operations are \emph{unconditionally logically reversible}.

\subsubsection{Fundamental Theorem of Generalized Reversible Computing}
\label{sssec:gen-rc}

The traditional theorem, above, is in essence about how we can avoid Landauer losses in the \emph{worst case}---that is, when we assume that we have no control over what the initial computational state $c_{\mathrm{I}}\in\mathbf{C}$ may be, and thus, any statistical mixture of initial states is possible. But, in a real computer, the initial state prior to a given computational operation may be (and usually is) a resultant state from a previous operation. Thus, it is frequently the case that we can, \emph{by design} in a computer, \emph{restrict} the set of possible initial states to a proper subset $\boldsymbol{A} \subset \mathbf{C}$ of \emph{allowed} states. This then makes it possible to design computing mechanisms that avoid Landauer losses by transforming just the \emph{subset} $\boldsymbol{A}$ of allowed states reversibly. This is, in fact, how typical real engineered technologies for reversible computing (including those described in \S\ref{ssec:tech} above) work---since it turns out, in general, to be much easier, in practice, to design mechanisms that only transform restricted subsets of computational states reversibly, rather than the full set of all potentially describable states. But to show why doing this is sufficient, we need a more general version of the fundamental theorem of RC, one that properly models the case where the set of initial states is restricted.

To do this, we also need to extend the concept of an entropy-ejecting operation from \S\ref{sssec:trad-rc} as follows:

\mypara{Entropy-ejecting computations.}\label{par:ent-ej-comps}

For purposes of the below theorem, let a (statistically contextualized) \emph{computation} $\mathcal{C}=\mathcal{C}(O,P)$ refer to the concept of performing a computational operation $O$ over its computational state space $\mathbf{C}$, \emph{given} a particular \emph{initial} probability distribution $P=P(C)$, where $C$ is a random variable ranging over computational states $c\in\mathbf{C}$. (The \emph{quantum contextualized computation} concept of \S\ref{sssec:corresp} is just a straightforward generalization of this concept to a quantum context.) We say that a (deterministic) computation $\mathcal{C}$ is \emph{(specifically) entropy-ejecting} if and only if the increase $\Delta S_\mathrm{nc}$ in the non-computational entropy required by the Fundamental Theorem of the Thermodynamics of Computation (\emph{i.e.}, due to a reduction in computational entropy $H(C)$) is greater than zero. If the computation $\mathcal{C}$ is not specifically entropy-ejecting, we call it non-entropy-ejecting.

As before, this then allows us to immediately state the corresponding theorem:

\paragraph{\textbf{Fundamental Theorem of Generalized Reversible Computing.} A deterministic computation $\mathcal{C}(O,P)$ is non-entropy-ejecting if and only if \emph{at least one} of its preconditions for reversibility is satisfied with probability 1 under the initial probability distribution $P$.} As with the traditional theorem, the proof of this is easy, but may be found in \href{https://arxiv.org/abs/1806.10183}{arXiv:1806.10183}.

Like with the traditional theorem, the generalized theorem has an immediate corollary, which is that if we wish to perform the computation $\mathcal{C}(O,P)$ in isolation (obliviously) on a subsystem bearing mutual information with other systems (such as a subsystem whose computational state was deterministically computed from those outside systems), then we can only avoid a global entropy increase from Landauer's Principle for that specific computation (assuming, as usual, that the non-computational state doesn't preserve information in a predictable form) if the operation $O$ is conditionally reversible, under (at least) the precondition that $c\in\boldsymbol{A}$, where $\boldsymbol{A}=\set{c_i\in\mathbf{C} | P_i > 0}$.

The significance of the two RC theorems together is that, in order to avoid the otherwise-necessary entropy increase resulting from Landauer’s Principle when performing isolated computational operations on subsystems in the context of larger deterministic computations, one must confine oneself to the above two cases (unconditionally reversible operations, and/or conditionally reversible operations that have a satisfied condition for reversibility). 

The significance of the \emph{generalized} reversible computing theorem, as opposed to the traditional one, is to observe that it is a sufficient logical-level requirement, to avoid requiring an entropy increase from Landauer’s Principle, if simply those initial states having \emph{nonzero probability} in the given statistical operating context $P(C)$ are mapped one-to-one onto final states.

Of course, in any event, even when these conditions for reversibility are satisfied, to avoid entropy increase in reality \emph{also} requires that the physical mechanisms \emph{implementing} the given computation must be designed to approach thermodynamic reversibility in practice---but, the import of the RC theorems is to say that, when the conditions of either theorem are satisfied, Landauer's Principle, at least, does not preclude doing this.

Additional discussion of these two theorems can be found in \cite{Fra17a}, with detailed proofs available in the associated postprint \href{https://arxiv.org/abs/1806.10183}{arXiv:1806.10183}. We should note that, although the particular proofs of these theorems presented in that earlier work did not yet explicitly utilize the quantum generalization of the concept of a statistical operating context that we presented in \S\ref{sssec:corresp} above, all of the constructions in \S\ref{sssec:corresp} were specifically designed to guarantee that the exact same proofs will go through essentially unmodified in the quantum version of the theory (given our assumptions about rapid decoherence of final states). Thus, the above theorems remain completely valid within the quantum framework of the present paper.

\subsection{Representations of Reversible Computing by Catalytic Thermal Operations}
\label{ssec:ctos-rc}

The results above follow from a \textit{static} analysis just based on the overall starting and ending states of a given computational process; however, obtaining more detailed results (\textit{e.g.}, about minimum energy dissipation as a function of speed) will require more detailed attention to the dynamics of computational transitions. This then requires engaging more detailed theoretical methods, such as the resource-theoretic tools we reviewed in~\ref{sssec:rtqt}.  In this section, we discuss how to think about reversible computational operations in terms of those more detailed methods.

\subsubsection{Reconsidering the Notion of a Catalyst}
\label{sssec:reversing-catalyst}
The nonequilibrium Landauer results in \S\S\ref{par:neal}--\ref{par:john} emphasize how essential it is to have a map as close to unital as possible in order to minimize the energy cost of the operation (\emph{i.e.}, the importance of a map that minimizes the entropy difference between $\rho_{\mathrm{in},\mathfrak{S}}$ and $\Lambda_{t}\left[\rho_{\mathrm{in},\mathfrak{S}}\right]$). Ideally, we'd like to connect these to the theory of thermal operations, so that we can begin to identify the thermal processes that are relevant for reversible computing. As discussed in \S\ref{par:neal}, the bound is zero when the Landauer reset protocol applied to the subsystem bearing computational degrees of freedom is conditioned on the state of that subsystem \emph{and} when the von Neumann entropies of the subsystem state before and after reset are equal. Although achieving this in practice is a nontrivial engineering challenge, there is no \emph{fundamental} (quantum mechnical) barrier to this constraint, so we can consider this case specifically (\emph{i.e.}, that $S\left(\rho_{i}\right) = S\left(\rho_{j}\right) = S\left(\rho_{\mathrm{r}}\right)$ where $\rho_{i}$ and $\rho_{j}$ can be any possible computational state).

In order to select the right TO, we again return to the Landauer reset protocols. These were described earlier as the process of resetting a state bearing some computational degrees of information, either with a conditioned or unconditioned potential. Without loss of generality, we can consider the system carrying these degrees of freedom as part of a larger system. Then, we can label the larger system $\mathfrak{S}$ and the subsystem carrying these degrees of freedom $\mathfrak{Q}$, as shown in Figure \ref{fig:subsyst-PQ-def}. In terms of $\mathfrak{Q}$, the Landauer reset protocol is the process of transforming the state $\rho_{\ell,\mathfrak{Q}}$ to the reset state $\rho_{\mathrm{r},\mathfrak{Q}}$. Until now, we've been content to consider the Hamiltonians acting on all of the systems and subsystems as background Hamiltonians. Ordinarily, this wouldn't pose any issues, and indeed helps us keep the properties of the system as general as possible.

However, as we saw in \S\ref{par:neal}, whether or not $\widehat{V}_{r,\mathfrak{Q}}$ can be conditioned on $\rho_{\ell,\mathfrak{Q}}$ plays a vital role in calculating the free energy bound on the reset process, and in particular on the information contribution to the bound. Thus, by contrast to the other contributions to the overall Hamiltonian, by having $\widehat{V}_{\mathfrak{Q}}$ as an ambient background potential we in fact lose a vital piece of context of the overall process. This context is relevant when trying to identify the correct TO we want to use to examine the process. Thus, we need to think of $\widehat{V}_{\mathfrak{Q}}$ as not an ambient potential that acts on $\mathfrak{Q}$, but rather a potential that acts on $\mathfrak{Q}$ from a \emph{different} subsystem $\mathfrak{P}$ of the overall system $\mathfrak{S}$. This echoes the discussion at the end of \S\ref{par:neal}, when we saw that the $\Delta I_{\mathrm{er}}$ term in the unconditional Landauer bound \eqref{eq:uncond-Land-bound} contained within it mutual information between the state of the system implementing the Landauer reset potential $\widehat{V}$ and the state of the subsystem $\mathfrak{S}^{\prime}$ bearing computational degrees of freedom.

We can immediately recognize $\mathfrak{S}^{\prime}$ in \S\ref{par:neal} as identical to $\mathfrak{Q}$ here, and $\mathfrak{P}$ as the system implementing $\widehat{V}$. If we require that the local state of $\mathfrak{P}$ (\emph{i.e.}, the state of $\mathfrak{P}$ when we trace out every other subsystem) be the same to within some value $\epsilon \in \mathbb{R}^{+}$ under the trace distance, then we can treat $\mathfrak{P}$ as a catalyst which is necessary to induce the local state transformation $\rho_{\ell,\mathfrak{Q}} \mapsto \rho_{\mathrm{r},\mathfrak{Q}}$. In this framework, the natural TO to examine this transformation is the general CTO given by \eqref{eq:Markus-CTO}. Indeed, we can recognize the $\Delta I_{\mathrm{er}}$ term in the unconditional Landauer bound \eqref{eq:uncond-Land-bound} as precisely the same as the QMI \eqref{eq:Markus-QMI-diff} built up between the transforming subsystem and the catalyst.

Thus, in terms of TOs, the correlation-preserving generalized CTOs discussed in \linebreak\S\ref{par:ctos} are precisely the conditioned Landauer reset protocols discussed in \S\ref{par:neal}. By contrast, the requirement in \eqref{eq:normal-CTO} that the final state of the catalyst remain uncorrelated with the final state of the transforming subsystem is identical to applying a single unconditioned potential for the Landauer reset. As was the case when comparing the most general CTO \eqref{eq:Markus-CTO} and the more traditional CTO \eqref{eq:normal-CTO}, the correlated information $\Delta I_{\mathrm{er}} = I\left(\mathfrak{Q}:\mathfrak{P}\right)$ is ejected from the overall system $\mathfrak{S} = \mathfrak{QP}$ after the unconditional protocol. Alternately, and equivalently, the unconditional Landauer reset protocol (resp., the more traditional CTO) can be realized by performing the conditional Landauer reset protocol (resp., the general CTO) and then ejecting the correlated information afterwards.

\begin{figure}[h] 
    \centerline{\includegraphics[width=10 cm]{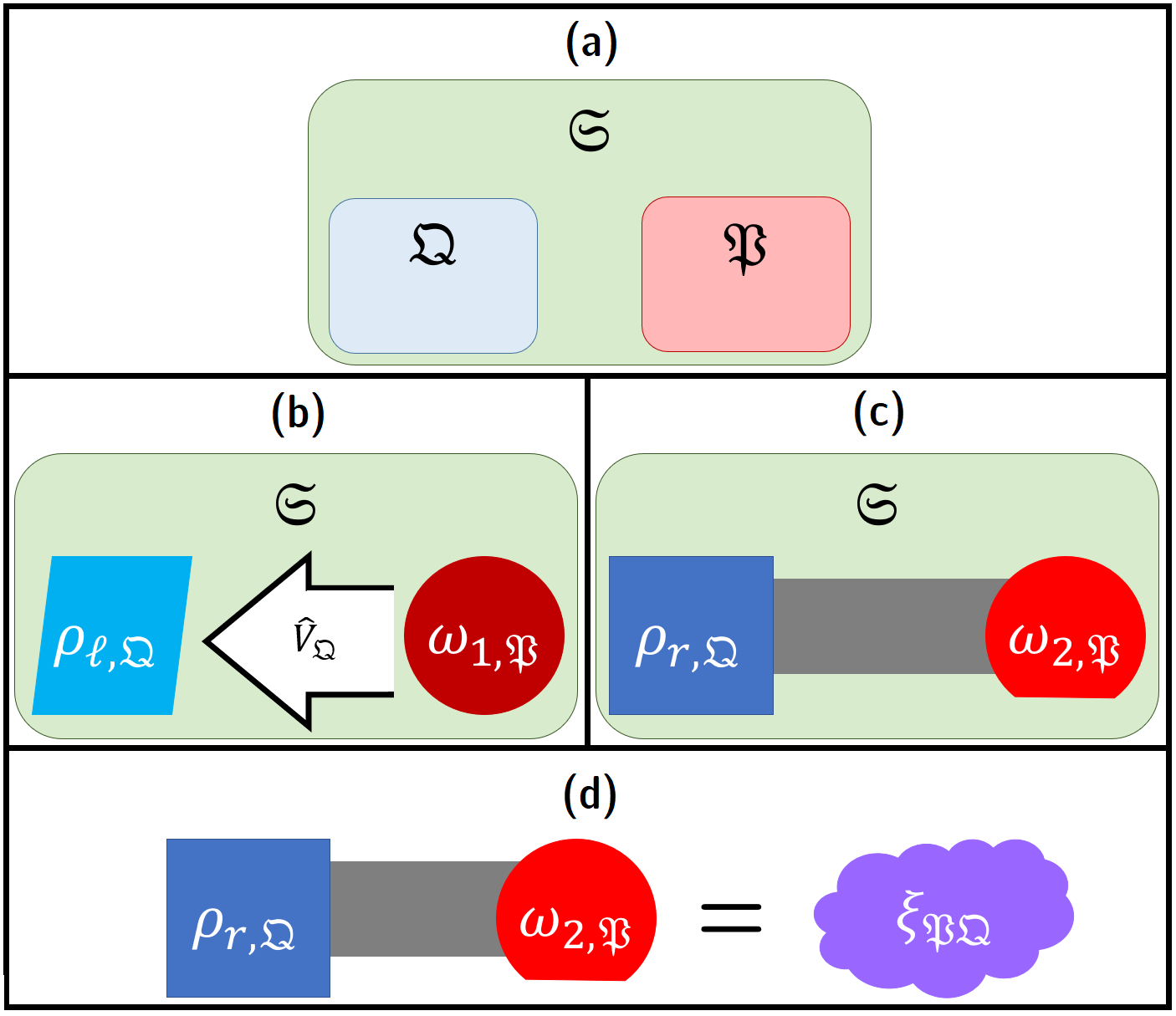}}
    \caption{\textbf{(a)} The embedding of a system $\mathfrak{Q}$ carrying computational degrees of freedom inside a larger system $\mathfrak{S}$, which also contains a subsystem $\mathfrak{P}$ used to induce state transformations on $\mathfrak{Q}$. \textbf{(b)} An example of the process discussed in \S\ref{par:neal}, in terms of this embedding. $\mathfrak{Q}$ is in one of the $N$ states $\left\{\rho_{\ell,\mathfrak{Q}}\right\}$. A potential $\widehat{V}_{\mathfrak{Q}}$ is applied to $\left\{\rho_{\ell,\mathfrak{Q}}\right\}$ to transform it into the reset state $\rho_{r,\mathfrak{Q}}$. Crucially, since $\widehat{V}_{\mathfrak{Q}}$ is part of the system, it must be contained in $\mathfrak{S}$ (\emph{i.e.}, it must \emph{not} be part of the environment); meanwhile, it also must be \emph{outside} of all of $\mathfrak{Q}$. We can without loss of generality consider the subsystem that applies $\widehat{V}_{\mathfrak{Q}}$ onto $\mathfrak{Q}$ to be part of (or all of) $\mathfrak{P}$. \textbf{(c)} The application of this potential transforms the local state of $\mathfrak{Q}$ as $\rho_{\ell,\mathfrak{Q}} \mapsto \rho_{r,\mathfrak{Q}}$. Assuming that the dynamics can be described by a CTO of the form \eqref{eq:Markus-CTO}, the transformation must also induce some state transformation $\omega_{1,\mathfrak{P}}\mapsto\omega_{2,\mathfrak{P}}$, to preserve unitarity of the overall dynamics on $\mathfrak{PQ}$. (The sole exception to this is the identity transformation.) The application of $\widehat{V}_{\mathfrak{Q}}$ by $\mathfrak{P}$ upon $\mathfrak{Q}$ also gives rise to a correleation between $\mathfrak{Q}$ and $\mathfrak{P}$, given by the QMI \eqref{eq:Markus-QMI-diff}. \textbf{(d)} The correlated states $\rho_{r,\mathfrak{Q}}$ and $\omega_{2,\mathfrak{P}}$ correspond to a single state $\xi_{\mathfrak{PQ}}$ over $\mathfrak{PQ}$ as a while, with $\mathrm{Tr}_{\mathfrak{P}}\left[\xi_{\mathfrak{PQ}}\right] = \rho_{r,\mathfrak{Q}}$ and $\mathrm{Tr}_{\mathfrak{Q}}\left[\xi_{\mathfrak{PQ}}\right] = \omega_{2,\mathfrak{P}}$. \label{fig:subsyst-PQ-def}}
\end{figure}

Thus far, we've used the general CTO to describe the process of a subsystem being transformed with the help of a catalyst, and to obtain the quantum thermodynamic restrictions on this process. However, the word ``catalyst'' is simply a convention based on our modelling; mathematically, the catalyst is simply another subsystem which (within an $\epsilon \in \mathbb{R}^{+}$) returns locally to the same state before and after the overall state transformation on $\mathfrak{S}$. Nothing \emph{a priori} tells us that the subsystem $\mathfrak{K}$ in \eqref{eq:Markus-CTO} must be a catalyst; as far as the mathematics is concerned, this is simply a statement about a specific type of transformation on the system $\mathfrak{S} = \mathfrak{TK}$, where $\mathfrak{T}$ and $\mathfrak{K}$ are subsystems and $\mathfrak{K}$ has the additional requirement \eqref{eq:Markus-CTO-trace-norm} that the state before and after the transformation is locally the same up to an infinitesimal $\epsilon$. This offers an interesting extension of the discussion above regarding the reset of $\mathfrak{Q}$ due to a subsystem $\mathfrak{P}$: what happens if we allow $\mathfrak{P}$ to change its state before and after the transition, but we require that the subsystem $\mathfrak{Q}$ bearing computational degrees of freedom must return to the same state? \emph{A priori} this seems like it would make no sense: how can we meaningfully talk about computation if the subsystem with computational degrees of freedom is left unchanged?

To answer this, we'll consider the decomposition of the transition $\rho_{\mathfrak{Q}} \otimes \omega_{\mathfrak{P}} \mapsto \xi_{\mathfrak{QP}}$ into two distinct (non-catalytic) TOs over the subsystems $\mathfrak{Q}$ and $\mathfrak{P}$. Here, we enforce $\mathrm{Tr}_{\mathfrak{PE}}\left[\xi_{\mathfrak{QP}}\right] = \rho_{\mathfrak{Q}}$ and $\lVert\mathrm{Tr}_{\mathfrak{QE}}\left[\xi_{\mathfrak{QP}}\right] - \Xi\left(\omega_{\mathfrak{P}}\right)\rVert_{1} < \epsilon$ for some infinitesimal $\epsilon \in \mathbb{R}^{+}$. In other words, the \emph{overall} transition $\rho_{\mathfrak{Q}} \otimes \omega_{\mathfrak{P}} \mapsto \xi_{\mathfrak{QP}}$ is an operation where now the state over $\mathfrak{Q}$ starts and ends in the same state. In this decomposition, one of these transformations locally takes the state of $\mathfrak{Q}$ away from $\sigma_{\mathfrak{Q}}$, and the other returns the state of $\mathfrak{Q}$ to $\sigma_{\mathfrak{Q}}$. Overall, then, we consider the pair of transformations:
\begin{equation}
    \label{eq:RC-CTO-composition-general}
    \rho_{\mathfrak{Q}}\otimes\omega_{\mathrm{in},\mathfrak{P}} \mapsto \chi_{\mathfrak{QP}} \mapsto \xi_{\mathfrak{QP}}.
\end{equation}
In addition to the properties for $\mathrm{Tr}_{\mathfrak{PE}}\left[\xi_{\mathfrak{QP}}\right]$ and $\mathrm{Tr}_{\mathfrak{QE}}\left[\xi_{\mathfrak{QP}}\right]$ stated above, we have $\mathrm{Tr}_{\mathfrak{PE}}\left[\chi_{\mathfrak{QP}}\right] = \gamma_{\mathfrak{Q}}$ for some $\gamma_{\mathfrak{Q}} \ne \rho_{\mathfrak{Q}}$.

The composed transition $\rho_{\mathfrak{Q}} \otimes \omega_{\mathfrak{P}} \mapsto \xi_{\mathfrak{QP}}$ is the CTO $\rho_{\mathfrak{Q}} \otimes \omega_{\mathfrak{P}} \mapsto \xi_{\mathfrak{QP}}$ discussed above, where now it is the state $\rho_{\mathfrak{Q}}$ of $\mathfrak{Q}$ that starts and ends in the same state under the composition. Thus, the composed transition must correspond to the constraints in \S\ref{par:ctos}. This has a direct interpretation in terms of computing if we think of $\sigma_{\mathfrak{Q}}$ as a reset state. The reset state is conventionally the state we use as our starting point to perform a computation on $\mathfrak{Q}$, which is why it's chosen as the reset state in the first place. Then, the process \eqref{eq:RC-CTO-composition-general} corresponds to starting with the standard reset state $\rho_{\mathfrak{Q}}$, using a different subsystem $\mathfrak{P}$ to manipulate the state of $\mathfrak{Q}$ (\emph{i.e.}, perform a computation on $\mathfrak{Q}$), and then using $\mathfrak{P}$ once more to perform a conditional Landauer reset on $\mathfrak{Q}$ to return it to the reset state. Remarkably, since these compose to yield an overall CTO of the form \eqref{eq:Markus-CTO}, this process can be achieved with an infinitesimal build-up of mutual information:  $I\left(\mathfrak{Q}:\mathfrak{P}\right) < \delta_{1} + \delta_{2}$ for $\delta_{1}, \delta_{2} \in \mathbb{R}^{+}$.

By contrast, we can also consider unconditional Landauer reset as a type of CTO, of the form \eqref{eq:normal-CTO}. In this case, we can consider the transformation over $\mathfrak{QP}$ given by: 
\begin{equation}
    \label{eq:not-RC-CTO-composition-general}
    \rho_{\mathfrak{Q}}\otimes\omega_{\mathrm{in},\mathfrak{P}} \mapsto \chi_{\mathfrak{QP}} \mapsto \sigma_{\mathfrak{Q}}\otimes\rho_{\mathrm{f},\mathfrak{P}}.
\end{equation}
Here, $\mathfrak{Q}$ and $\mathfrak{P}$ are uncorrelated at the end of the transformation. As before, we can think of $\sigma_{\mathfrak{Q}}$ as the reset state. We can think of this pair of transformations corresponding to a decomposition of a CTO of the form \eqref{eq:normal-CTO}, and, \emph{simultaneously}, the unconditional Landauer protocol in \eqref{eq:uncond-Land-TO}. As discussed in \S\ref{par:ctos} and \S\ref{par:markus-vs-2ndlaws}, the fact that the final states of $\mathfrak{Q}$ and $\mathfrak{P}$ are uncorrelated corresponds to an intrinsic asymmetry between the work of formation $F_{\infty}$ and the extractable work $F_{0}$; this loss of energy represents the energy lost as a result of the expulsion of the QMI between $\mathfrak{Q}$ and $\mathfrak{P}$ into the environment.

\subsubsection{Transformations on Computational States and Catalytic Thermal Operations}
\label{sssec:CTOs-catalysts-computers}
For the sake of clarity, we explicitly restate this way of viewing computational operations, now referencing the computational subspace $\mathfrak{C}$ and a control subspace $\mathfrak{K}$, combining the framework of \S\ref{par:ctos} and the current section with the notation and viewpoint of \S\ref{par:neal}. As discussed in \S\ref{par:neal}, for a subspace $\mathfrak{C}$ bearing computational degrees of freedom, we define the reset state $\rho_{\mathrm{r},\mathfrak{C}}$ as a standard, known reference state upon which operations can be performed. These operations correspond to known computations, which transform the state of the system from the reset state to one of $N$ known final computational states $\rho_{\ell,\mathfrak{C}}$. Then, in the Landauer protocol, we reset the final state back to the reset state, either conditioning the reset protocol on the final state or not conditioning the reset protocol on the final state. These correspond to the conditional and unconditional Landauer protocols, respectively, with the lower bound on the energy cost of each given in \S\S\ref{par:neal}--\ref{par:john}. Specifically, we saw that the conditional Landauer protocol was bounded below by zero, whereas the unconditional Landauer protocol was bounded below by the amount of correlated information between the computational state and the subsystem applying the reset potential onto the computational state.

We can understand the process of repeatedly resetting the computational subspace $\mathfrak{C}$ to $\rho_{\mathrm{r},\mathfrak{C}}$, evolving the state of $\mathfrak{C}$ to a final computational state $\rho_{\ell,\mathfrak{C}}$, again resetting, again evolving to a final computational state $\rho_{m,\mathfrak{C}}$ (which may be the same or different), and continuing in this fashion as a sequence of CTOs as discussed earlier in \S\ref{sssec:reversing-catalyst}. In particular, since the local state of $\mathfrak{C}$ is constantly reset, evolved, and then reset and evolved again and again all under the influence of a secondary operator $\mathfrak{K}$; we can consider $\mathfrak{C}$ as the catalyst subsystem in the sense of the discussion earlier in this section. A series of computational operations, performed by a subsystem $\mathfrak{K}$ of $\mathfrak{S}$ that's distinct from $\mathfrak{C}$ but contained entirely within $\mathfrak{S}$, transforms the state of $\mathfrak{C}$ from $\rho_{\mathrm{r},\mathfrak{C}}$ to some $\rho_{\ell,\mathfrak{C}}$, which corresponds to our computational operation. $\mathfrak{K}$ then performs the Landauer reset $\rho_{\ell,\mathfrak{C}} \mapsto \rho_{\mathrm{r},\mathfrak{C}}$ of $\mathfrak{C}$, following either the conditional Landauer protocol \eqref{eq:cond-Land-U} or the unconditional Landauer protocol \eqref{eq:uncond-Land-U}, with the corresponding energy costs given by \eqref{eq:cond-Land-bound} and \eqref{eq:uncond-Land-bound} respectively.

\emph{A priori}, it may not be clear why we insist that $\mathfrak{K}$ must be the same subsystem that performs the local transformation $\rho_{\mathrm{r},\mathfrak{C}}\mapsto\rho_{\ell,\mathfrak{C}}$ and the Landauer reset $\rho_{\mathrm{r},\mathfrak{C}}\mapsto\rho_{\ell,\mathfrak{C}}$. Indeed, these transitions may well be performed by different machines within $\mathfrak{S}$. However, without loss of generality, we can lump the set of \emph{all} machines which perform operations on $\mathfrak{C}$ collectively into a single subspace $\mathfrak{K}$, and subsequently examine the set of operations that $\mathfrak{K}$ in its entirety performs on $\mathfrak{C}$. In particular, from the decomposition of $\mathfrak{S}$ and $\mathfrak{E}$ given in \S\ref{sssec:open-q}, we note that all of these individual machines, as well as their combined collection $\mathfrak{K}$, \emph{must} correspond to a subspace of $\mathfrak{S}$.

\begin{figure}[h] 
    \centerline{\includegraphics[width=10 cm]{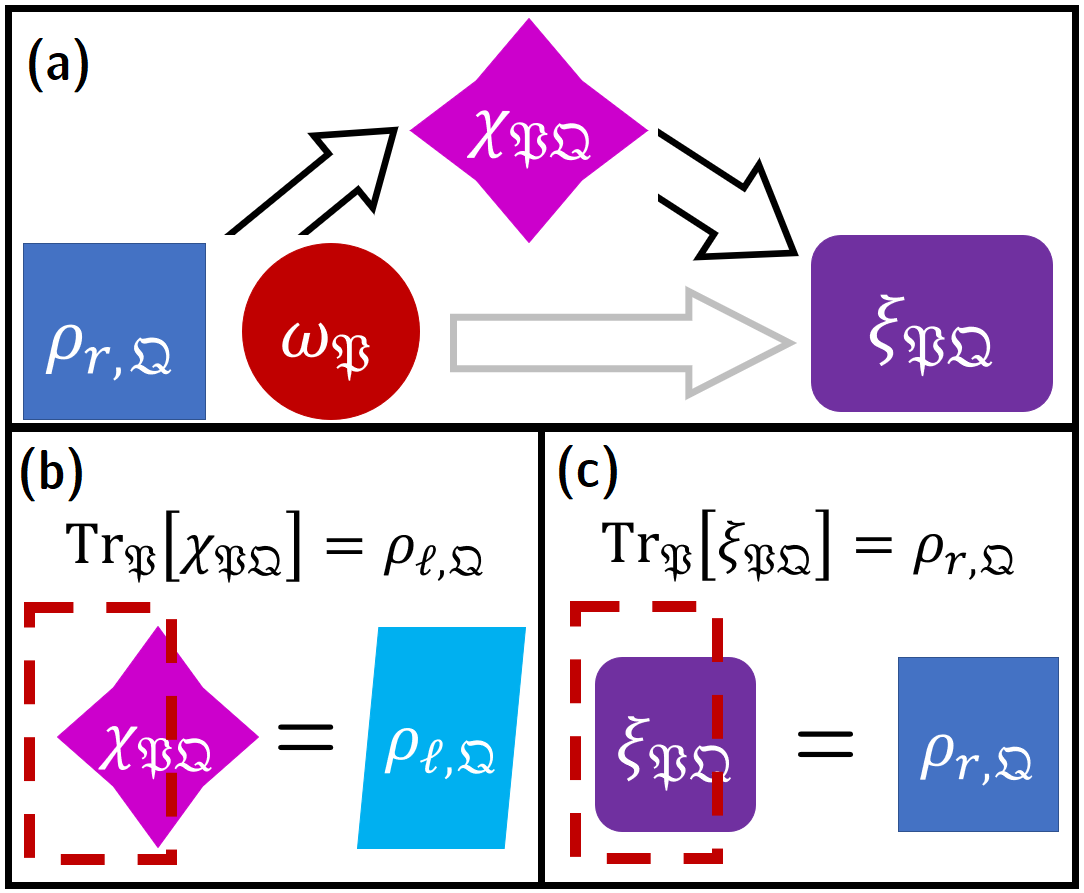}}
    \caption{A representation of a computational operation cycle in terms of a CTO, with conditional Landauer reset. \textbf{(a)} The computational system $\mathfrak{C}$ starts in a standard reset state $\rho_{\mathrm{r},\mathfrak{C}}$. An auxiliary system $\mathfrak{K}$ starting in the state $\sigma_{\mathfrak{K}}$ performs a series of operations on $\rho_{\mathrm{r},\mathfrak{C}}$, causing $\mathfrak{CK}$ to jointly evolve into a state $\chi_{\mathfrak{CK}}$, with $\mathrm{Tr}_{\mathfrak{K}}\left[\chi_{\mathfrak{CK}}\right] = \rho_{\ell,\mathfrak{C}}$ as one of the possible known final computational states. (The series of operations that $\mathfrak{K}$ performs on $\mathfrak{C}$ corresponds to computation.) Then, $\mathfrak{K}$ performs a Landauer reset of $\mathfrak{C}$, returning it locally to the standard reset state. In the case of the conditional Landauer reset, pictured here, the reset protocol corresponds to $\mathfrak{CK}$ jointly evolving into the final state $\xi_{\mathfrak{CK}}$, with $\mathrm{Tr}_{\mathfrak{K}}\left[\xi_{\mathfrak{CK}}\right] = \rho_{\mathrm{r},\mathfrak{C}}$. The composition of operations $\rho_{\mathrm{r},\mathfrak{C}}\otimes\sigma_{\mathfrak{K}}\mapsto\chi_{\mathfrak{CK}}\mapsto\xi_{\mathfrak{CK}}$ is a composition of the form \eqref{eq:RC-CTO-composition-general}, and thus corresponds to a CTO of the form \eqref{eq:Markus-CTO} (shown as the gray arrow). \textbf{(b)} Locally in $\mathfrak{Q}$, the state of $\chi_{\mathfrak{PQ}}$ is given by one of the $N$ final computational states $\left\{\rho_{\ell,\mathfrak{Q}}\right\}_{\ell\:=\:1}^{N}$; \emph{i.e.}, we have $\mathrm{Tr}_{\mathfrak{P}}\left[\chi_{\mathfrak{PQ}}\right] = \rho_{\ell,\mathfrak{Q}}$. \textbf{(c)} Locally in $\mathfrak{Q}$, the state of $\xi_{\mathfrak{PQ}}$ is once again given by the reset state $\rho_{r,\mathfrak{Q}}$; \emph{i.e.}, we have $\mathrm{Tr}_{\mathfrak{P}}\left[\xi_{\mathfrak{PQ}}\right] = \rho_{r,\mathfrak{Q}}$. \label{fig:cto-rc}}
\end{figure}

We can use the techniques of CTOs to examine these transformations, in particular using the argument in \S\ref{sssec:reversing-catalyst} to examine the conditional Landauer bound. We start with $\mathfrak{C}$ in a reset state, have $\mathfrak{K}$ perform some operations to transform it into a final computational state, and then reset $\mathfrak{C}$ to the reset state once again. As in \S\ref{sssec:reversing-catalyst}, this chain of operations permits us to think of $\mathfrak{C}$ as the ``catalyst'' in a CTO, despite $\mathfrak{C}$ being the actual computational system of interest. Then, the means by which we transform from $\rho_{\mathrm{r},\mathfrak{C}}$ to $\rho_{\ell,\mathfrak{C}}$ and then back to $\rho_{\mathrm{r},\mathfrak{C}}$ tells us whether we have a CTO of the form \eqref{eq:Markus-CTO} or of the form \eqref{eq:normal-CTO}; equivalently, the means by which this pair of transformations takes place tells us whether we have a conditional Landauer reset protocol \eqref{eq:cond-Land-TO} or an unconditional Landauer reset protocol \eqref{eq:uncond-Land-TO}.

In the case of the unconditional Landauer reset protocol, we have the pair of transformations $\rho_{\mathrm{r},\mathfrak{C}}\otimes\sigma_{\mathrm{in},\mathfrak{K}}\mapsto\chi_{\mathfrak{CK}}\mapsto\rho_{\mathrm{r},\mathfrak{C}}\otimes\sigma_{f,\mathfrak{K}}$ with $\mathrm{Tr}_{\mathfrak{K}}\left[\chi_{\mathfrak{CK}}\right] = \rho_{\ell,\mathfrak{C}}$. This is of the exact same form as the transformation \eqref{eq:not-RC-CTO-composition-general}. As such, the same conclusion applies: in this case, this transformation corresponds to the CTO described in \eqref{eq:normal-CTO}. The final correlation between $\mathfrak{C}$ and $\mathfrak{K}$ is ejected into the environment, yielding the irreversible energy difference $F_{\infty} - F_{0}$. Conversely, in the case of the conditional Landauer reset protocol, we have the pair of transformations $\rho_{\mathrm{r},\mathfrak{C}}\otimes\sigma_{\mathrm{in},\mathfrak{K}}\mapsto\chi_{\mathfrak{CK}}\mapsto\xi_{\mathfrak{CK}}$ with $\mathrm{Tr}_{\mathfrak{K}}\left[\chi_{\mathfrak{CK}}\right] = \rho_{\ell,\mathfrak{C}}$ and $\mathrm{Tr}_{\mathfrak{K}}\left[\xi_{\mathfrak{CK}}\right] = \rho_{\mathrm{r},\mathfrak{C}}$. Here, we permit the QMI \eqref{eq:Markus-QMI-diff} between $\mathfrak{C}$ and $\mathfrak{K}$ to build up in both transformations. As before, the QMI in each transformation can be made as small as possible, but cannot in general be zero. The representation of the conditional Landauer reset protocol as a CTO in which the computational subsystem $\mathfrak{C}$ is thought of as the ``catalyst'' (inasmuch as it returns to the starting state) after two successive operations is given in Figure \ref{fig:cto-rc}.

\subsection{Subspace Representations of Computational and Noncomputational Operations}
\label{ssec:rc-op-rep}
In \S\ref{par:gksl-mas}, we discussed some of the basic properties of open quantum systems with multiple asymptotic states evolving under the GKSL approximation. A key aspect of the asymptotic subspace $\mathsf{As}\left(\mathcal{H}_{\mathfrak{S}}\right)$ is that the evolution after GKSL relaxation supports the \emph{full} dynamics available to closed quantum systems. In other words, $\mathsf{As}\left(\mathcal{H}_{\mathfrak{S}}\right)$ supports any dynamics that can be expressed by a Hilbert space of states evolving under a Hamiltonian; in this case, the Hamiltonian $\widehat{H}_{\infty}$ governs the dynamics of $\mathsf{As}\left(\mathcal{H}_{\mathfrak{S}}\right)$ after the GKSL relaxation. This provides an extremely powerful framework to represent reversible computing operations: if we can represent RC operations for closed system dynamics, we can automatically get a representation for RC operations in GKSL dynamics. As we saw in \S\ref{par:c-vs-nc}, the most general framework for representing a computational subsystem is with the DFS sum $\mathcal{H}_{\mathfrak{S}} = \bigoplus_{i}\mathcal{H}_{\mathsf{DFS},i}$, where each DFS $\mathcal{H}_{\mathsf{DFS},i}$ represents a computational basis element $c_{i}$.

When examining the dynamics on $\mathcal{H}_{\mathfrak{S}}$, this immediately gives us a way of distinguishing computational and noncomputational operations. In particular, we note that since each DFS corresponds to a specific computational basis element, a computational operation must \emph{transfer} states from one subspace to another. Conversely, a noncomputational operation \emph{cannot} transfer states from one subspace to another; therefore, it must only be able at most to rearrange states within each subspace. A direct consequence of this is that noncomputational operations must commute with the DFS structure of the computational system, whereas computational operations in general have no such restrictions and permit coherences between different computational subspaces during the immediate period of computational operation. A visual representation of these two different kinds of operations is provided in Figure \ref{fig:c-vs-nc-vN}.

\begin{figure}[h]
    \centerline{(a) \includegraphics[width=5 cm]{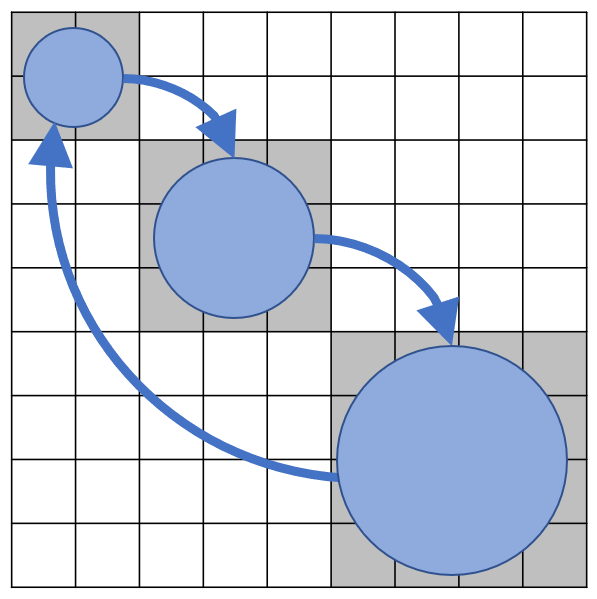}
    \hspace{1 cm}(b) \includegraphics[width=5 cm]{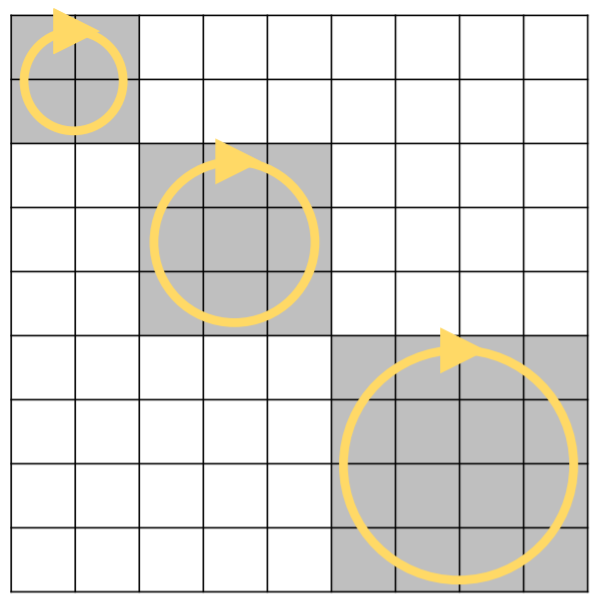}}
    \caption{The representation of computational and noncomputational operations in a Hilbert space, such as $\mathsf{As}\left(\mathcal{H}_{\mathfrak{S}}\right)$. (\textbf{a}) Computational operations (blue arrows), which transfer between different computational states (blue circles). As discussed in \S\ref{par:basis-sets}, each computational state corresponds to a distinct, orthogonal DFS (gray), with the overall Hilbert space corresponding to the direct sum of these. (\textbf{b}) Noncomputational operations (yellow arrows), which cannot transfer between different computational states and thus can only transfer protocomputational states within the same DFS. Note that a direct consequence of this is that noncomputational operations must commute with the DFS structure.
    \label{fig:c-vs-nc-vN}}
\end{figure}

Our interest here is in classical computing, rather than quantum computing. As a result, we expect no quantum coherences to develop between different computational subspaces; quantum coherences may only exist within a DFS representing a single computational state. However, computational operations of the type shown in Figure \ref{fig:c-vs-nc-vN}\textbf{(a)} \emph{necessarily} induce coherences; these are, indeed, characteristic of the transfer between one subspace and another. As a result, immediately after a computational operation, $\mathsf{As}\left(\mathcal{H}_{\mathfrak{S}}\right)$ will appear as a single space. For our computer to remain a \emph{classical} computer, then, we require that this space dephase into the DFS sum we expect \emph{faster} than the computer's ability to resolve distinct times; this is showing in Figure \ref{fig:dephasing}. This provides us with a dephasing timescale, which can in fact offer a way to distinguish between classical, quantum, and ``approximately classical'' computing representations as we tune the dephasing timescale. (Here, by ``approximately classical'', we mean those operations where the dephasing timescale is on the same order as the computer resolution timescale.) The relative strength of the dephasing and computer resolution timescales, and the consequences of tuning this relative strength, will be of significant interest for future work.

\begin{figure}[h] 
    \centerline{\includegraphics[width=11 cm]{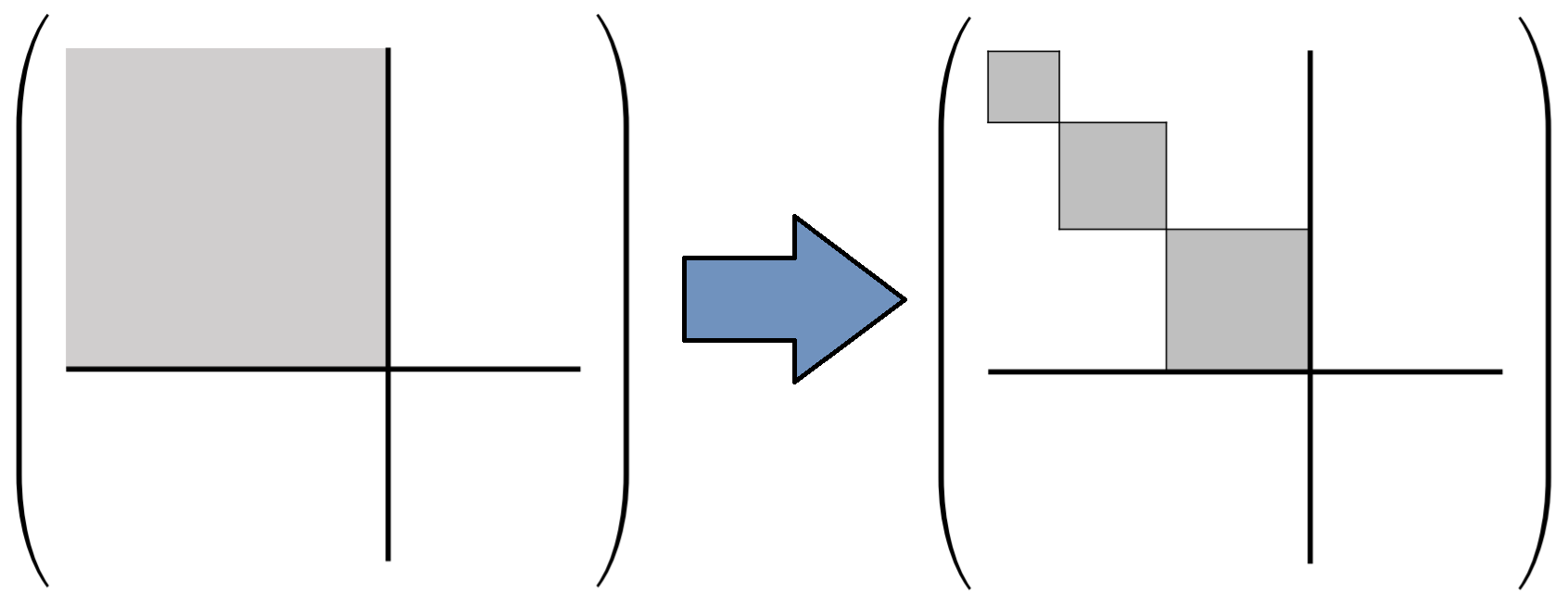}}
    \caption{The dephasing of $\mathsf{As}\left(\mathcal{H}_{\mathfrak{S}}\right)$ from a single subspace to a DFS direct sum. For classical reversible operations, this dephasing must occur faster than the computer's ability to resolve distinct times (\emph{i.e.}, on a timescale faster than the computer can see). Classical and quantum computing operations can be distinguished by the relation between this timescale and the computer resolution timescale. \label{fig:dephasing}}
\end{figure}

It's an almost trivial statement to note that any matrix can be represented as the sum of two other matrices. However, the distinction between computational and noncomputational operations discussed above, and the point that a computational operation necessarily mixes the different computational DFSs to temporarily make a single large space, provides us with an interesting decomposition of such operations. In particular, we note that a computational operation which appears as an operator mixing all of the states in $\mathsf{As}\left(\mathcal{H}_{\mathfrak{S}}\right)$ can be decomposed into a `noncomputational part', which commutes with the DFS structure, and a `pure computational part', which contains all of the information involving transfer of states between DFS blocks (and thus, \emph{all} of the information regarding the actual computational content of the operation); this decomposition is shown in Figure \ref{fig:c-nc-decomp}. A central property of GKSL dynamics with multiple asymptotic states, derived in \cite{ABFJ16} and discussed there and in \cite{Albert18}, is the nontrivial quantum geometric tensor over $\mathsf{As}\left(\mathcal{H}_{\mathfrak{S}}\right)$ that emerges, and the dependence that the dynamics exhibits on the QGT. Notably, different shapes of the asymptotic subspace exhibit different geometric signatures; thus, computational operations which mix different DFS states will have a different quantum geometric signature than noncomputational operations. In light of the decomposition of computational operations into noncomputational and pure computational operations, this also means that each of these parts will exhibit distinct quantum geometric signatures which can identify the computational and noncomputational part. As an added benefit, we expect this decomposition to provide additional intuition for the distinction mentioned above between Landauer's Principle and the Fundamental Theorem. The discussion of this will be provided in much greater detail in the forthcoming work examining the quantum geometric properties of RC operations in general.

\end{paracol}
\begin{figure}[h]
\widefigure
    \centerline{\includegraphics[width=18 cm]{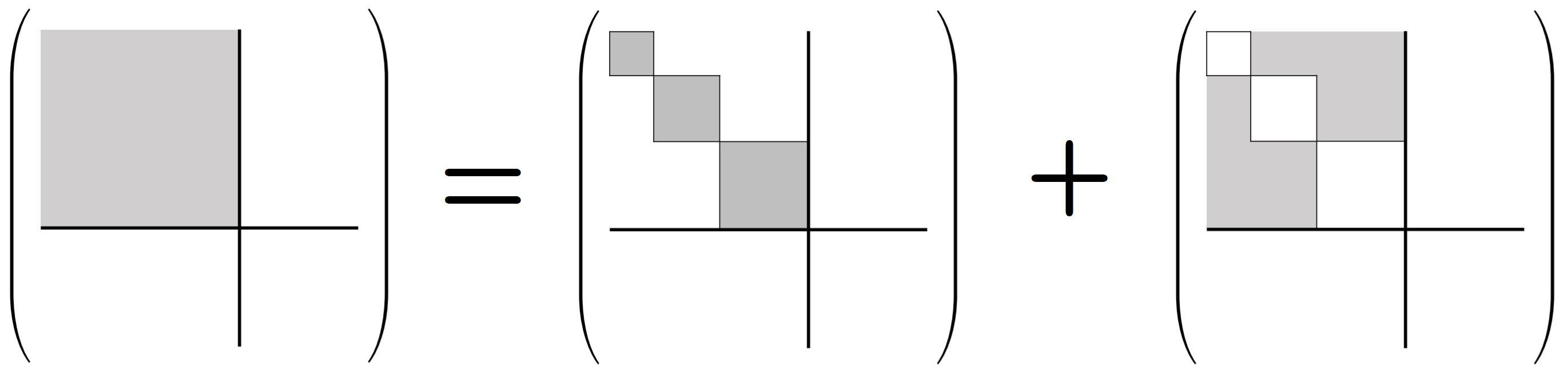}}
    \caption{Decomposition of a computational operation (left) into a noncomputational part (middle), which commutes with the DFS structure that distinguishes different computational states, and a pure computational part (right), which contains \emph{all} of the information regarding the transfer of states between different DFS blocks, and thus all of the information regarding the computational part of the operation. Notably, because the quantum geometric tensor of $\mathsf{As}\left(\mathcal{H}_{\mathfrak{S}}\right)$ as a single space has a different shape than that of $\mathsf{As}\left(\mathcal{H}_{\mathfrak{S}}\right)$ as a DFS sum, the QGT of each of these will naturally be distinct as well. As such, the noncomputational part of a computational operation will have a different, measurable quantum geometric signature to the pure computational part of a computational operation. \label{fig:c-nc-decomp}}
\end{figure}
\begin{paracol}{2}
\switchcolumn

Beyond simply distinguishing between computational and noncomputational operations, the discussion in \S\ref{sssec:comp-oper} highlights how essential it is to distinguish between the different types of computational operations---deterministic irreversible, deterministic reversible, stochastic irreversible, and stochastic reversible. As further discussed in \S\ref{sssec:comp-trans}, these types of operations are themselves comprised of a set of primitive computational state transitions; namely the bijections, merges, and splits as discussed in Figure \ref{fig:transitions}. As such, in order to understand the representations of the different types of reversible (and, indeed, irreversible) classical operations, we must first find a representation of these different types of operations. Although the nature of bijections and merges is somewhat self-evident, the case of splits must be handled with slightly more care, since they can be generally expected to result in temporary coherences (which will, typically, quickly decohere). As with the decomposition of computational operations, these will in all likelihood exhibit distinct quantum geometric signatures in lieu of the quantum geometric signatures of different asymptotic subspaces discussed in \cite{ABFJ16, Albert18}. Along with the previously-mentioned issues regarding quantum geometric signatures of RC operations, the discussion of this will be addressed in the forthcoming work centering on the quantum geometric properties of reversible computing operations more broadly.

\section{Discussion}
\label{sec:disc}

\subsection{Essential Consistency of the Classic RC Formulation with NEQT}
\label{ssec:consis}
An important high-level conclusion supported by this paper is that there is no inconsistency between the simple, classic formulation of Landauer's Principle and reversible computing that we reviewed in \S\S\ref{ssec:lan-prin}--\ref{ssec:rc-thms}, and a more detailed treatment based on NEQT.  Indeed, no such inconsistency is possible, since, as we showed, the classic formulation can be presented in a form that makes no equilibrium assumptions whatsoever. The only assumptions we made there were the fundamental unitarity of the underlying quantum evolution, which is also assumed by all of quantum thermodynamics, and the treatment of the environment as immediately thermalizing all ejected information, which is equivalent to the Markov assumption underlying GKSL dynamics as discussed in \S\ref{ssec:rc-op-rep}. Therefore, the more detailed NEQT formulations \textit{cannot} negate the basic results of \S\S\ref{ssec:lan-prin}--\ref{ssec:rc-thms}. Indeed, we showed how to draw explicit correspondences between a more detailed treatment of classical reversible computing based on generalized CTOs and multiple asymptotic states in GKSL, and the simpler model of \S\ref{ssec:found} on which the basic results of \S\S\ref{ssec:lan-prin}--\ref{ssec:rc-thms} can be based.  We next discuss a few specific aspects of this correspondence in more detail.

\subsection{CTO Representations of Reversible Computing and System Boundaries}
\label{ssec:cto-rc-sys}
In \S\ref{ssec:ctos-rc}, we discussed the representation of quantum mechanical models of reversible computing in terms of the general CTO \eqref{eq:Markus-CTO}. Specifically, we identified the Landauer reset potential(s) $\widehat{V}_{\ell r,\mathfrak{Q}}$ applied to a subsystem $\mathfrak{Q}$ as coming from the interaction between $\mathfrak{Q}$ and a distinct separate subsystem $\mathfrak{P}$, where $\mathfrak{Q}$ and $\mathfrak{P}$ are both subsystems of the overall system $\mathfrak{S}$ we're examining in our TO. This analysis reinforces the importance of properly drawing the system boundaries, as discussed in \cite{Anderson19}.

The specifics of where we choose to draw the boundaries between each subsystem, or between the overall system and the environment, plays a vital role in being able to properly identify the reset protocol of a system bearing computational degrees of freedom. Properly drawing the boundaries makes the distinction as to whether a given reset protocol is a \emph{Landauer} reset protocol \emph{per se} or not. In \cite{Anderson19}, we have the example of the reset of a system $\mathfrak{P}$ coupled to a copy / referent subsystem $\mathfrak{Q}$ which stores the same computational state as $\mathfrak{P}$ before reset. The set of transformations on $\mathfrak{S}$ in this case are given by $\left\{\rho_{\mathrm{in},\mathfrak{P}}\otimes\sigma_{\mathrm{in},\mathfrak{Q}}\right\}_{i\:=\:1}^{N}\mapsto\left\{\rho_{\mathrm{r},\mathfrak{P}}\otimes\sigma_{i,\mathfrak{Q}}\right\}_{i\:=\:1}^{N}.$ Comparing these to expressions \eqref{eq:cond-Land-TO}, \eqref{eq:cond-Land-final-global-state}, and \eqref{eq:uncond-Land-TO}, we see that identifying $\mathfrak{P}$ as the reset system is commensurate with the Landauer reset definitions, whereas identifying $\mathfrak{PQ}$ as the reset system doesn't count as Landauer reset. (Specifically, identifying $\mathfrak{P}$ as the reset system, this corresponds to conditional Landauer reset.)

In precisely the same way, identifying the system boundaries is essential to representing reset processes as CTOs. The CTO expression \eqref{eq:Markus-CTO} requires a specific shape for the starting and ending states (namely, that the overall system $\mathfrak{S}$ be subdivided into a subsystem $\mathfrak{P}$ which locally undergoes the state transformation $\rho_{\mathrm{in},\mathfrak{P}}\mapsto\rho_{\mathrm{f},\mathfrak{P}}$ and a subsystem $\mathfrak{Q}$ which locally returns to the same state). Thus, properly identifying the system boundaries plays a vital role in identifying the effect of the global universal CPTP map as a catalytic thermal operation, or some other kind of operation. This dependence on system boundaries is, indeed, a \emph{general} feature of resource theories \cite{CG19}. Thus, when applying the results of RTQT to analyzing Landauer reset protocols and formulating bounds on quantities of interest, it is vital to specify the subsystem boundaries properly: these boundaries affect whether or not we can properly classify a reset as a Landauer reset, whether or not we can properly classify a CPTP map as a CTO, and whether or not we can properly relate these two. (Incidentally, we saw a specific example of this earlier: in \S\ref{ssec:ctos-rc}, this dependence is what constrained the reset potential(s) to be implemented by a separate cataylst subsystem within $\mathfrak{S}$ as a whole.)

\subsection{Applicability of the Markov Approximation to Reversible Computing}
\label{ssec:markov-rc}

As mentioned in \S\ref{par:gksl-intro}, the Markov approximation (and, especially, the Markov approximation for systems with multiple asymptotic states) involves the separation of several different time scales, given by \eqref{eq:Markov-time-scales} and \eqref{eq:multiplte-As-Markov-time-scales} respectively. As previously mentioned, a direct consequence of this approximation is that fluctuations that take information from $\mathfrak{S}$ to $\mathfrak{E}$ during an intermediate time period and return it to $\mathfrak{S}$ happen much faster than our ability to resolve the dynamics; instead, any information that is ejected from the system to the environment cannot be returned to the system. We might be concerned that this might represent an ``artificial'' type of information loss in our model that comes from imposing a limitation on the types of processes we consider, which does not reflect computational systems in the real world. In fact, the opposite is true: this assumption matches perfectly with our understanding of the thermalization of ejected correlated information from $\mathfrak{S}$.

In \S\ref{ssec:lan-prin} and \cite{Fra18}, we saw that the entropy increase due to Landauer reset was a consequence of the thermalization of mutual information ejected into the environment. In particular, the thermalization of entropy ejected into the environment occurs at a time frame much faster than our ability to capture the dynamics of the system. Thus, in our model, the role of fluctuations between $\mathfrak{S}$ and $\mathfrak{E}$ originating from $\mathfrak{S}$ is suppressed: the effect of perturbative fluctuations looks identical to the effect of environment perturbations. This is exactly the kind of behavior we expect for a system $\mathfrak{S}$ bearing computational degrees of freedom in a larger open system evolution: the system has no practical way of tracking mutual information that is transferred to the environment at a given time and then is transferred from the environment back to the system at a later time. Any perturbations arising from information that originated from $\mathfrak{S}$ at an earlier time appears to $\mathfrak{S}$ as indistinguishable from perturbations due to $\mathfrak{E}$. This is also true for models of computation where the computational degrees of freedom reside in a subsystem $\mathfrak{P}$ being acted on by one or more orthogonal subsystems $\mathfrak{Q}_{i}$ of $\mathfrak{S}$: a crucial part of the model is that neither $\mathfrak{P}$ nor any of the $\mathfrak{Q}_{i}$s are able to track information after it has been ejected into $\mathfrak{E}$; and, indeed, this is a completely physically realistic framework. Thus, the Markov assumption, in addition to being a vital calculational tool to retrieve closed form expressions, also represents the real-world dynamics of the system.

\subsection{Relationship to the Stochastic Thermodynamics of Computation}
\label{ssec:wolpert}
The \emph{stochastic thermodynamics of computation} \cite{Wol19a, Wol19b} is a framework for examining the entropy cost of classical computational systems that has gained substantial prominence in recent years. As such, an important question is the relationship between this framework and the NEQT framework and results given in \S\S\ref{par:ctos}--\ref{par:john}, which we discuss here.

In the stochastic thermodynamics of information, the thermodynamic properties of classical computational systems are examined \cite{Wol19a} from the point of view of purely classical information theory, relying on the properties of the continuous probability distributions of classical random variables that arise therein. In this approach, classical computation is represented as a continuous-time Markov chain (CTMC), represented by a directed acyclic graph (DAG) which in turn represents a set of functions over a given Boolean string $s \in \left\{0,1\right\}^{n}$ of length $n$. Then, in the stochastic thermodynamics of information framework, the entropy cost of transitions is calculated by examining the difference in the classical relative entropies (\emph{i.e.}, the Kullback-Leibler divergences) of the CTMC distribution relative to an arbitrary distribution before and after a single set of state evolutions along the graph. This framework explicitly does not consider the perspective from quantum information theory,\footnote{In the words of \cite{Wol19a} directly, ``[this paper] will not consider the thermodynamics of quantum mechanical information processing, or indeed any aspect of quantum computation.''} in contrast with the representation of classical computation in terms of quantum channels (\emph{e.g.}, as in \cite{Lan61,Ben73,Ben82,BL85,Lan87,Ben88,NC00,NRS07,Wolf12,Attal14,Wilde17}). Instead, the thermodynamic properties of computation are derived solely from examining the entropy production and entropy flow rates of probability distributions which evolve under this CTMC representation.

Despite avoiding the quantum information representation for classical computation (and indeed for nonequilibrium dynamics), we would expect \emph{no} differences whatsoever in the conclusions we get from this framework compared to the framework focusing on quantum information theory. At its core, the technique of correlation engineering \cite{Fra18,Mueller18} relies on performing operations on correlated systems in such a way as to minimize the mutual information build-up and to make the Helmholtz free energy cost of the transition arbitrarily small, as discussed in \S\ref{sssec:rtqt}. Although \cite{Fra18,Mueller18} and the discussion in \S\ref{sssec:rtqt} have focused on quantities such as the quantum mutual information, the principles of correlation engineering are in fact completely oblivious to whether the correlations and mutual information quantities are classical or quantum in nature. Indeed, since the examination in these works of the effect of correlations on overall entropic cost do \emph{not} rely on whether or not the system state commutes with the thermal reference state or the Hamiltonian, we can freely substitute classical information quantities (such as the Kullback-Leibler divergence) and retrieve valid statements for correlation engineering of classical thermodynamic systems.\footnote{Indeed, this precise property is the insight underlying the $\beta$-ordering and thermomajorization curve technique in \cite{HO13}.}

For individual systems in isolation, this is precisely what we find: the stochastic thermodynamics of computation is completely able to reproduce the results regarding the distinctions between logical and thermodynamic reversibility found earlier in \cite{Ben73,Ben82,Ben03,Sagawa13}. Specifically, the now-famous result in those works that logically irreversible operations (such as erasure of a single, isolated bit or computational system) can be nevertheless thermodynamically reversible has also been reproduced in \cite{Wol19a,Wol19b}. This reflects our exact point about the proper interpretation of Landauer’s principle that we elaborated upon in \S\ref{ssec:lan-prin}, which is that erasing bits in isolation when they contain mutual information with other bits results in thermodynamic irreversibility, and an amount of entropy increase corresponding to the loss of mutual information.

Nevertheless, the subject of extending the framework of the stochastic thermodynamics of computation to \emph{correlated} systems appears to remain a matter of controversy. In particular, the example of the thermodynamic reversibility of erasure of a single bit \emph{in isolation} has been used to argue that a \emph{correlated} system cannot realize operations that are both logically and thermodynamically reversible to within a mutual information difference of $\delta\in\mathbb{R}^{+}$, as given in \eqref{eq:Markus-QMI-diff}. Here, it appears that the misunderstanding has persisted due to confusing the matter of isolated versus correlated systems: the example in \cite{Ben73,Ben82,Sagawa13} and reproduced in \cite{Wol19a,Wol19b} continues to focus on the case of an \emph{isolated} system, whereas \cite{Mueller18,Goold15,Anderson19,Fra18} highlight that correlation engineering relies entirely on reducing the correlation buildup between systems and applying operations in a way that does not increase the net entropy flow from the overall system.\footnote{Somewhat more confusingly, \S IX-B in \cite{Wol19a} appears to use this example to claim that correlated systems cannot realize such operations; however, Example 6 in \S V-B in \cite{Wol19a} appears to use a general argument to  verify that logically irreversible operations on \emph{correlated} systems cannot be made thermodynamically reversible when each irreversible operation is at the subsystem level. This is precisely the same principle used in \cite{Mueller18,Goold15,Anderson19,Fra18}. The source of the internal disagreement in \cite{Wol19a} remains unclear.}

It must be emphasized again, there is no fundamental disagreement between the results calculated in \cite{Mueller18,Goold15,Anderson19,Fra18}. Indeed, there \emph{cannot} be, since all of these paths start with precisely the same set of fundamental assumptions about the underlying dynamics of the system; \emph{i.e.}, that the dynamics are described by some CPTP map which we can express in terms of a larger subspace as per the Stinespring dilation theorem. Thus, although some of the \emph{conclusions} in \cite{Wol19a, Wol19b} seem to disagree with the consensus of \cite{Mueller18,Goold15,Anderson19,Fra18,Sagawa19}, this is only a matter of mistaken identity: \cite{Wol19a, Wol19b} has erroneously drawn conclusions about \emph{correlated} systems from a valid calculation about the thermodynamic reversibility of a logically irreversible process applied to an \emph{independent} system (see also footnote \ref{foot:bad-Sagawa}). Furthermore, although some of the conclusions of \cite{Wol19a, Wol19b} may be drawn from a mistaken conflation between the properties of independent and correlated systems, it is worth re-emphasizing that the underlying techniques in these works remain completely valid. Indeed, the stochastic thermodynamics of computation may nevertheless serve as a useful additional tool to examine correlation engineering of correlated systems.\footnote{This further reinforces how delicate the issue of system boundaries is, as discussed in \cite{Anderson19}.}

\subsection{Thermodynamically Reversible Transformations of Extended Systems}
\label{ssec:extend}

An appropriate caveat to Landauer's principle could be to mention that 
even a \emph{correlated} state in a computer could in principle be erased in a thermodynamically reversible way if, rather than erasing bits in isolation, we instead considered the full space of \emph{all possible} thermodynamic transformations.  But what, in detail, would a concrete protocol for accomplishing such a transformation look like?  One such protocol would be to simply \emph{unwind the correlations} by running, in reverse, a reversible computation that could have computed those correlations in the first place, thereby leaving us, potentially, with an array of independently random input bits, that could then each be thermodynamically reversibly erased in isolation from each other.  But, note that the existence of protocols such as this one doesn't refute the need for reversible computing, since this protocol itself uses reversible computing to begin with.  

Fundamentally, if we wish to be able to construct complex computational processes by composing them out of primitive transformations which operate \textit{locally} on part of the full computational state, then by definition, those primitives \emph{cannot} operate monolithically on the state of the entire extended system, and thus they \emph{must} be logically reversible if we wish to avoid entropy increase in the face of the kind of non-local correlations which naturally arise, as a matter of course, in typical large computations.  This is not to say that some radically different alternative basis for computation that was \emph{not} based on local computational primitives \emph{at all} would necessarily be impossible to develop, but until that has been done, reversible computing remains the most promising and well-developed available avenue towards performing general digital computing in a thermodynamically efficient way.

That being said, exploration of alternative protocols for thermodynamically efficient computing may someday also be worthwhile. One conceptual example of such an alternative is illustrated by \cite{FD16}, in which a chaotic dynamical system whose strange attractor is engineered to encode the state of an extended Boolean circuit is \textit{monolithically} transformed adiabatically from an old state to a new state of the \textit{entire circuit} all at once. This example illustrates that thermodynamically efficient alternatives to performing computations via a sequence of \textit{local} logically reversible transformations \textit{do} in fact exist. This example even still preserves a kind of compositionality, albeit at a different level, in the sense that the \textit{structure} of the extended system is still composed out of local interaction Hamiltonian terms representing individual Boolean logical constraints, even though the \textit{transformation} of the system is done monolithically.

So, the existence of the example of \cite{FD16} could be cited as partial vindication of assertions 
that reversible computing via \textit{local} transformations is not \textit{necessarily} the only way to accomplish digital computation in a thermodynamically efficient way. However, the results of \cite{FD16} appear to suggest that the penalty for doing \textit{non-local} thermodynamically reversible transformations of digital machines is generally to incur an exponential increase in time complexity---which, in retrospect, is not surprising, since otherwise, one could use methods similar \cite{FD16} to solve NP-complete problems using only polynomial resources (by constraining the output of the computation, rather than the input). So, it may still be the case that reversible computing remains the only physically possible way to achieve thermodynamically efficient digital computation with only modest (\textit{i.e.}, polynomial) resource overheads.

We should also note in passing that, apart from \cite{FD16}, many other, more ``analog'' approaches to physical computing with dynamical systems also exist; see \S4.2 of \cite{BC20} for a survey. As with \cite{FD16}, the time-evolution of other, more analog kinds of conservative dynamical systems that may be useful for computing could also conceivably be engineered to approach thermodynamic reversibility, although most existing analog computing schemes have not been specifically designed to do so.

\subsection{Future Directions}
\label{ssec:future-dir}
Although providing a valuable framework both theoretically and for our purposes in modeling reversible computation, the structure of GKSL systems with multiple asymptotic states \cite{Albert14, ABFJ16, Albert18} leaves several questions remaining. These questions are of great interest to the theory of open quantum systems generally, and also offer substantial insights for open quantum systems models of reversible computing. One important question is the question of energy dissipation for generic nonequilibrium protocols. The notion of thermodynamic length has been developed \cite{SPL19} for GKSL systems with \emph{single} asymptotic states. Minimization of this length provides a characterization of the minimal dissipation of the time evolution of a Hamiltonian in an open quantum system. Meanwhile, general thermodynamic uncertainty relations have also been developed \cite{GCLG19} for single asymptotic states. These provide a general set of uncertainty relations between the currents of a system in an asymptotic state and the entropy production rate of the system, in terms of the information geometric metric (\emph{i.e.}, the Fisher information) on nonequilibrium asymptotic states.

Extending these notions to multiple asymptotic states can help us develop expressions of energy dissipation for time evolving systems for multiple asymptotic states. In order to characterize the efficiency of reversible information processing operations, a key figure of merit that we're interested in is the dissipation as a function of delay, $D\left(d\right)$. For reversible computing in an open quantum system, multiple asymptotic state framework, we want to be able to characterize the minimal dissipation for any Hamiltonian we might want to write that can represent the model we’re using for a reversible computer\footnote{This is not the only energetic cost of interest when examining models of computation: we may also be interested in the minimum energy required to perform a computation \cite{GB02} or the \emph{maximum} information cost that an operation can take \cite{Deffner21}. Since we intend to develop our expression for $D\left(d\right)$ for classical RC operations using NEQT (and in particular its quantum information formulation), the expression for $D\left(d\right)$ will serve in concert with these other energetic costs to provide a strong characterization of the energetic constraints of classical reversible and quantum computations.}. This will, in general, be a function of the amount of time (\emph{i.e.}, the \emph{delay}) of the operation.

Fundamentally, both the dissipation and delay will depend quite fundamentally and intrinsically on the underlying geometry of the asymptotic states. From the derivation of an adiabatic (\emph{i.e.}, Berry-like) curvature on the space induced by asymptotic states \cite{ABFJ16, Albert18}, we can immediately anticipate that the expression for thermodynamic length in open quantum systems will depend intricately on this induced curvature. Given that quantum speeds for GKSL systems are expressed through a suitable metric on the information geometry of states \cite{DC17}, we can expect that the delay can be derived similarly.\footnote{To elaborate slightly, note that the dissipation-delay relation is not necessarily directly bounded by the quantum speed limit, which is defined in terms of dynamical energy \textit{invested}, not energy \textit{dissipated}; however, we can expect that the derivation of the delay will still involve considerations of the speed limit, to the extent that dissipation can be bounded as a fraction of the dynamical energy.} By relating the information geometry metric tensor with the QGT, we can derive expressions for the delay and in turn the dissipation in terms of the QGT. This can then provide us with dissipation as a function of delay. 

The development of $D\left(d\right)$ will involve as intermediate steps several quantities which will be of interest to the open quantum systems community as a whole, such as the thermodynamic length and the thermodynamic uncertainty relations for GKSL systems with multiple asymptotic states. These will be essential for understanding properties of open quantum systems which have direct bearing on classical reversible computing operations. The bearing that these properties will have on RC operations \emph{per se} also depends on the particulars of how RC operations are represented in open quantum systems. We have here provided some of the initial groundwork for these representations, but several specific details which are vital for understanding RC operations and the application of open quantum system properties to RC remain to be developed in full. Here, we have identified several key remaining details---specifically, the quantum geometric signature that different RC operations will have, the representations of merges and splits in the framework of GKSL dynamics with multiple asymptotic states, and the specific timescale under which the result of a computational operation dephases to retrieve the DFS sum structure characteristic of classical RC operations. These details are currently being developed in a forthcoming work, but they are likely not an exhaustive list of the specific characteristics of open quantum systems representations of RC operations.


\section{Conclusions}
\label{sec:conc}

At this time, much work remains to complete the task of fully fleshing out a useful physical theory of classical reversible computing based on the tools of modern quantum thermodynamics and quantum information. Our goal, in this article, was to lay some key conceptual foundations for that effort, point the way towards further progress that can be made, and review some important preliminary results.

Our primary conclusion thus far, from this line of work, is that the core insights from the classic theory of the thermodynamics of computing which originally motivated the field of reversible computing, rather than being contradicted by the modern non-equilibrium quantum thermodynamics perspective, are, to the contrary, supported by it.  We argued that the most appropriate understanding of Landauer's Principle is as a statement about the absolute \textit{increase} in total entropy that is required when correlated information is lost, and reviewed the theorems showing that \emph{only} computational operations that are (fully or conditionally) logically reversible can avoid such increases, when applied to subsystems that exhibit correlations to other subsystems, as is normal for subsystems that contain computed information. In addition, we provided a complementary way of seeing conditional and unconditional Landauer reset in terms of catalytic thermal operations, which helps shed some light on the underlying nonequilibrium quantum thermodynamic principles at play distinguishing the reset processes.

Even in its early stages of development, the GKSL dynamical perspective on computational operations in open quantum systems is already showing quite surprising implications, of interest both on a purely theoretical level as well as for applications to reversible computing models. In particular, we see that the quantum geometric properties of the space supporting RC operations play a central role in governing these operations and the dynamics of systems supporting reversible computation. This offers a tantalizing glimpse into the rich geometrical structure which underpins and can help support RC operations, and suggests that the discovery of these signatures \emph{generally} can support reversible computational operations in exotic structures. Much more work remains to be done in teasing out the geometric structure of RC operations, with implications of substantial interest both to reversible computing engineering and the theory of open quantum systems generally.

We should note, in passing, that there is a useful concept of \textit{effective} physical entropy that can be considered to include not just the statistical form of entropy considered here, but also measures of information complexity or \textit{algorithmic randomness} \cite{Zur89}.  One can summarize such concepts by saying that, for pragmatic purposes, physical entropy effectively includes both \textit{unknown} and \textit{known but incompressible} information.  However, such expanded conceptions of entropy do not affect the particular concerns of this paper, since algorithmically random, incompressible information remains equally incompressible regardless of whether the use of logically reversible algorithms is considered.

In conclusion, we see that there is a potentially enormous long-term practical value to be gained through seriously studying the limits and potentialities of physical mechanisms designed to efficiently implement reversible computational processes, with an eye towards making technologies for general digital computing far more efficient. In this article, we have reviewed a number of key theoretical tools from modern non-equilibrium quantum thermodynamics which we believe will be useful for continuing this line of work and investigating the physics of reversible computing in more depth. We intend to continue this effort in future papers, and we invite other interested researchers to join us.




\vspace{6pt} 



\authorcontributions{Conceptualization, Michael Frank and Karpur Shukla; Formal analysis, Michael Frank and Karpur Shukla; Funding acquisition, Michael Frank and Karpur Shukla; Investigation, Michael Frank and Karpur Shukla; Methodology, Michael Frank and Karpur Shukla; Project administration, Michael Frank; Supervision, Michael Frank; Visualization, Michael Frank and Karpur Shukla; Writing – original draft, Michael Frank and Karpur Shukla; Writing – review \& editing, Michael Frank and Karpur Shukla.}

\funding{This research was funded in part by the Laboratory Directed Research and Development (LDRD) and Advanced Simulation and Computing (ASC) programs at Sandia National Laboratories, a multimission laboratory managed and operation by National Technology and Engineering Solutions of Sandia, LLC., a wholly owned subsidiary of Honeywell International, Inc., for the U.S. Department of Energy's National Nuclear Security Administration under contract DE-NA-0003525.
It was also supported in part by the U.S. Army Research Office (ARO) under cooperative agreement W911NF-14-2-0075 and BAA W911NF-19-S-0007, and in part by the U.S. Air Force Office of Scientific Research (AFOSR) under grant number FA9550-19-1-0355. This document describes objective technical results and analysis. Any subjective views or opinions that might be expressed in this document do not necessarily represent the views of the U.S. Department of Energy or the United States Government.
Approved for public release, SAND2021-6489 J.}

\acknowledgments{The authors would like to thank Victor Albert, Neal Anderson, Gavin Crooks, Ed Fredkin, John Goold, Giacomo Guarnieri, David Gu\'ery-Odelin, Norm Margolus, Markus M\"uller, Kevin Osborn, Subhash Pidaparthi, Greg Snider, David Wolpert, and Noboyuki Yoshikawa for helpful discussions, and Jimmy Xu for his support. M.F. would also like to thank Rudro Biswas, Robert Brocato, Erik DeBenedictis, Rupert Lewis, Nancy Missert, and Brian Tierney for their contributions to the reversible computer engineering efforts at Sandia, and Gladys Eden for her love and encouragement. K.S. would also like to thank Hannah Watson for her boundless love and emotional support.}

\conflictsofinterest{The authors declare no conflict of interest. The funders had no role in the design of the study; in the collection, analyses, or interpretation of data; 
in the writing of the manuscript, or in the decision to publish the~results.} 


\pagebreak
\abbreviations{The following abbreviations are used in this manuscript:\\

\noindent 
\begin{tabular}{@{}ll}
AQFP & Adiabatic quantum flux parametron \\
ASIC & Application-specific integrated circuit \\
BARC(S) & Ballistic asynchronous reversible computing (in superconductors) \\
CMOS & Complementary metal-oxide-semiconductor (circuit/technology) \\
CTMC & Continuous-time Markov chain \\
CPTP & Completely positive trace-preserving (map/channel) \\
CTO & Catalytic thermal operation(s) \\
DAG & Directed acyclic graph \\
DPI & Data processing inequality \\
FET & Field-effect transistor \\
GKSL & Gorini-Kossakowski-Sudarshan-Lindblad (operator/theory) \\
LvN & Liouville-von Neumann (equation)\\
NEQT & Non-equilibrium quantum thermodynamics\\
nFET & n-type FET \\
PES & Potential energy surface \\
QC & Quantum computation \\
QCA & Quantum-dot cellular automaton \\
QRD & Quantumr relative divergence \\
QRT & Quantum resource theory \\
QGT & Quantum geometric tensor \\
RA-CMOS & Reversible adiabatic CMOS\\
RC & Reversible computing\\
RNRL & Reversible nanomechanical rod logic \\
R-QCA & Reversible QCA \\
RRE & Relative Rényi entropy \\
RTQT & Resource theory of quantum thermodynamics \\
RQFP & Reversible quantum flux parametron \\
SPICE & Simulation Program with Integrated Circuit Emphasis \\
TO & Thermal operation(s)
\end{tabular}}

\appendixtitles{yes} 
\appendixstart
\appendix
\section{Minimum-Energy Scaling for Classical Adiabatic Technologies}
\label{app:adia-minE}

In this appendix, we briefly present the derivation for the scaling of minimum energy dissipation for reversible technologies such as RA-CMOS (\S\ref{sssec:ra-cmos}) that obey classic adiabatic scaling and that can be characterized in terms of relaxation and equilibration timescales.\footnote{Note that this particular scaling analysis does not extend to families of technologies that may potentially offer some approximation to a Landau-Zener type of \textit{exponential} quantum adiabatic scaling, such as R-QCA (see \S\ref{sssec:r-qca}).}

First, we assume (as is the case for ``perfectly adiabatic'' technologies such as \cite{Fra+20b}) that the total energy dissipation per clock cycle $E_\mathrm{diss}$ in a reversible circuit can be expressed as a sum of \textit{switching losses} and \textit{leakage losses},
\begin{equation}\label{eq:Ediss}
    E_\mathrm{diss} = E_\mathrm{sw} + E_\mathrm{lk},
\end{equation}
and further, that switching and leakage losses depend on the signal energy $E_\mathrm{sig}$ and transition time $t_\mathrm{tr}$ approximately as follows:
\begin{gather}
    E_\mathrm{sw} \simeq E_\mathrm{sig}\cdot c_\mathrm{sw}\cdot\frac{\tau_\mathrm{r}}{t_\mathrm{tr}},\label{eq:Esw}\\
    E_\mathrm{lk} \simeq E_\mathrm{sig}\cdot c_\mathrm{lk}\cdot\frac{t_\mathrm{tr}}{\tau_\mathrm{e}},\label{eq:Elk}
\end{gather}
where $\tau_\mathrm{r},\tau_\mathrm{e}$ are the relaxation and equilibration timescales, respectively, and $c_\mathrm{sw},c_\mathrm{lk}$ are small dimensionless constants characteristic of a particular reversible circuit in a specific family of technologies, such as \cite{Fra+20b}. In practice, although these specific formulas are only approximate, they approach exactness in the regime $\tau_\mathrm{r} \ll t_\mathrm{tr} \ll \tau_\mathrm{e}$.

Then, now treating \eqref{eq:Esw},\eqref{eq:Elk} as exact, we can write:
\begin{equation}\label{eq:Ediss-ttr}
    E_\mathrm{diss} = E_\mathrm{sig} \left( c_\mathrm{sw}\tau_\mathrm{r}\cdot \frac{1}{t_\mathrm{tr}} + \frac{c_\mathrm{lk}}{\tau_\mathrm{e}}\cdot t_\mathrm{tr} \right).
\end{equation}
We can collect the constants, absorbing them into adjusted timescales $\tau_\mathrm{r}^\prime = c_\mathrm{sw}\tau_\mathrm{r}$ and $\tau_\mathrm{e}^\prime = \tau_\mathrm{e}/c_\mathrm{lk}$, so
\begin{equation}\label{eq:Ediss-prime}
    E_\mathrm{diss} = E_\mathrm{sig}\left( \tau_\mathrm{r}^\prime\cdot\frac{1}{t_\mathrm{tr}} + \frac{1}{\tau_\mathrm{e}^\prime}\cdot t_\mathrm{tr} \right).
\end{equation}
Setting the derivative of \eqref{eq:Ediss-prime} with respect to $t_\mathrm{tr}$ equal to zero, we find that $E_\mathrm{diss}$ is minimized when
\begin{equation}
    \tau_\mathrm{r}^\prime\frac{1}{t_\mathrm{tr}^2} =
    \frac{1}{\tau_\mathrm{e}^\prime},
\end{equation}
or in other words, when
\begin{equation}
    t_\mathrm{tr} = \sqrt{\tau_\mathrm{r}^\prime\tau_\mathrm{e}^\prime},
\end{equation}
at which point $E_\mathrm{sw}$ and $E_\mathrm{lk}$ are equal. The minimum energy dissipation per cycle is then
\begin{equation}
    E_\mathrm{diss}=2E_\mathrm{sig}\sqrt{\frac{\tau_\mathrm{r}^\prime}{\tau_\mathrm{e}^\prime}}.
\end{equation}

Thus, for any given reversible circuit design in a family of technologies with given values of the constants $c_\mathrm{sw},c_\mathrm{lk}$, in order for $E_\mathrm{diss}$ to approach 0 as the technology develops, we must have that the ratio of equilibration/relaxation timescales $\tau_\mathrm{e}/\tau_\mathrm{r}\rightarrow\infty$, and, if the relaxation timescale $\tau_\mathrm{r}$ is fixed, this implies that also the (minimum-energy) value of the transition time $t_\mathrm{tr}\rightarrow\infty$. These requirements were mentioned in \S\ref{sssec:ra-cmos}.

More specifically, in order to increase the peak energy efficiency of a reversible circuit by a factor of $N\times$, in a given family of technologies obeying classic adiabatic scaling, this requires that the timescale ratio $\tau_\mathrm{e}/\tau_\mathrm{r}$ must be increased by $N^2\times$, and (assuming $\tau_\mathrm{r}$ is fixed) the transition time $t_\mathrm{tr}$ for minimum energy will increase by $N\times$.

\section{Vectorization of the Operator Algebra on Quantum States}
\label{app:vectorization}
In ordinary quantum mechanics, expressions such as \eqref{eq:LvN-formal-sol}, \eqref{eq:GKSL-def}), and \eqref{eq:exp-Madhava} can be easily solved using operator algebra techniques. This same principle holds for operators which operate on other operators (\emph{e.g.}, $\hat{\hat{\mathcal{L}}}$ operates on density matrices $\rho\in\mathcal{D}\left(\mathcal{H}_{\mathfrak{S}}\right)$), known as \emph{superoperators}. Since the space of $L^{2}$-bounded operators $\mathcal{B}\left(\mathcal{H}_{\mathfrak{S}}\right)$ forms a Hilbert space in its own right under the Hilbert-Schmidt inner product $\left\langle\widehat{A},\widehat{B}\right\rangle\coloneqq\mathrm{Tr}\left[\widehat{A}^{\dagger}\widehat{B}\right]$, the exact same operator algebra techniques can be applied to examining superoperators.

A simple way of explicitly writing down these techniques for the $L^{2}$-bounded operators is using the process of \emph{vectorization}, which we very briefly discuss here following the excellent presentation in \cite{Albert18}. Succinctly, vectorization is the process of rewriting matrices in $\mathbb{F}^{m\:\times\:n}$ as vectors in $\mathbb{F}^{mn}$ via the mapping $\ket{v}\bra{w}\mapsto\ket{v}\otimes\ket{w}$. (Here, $\mathbb{F}$ is the field ($\mathbb{R}$ or $\mathbb{C}$) that the matrices live over.) In terms of a basis $\left\{\ket{b_{i}}\right\}_{i\:=\:1}^{N}$ of $\mathcal{H}_{\mathfrak{S}}$, the vectorization of an operator $\widehat{A}\in\mathcal{B}\left(\mathcal{H}_{\mathfrak{S}}\right)$ appears as:
\begin{equation}
    \label{eq:vectorization}
    \widehat{A} = \sum\limits_{i}c_{i}\ket{b_{i}} \mapsto \left. \ket{A}\!\right\rangle \coloneqq \bigotimes\limits_{i}c_{i}\ket{b_{i}}.
\end{equation}
As a concrete example, the vectorization of a $2\times2$ matrix appears as:
\begin{equation*}
    \begin{pmatrix}
        a & b \\
        c & d
    \end{pmatrix}
    \mapsto
    \begin{pmatrix}
        a \\
        c \\
        b \\
        d \\
    \end{pmatrix}.
\end{equation*}
The ``double-ket'' notation for vectorized matrices acts largely the same as the familiar Dirac notation:\footnote{Although our intuitions from Dirac notation directly carry over to the ``double-ket'' notation, translating back and forth from the vectorized expressions to the operator expressions can be somewhat nontrivial. Care must be taken when doing so, although discussing these difficulties is beyond the scope of this paper.}
\begin{itemize}[labelindent=0mm,labelsep=1.25mm,leftmargin=*]
    \item Superoperators $\hat{\hat{\mathcal{O}}}$ act on operators $\widehat{A}$ as $\hat{\hat{\mathcal{O}}}\left.\ket{\widehat{A}}\!\!\right\rangle = \left.\ket{\mathcal{O}\left[\widehat{A}\right]}\!\!\right\rangle$. 
    \item The Hermitian adjoint of $\left.\ket{\widehat{A}}\!\!\right\rangle$ is $\left.\ket{\widehat{A}}\!\!\right\rangle^{\ddagger} = \left\langle\!\!\bra{\widehat{A}}\right.$.
    \item The Hermitian adjoint of $\hat{\hat{\mathcal{O}}}\left.\ket{\widehat{A}}\!\!\right\rangle$ is given by $\hat{\hat{\mathcal{O}}}\left.\ket{\widehat{A}}\!\!\right\rangle = \left\langle\!\!\bra{\widehat{A}}\right.\hat{\hat{\mathcal{O}}}^{\ddagger} = \left\langle\!\!\bra{\hat{\hat{\mathcal{O}}}^{\ddagger}\left[\widehat{A}\right]}\right.$.
    \item The Hilbert-Schmidt inner product is given by $\left\langle\!\!\Braket{\widehat{A}|\widehat{B}}\!\!\right\rangle \coloneqq \mathrm{Tr}\left[\widehat{A}^{\dagger}\widehat{B}\right]$.
    \begin{itemize}[labelindent=0mm,labelsep=1.25mm,leftmargin=*]
        \item Thus, the trace of $\widehat{A}$ is given by $\left\langle\!\!\Braket{\mathbbm{1}|\widehat{A}}\!\!\right\rangle$.
    \end{itemize}
    \item The basis $\left\{\ket{b_{i}}\right\}_{i\:=\:1}^{N}$ of $\mathcal{B}\left(\mathcal{H}_{\mathfrak{S}}\right)$ gives a corresponding basis of $\mathcal{B}\left(\mathcal{H}_{\mathfrak{S}}\right)$: $\left.\ket{\mathsf{b}_{ij}}\!\!\right\rangle = \ket{b_{i}}\bra{b_{j}}$.
        \begin{itemize}[labelindent=0mm,labelsep=1.25mm,leftmargin=*]
        \item From this structure, changing the basis of $\mathcal{H}_{\mathfrak{S}}$ changes the basis of $\mathcal{B}\left(\mathcal{H}_{\mathfrak{S}}\right)$, and thus, the explicit decompositions of the vectorized operators $\left.\ket{\widehat{A}}\!\!\right\rangle \in \mathcal{B}\left(\mathcal{H}_{\mathfrak{S}}\right)$ and the superoperators $\hat{\hat{\mathcal{O}}}$.
        \\[8pt]
        However, the basis change in $\mathcal{H}_{\mathfrak{S}}$ directly reflects a basis change in $\mathcal{B}\left(\mathcal{H}_{\mathfrak{S}}\right)$: transforming $\ket{b_i} \mapsto \ket{c_i}$ directly corresponds to the transformation $\left.\ket{\mathsf{b}_{ij}}\!\!\right\rangle = \ket{b_{i}}\bra{b_{j}} \mapsto \left.\ket{\mathsf{c}_{ij}}\!\!\right\rangle = \ket{c_{i}}\bra{c_{j}}$. Thus, we don’t need any ``extra'' information in the transformation: everything can be expressed entirely in terms of what lives in $\mathcal{B}\left(\mathcal{H}_{\mathfrak{S}}\right)$, without needing to further reference $\mathcal{H}_{\mathfrak{S}}$.
        \end{itemize}
        \item An additional complication that doesn't appear with ordinary Hilbert spaces is the operator algebra structure $\mathcal{B}\left(\mathcal{H}_{\mathfrak{S}}\right)\times\mathcal{B}\left(\mathcal{H}_{\mathfrak{S}}\right)\rightarrow\mathcal{B}\left(\mathcal{H}_{\mathfrak{S}}\right)$; thus, we need to describe the vectorized version of matrix multiplication. Explicitly, the vectorized product of the operators $\widehat{A}$, $\widehat{B}$, and $\widehat{C}$ is $\left.\ket{\widehat{A}\widehat{B}\widehat{C}}\!\!\right\rangle = \left(\widehat{C}^{\top}\otimes{\widehat{B}}\right)\left.\ket{\widehat{A}}\!\!\right\rangle$.
\end{itemize}

In vectorized form, the GKSL equation \eqref{eq:GKSL-def} appears as:
\end{paracol}
\begin{equation}
    \label{eq:vectorized-GKSL}
    \hat{\hat{\mathcal{L}}}\left.\ket{\rho_{\mathfrak{S}}}\!\right\rangle = \left[-\mathrm{i}\left(\mathbbm{1}\otimes\widehat{H}_{\mathfrak{S}}-\widehat{H}_{\mathfrak{S}}^{\top}\otimes\mathbbm{1}\right)+\frac{1}{2}\sum_{a,\:b\:>\:0}\kappa_{ab}\left(2\widehat{F}_{ab}^{\ast}\otimes\widehat{F}_{ab}-\mathbbm{1}\otimes\widehat{F}_{ab}^{\dagger}\widehat{F}_{ab}-\left(\widehat{F}_{ab}^{\dagger}\widehat{F}_{ab}\right)^{\top}\otimes\mathbbm{1}\right)\right]\left.\ket{\rho_{\mathfrak{S}}}\!\right\rangle.
\end{equation}
\begin{paracol}{2}
\switchcolumn
\noindent Meanwhile, the vectorized form of the adjoint GKSL equation \eqref{eq:GKSL-adj-def}) is given by:
\end{paracol}
\begin{equation}
    \label{eq:vectorized-adj-GKSL}
    \hat{\hat{\mathcal{L}}}^{\ddagger}\left.\ket{\widehat{A}}\!\right\rangle = \left[i\left(\mathbbm{1}\otimes\widehat{H}_{\mathfrak{S}}-\widehat{H}_{\mathfrak{S}}^{\top}\otimes\mathbbm{1}\right)+\frac{1}{2}\sum_{a,\:b\:>\:0}\kappa_{ab}\left(2\widehat{F}_{ab}\otimes\widehat{F}_{ab}^{\ast}-\mathbbm{1}\otimes\widehat{F}_{ab}^{\dagger}\widehat{F}_{ab}-\left(\widehat{F}_{ab}^{\dagger}\widehat{F}_{ab}\right)^{\top}\otimes\mathbbm{1}\right)\right] \left.\ket{\widehat{A}}\!\!\right\rangle.
\end{equation}
\begin{paracol}{2}
\switchcolumn
\noindent We can also derive the formal solution to the adjoint evolution equation by examining $\mathrm{Tr}\left[\widehat{A}^{\dagger}\left(0\right)\,\rho_{\mathfrak{S}}\left(t\right)\right] = \left\langle\!\!\Braket{\widehat{A}^{\dagger}\left(0\right)|\rho_{\mathfrak{S}}\left(t\right)}\!\!\right\rangle$:
\begin{equation}
    \label{eq:vect-op-ev}
    \begin{split}
        \mathrm{Tr}\left[\widehat{A}^{\dagger}\left(0\right)\:\rho_{\mathfrak{S}}\left(t\right)\right] &= \left\langle\!\!\Braket{\widehat{A}^{\dagger}\left(0\right)|\rho_{\mathfrak{S}}\left(t\right)}\!\!\right\rangle = \left\langle\!\!\!\Braket{\widehat{A}^{\dagger}\left(0\right)|\mathrm{e}^{t\hat{\hat{\mathcal{L}}}}|\rho_{\mathfrak{S}}\left(t\right)}\!\!\!\right\rangle \\[4pt]
        &= \left\langle\!\!\!\Braket{\mathrm{e}^{t\hat{\hat{\mathcal{L}}}^{\ddagger}}\widehat{A}^{\dagger}\left(0\right)|\rho_{\mathfrak{S}}\left(t\right)}\!\!\!\right\rangle = \left\langle\!\!\Braket{\widehat{A}^{\dagger}\left(t\right)|\rho_{\mathfrak{S}}\left(t\right)}\!\!\right\rangle.
    \end{split}
\end{equation}
Here, since we have $\left\langle\!\!\bra{\widehat{A}^{\dagger}\left(t\right)}\right. = \left\langle\!\!\bra{\widehat{A}^{\dagger}\left(0\right)}\right.\mathrm{e}^{t\hat{\hat{\mathcal{L}}}}$ and $\mathrm{e}^{t\hat{\hat{\mathcal{L}}}}\,\left.\ket{\widehat{A}\left(0\right)}\!\!\right\rangle = \left.\ket{\widehat{A}\left(t\right)}\!\!\right\rangle$, we have $\left.\ket{\widehat{A}\left(t\right)}\!\!\right\rangle$ satisfying the differential equation:
\begin{equation}
    \label{eq:adjoint-LvN-eq}
    \frac{\mathrm{d}\left.\ket{\widehat{A}\left(t\right)}\!\!\right\rangle}{\mathrm{d}t} = \hat{\hat{\mathcal{L}}}\,\left.\ket{\widehat{A}\left(t\right)}\!\!\right\rangle
\end{equation}
Finally, the vectorized solutions to the differential equation \eqref{eq:GKSL-def} for time-dependent and time-independent GKSL superoperators are, respectively:
\begin{subequations}
    \label{eq:GKSL-vec-sol}
    \begin{align}
        \label{eq:GKSL-vec-sol-time-dep}
        \left.\ket{\rho_\mathfrak{S}\left(t\right)}\!\right\rangle &= \mathcal{T}\:\mathrm{exp}\left\{\int\limits_{t_{0}}^{t}\mathrm{d}t'\:\hat{\hat{\mathcal{L}}}\left(t'\right)\right\} \left.\ket{\rho_\mathfrak{S}\left(t_{0}\right)}\!\right\rangle \\[4pt]
        \label{eq:GKSL-vec-sol-time-ind}
        \left.\ket{\rho_\mathfrak{S}\left(t\right)}\!\right\rangle &= \mathrm{e}^{t\hat{\hat{\mathcal{L}}}} \left.\ket{\rho_\mathfrak{S}\left(t_{0}\right)}\!\right\rangle.
    \end{align}
\end{subequations}
These expressions are the same as \eqref{eq:LvN-formal-sol} and $\rho\left(t\right) = \mathrm{e}^{t\hat{\hat{\mathcal{L}}}}\rho\left(t_{0}\right)$, but crucially they now allow us to examine the spectral decomposition of (and analytic functions of) $\hat{\hat{\mathcal{L}}}$.

\end{paracol}

\reftitle{References}




%

\end{document}